\documentclass[onecolumn,superscriptaddress,nofootinbib,notitlepage,aps]{revtex4-1}
\pdfoutput=1

\usepackage{graphicx}
\usepackage{amsmath}
\usepackage{amssymb}
\usepackage{amsfonts}
\usepackage{dcolumn}
\usepackage{bm}
\usepackage[linktoc=none]{hyperref}
\usepackage[utf8]{inputenc}
\usepackage{subfigure}

\usepackage[T1]{fontenc}
\usepackage{pstricks}
\usepackage{color}
\usepackage{multirow}
\usepackage{slashed}
\usepackage{mathtools}

\usepackage{feynmp}
\usepackage{feynmp-auto}
\hypersetup{colorlinks,linkcolor={blue},citecolor={teal},urlcolor={violet}}

\newcommand{\GRAPPA}{%
Gravitation Astroparticle Physics Amsterdam (GRAPPA),\\
Institute for Theoretical Physics Amsterdam
and Delta Institute for Theoretical Physics,\\
University of Amsterdam, Science Park 904, 1098 XH Amsterdam, The Netherlands}

\newcommand{\SOTON}{Department of Physics and Astronomy, University of Southampton, SO17 1BJ Southampton, United Kingdom}

\begin{document}

\title{Interplay between neutrino and gravity portals for FIMP dark matter}

\author{Marco Chianese}
\email{m.chianese@uva.nl}
\affiliation{\GRAPPA}

\author{Bowen Fu}
\email{B.Fu@soton.ac.uk}
\affiliation{\SOTON}

\author{Stephen F. King}
\email{king@soton.ac.uk}
\affiliation{\SOTON}

\date{\today}

\begin{abstract}
In the classic type I seesaw mechanism with very heavy right-handed (RH) neutrinos, it is possible to account for dark matter via RH neutrino portal couplings to a feebly interacting massive particle (FIMP) dark sector. However, for large RH neutrino masses, gravity can play an important role. We study the interplay between the neutrino portal through the right-handed neutrinos and the gravity portal through the massless spin-2 graviton in producing dark matter particles in the early universe. As a concrete example, we consider the minimal and realistic Littlest Seesaw model with two RH neutrinos, augmented with a dark scalar and a dark fermion charged under a global $U(1)_D$ dark symmetry. In the model, the usual seesaw neutrino Yukawa couplings and the right-handed neutrino masses (the lightest being about $5\times 10^{10}$ GeV) are fixed by neutrino oscillations data and leptogenesis. Hence, we explore the parameter space of the two RH neutrino portal couplings, the two dark particle masses and the reheating temperature of the universe, where the correct dark matter relic abundance is achieved through the freeze-in mechanism. In particular, we highlight which class of processes dominate the dark matter production. We find that, despite the presence of the gravity portal, the dark matter production relies on the usual seesaw neutrino Yukawa coupling in some regions of the parameter space, so realising a direct link between dark matter and neutrino phenomenology. Finally, we report the threshold values for the neutrino portal couplings below which the neutrino portal is irrelevant and the Planckian Interacting Dark Matter paradigm is preserved.
\end{abstract}

\maketitle

\tableofcontents

\section{Introduction \label{sec:intro}}

One of the most convincing indications of new particle physics beyond the Standard Model (SM) are the masses of neutrinos and their mixing, which is evidenced by the neutrino oscillation experiments~\cite{2016NuPhB.908....1O}. Many theories have been developed to explain the theoretical origin of the neutrino masses, among which the classic type I seesaw mechanism with very heavy right-handed (RH) neutrinos is the most well-known one~\cite{Minkowski:1977sc,Yanagida:1979as,GellMann:1980vs,Schechter:1980gr,Mohapatra:1979ia,Mohapatra:1980yp}. However, there are so many free parameters in the most general version of the type I seesaw model that it is difficult to be tested by experiments due to lack of enough physical constraints. An effective solution to reduce the number of free parameters is provided by the Littlest Seesaw (LS) model with one texture zero, two RH neutrinos and a particular version of constrained sequential dominance (CSD)~\cite{King:2013iva,Bjorkeroth:2014vha,King:2015dvf,Bjorkeroth:2015ora,Bjorkeroth:2015tsa,King:2016yvg,Ballett:2016yod,King:2018fqh}. Indeed, it has been shown that the neutrino sector defined by the usual seesaw Yukawa couplings and the right-handed neutrino masses is completely determined by the neutrino oscillation data and the baryon asymmetry of the universe accounted for by leptogenesis~\cite{King:2018fqh}.

Another indication of new physics beyond the SM is the existence of dark matter (DM), which accounts for about 25\% of the energy density of the universe~\cite{Aghanim:2018eyx}. Many works have explored the possibility to include the DM and the neutrino mass and mixing into one framework~\cite{Caldwell:1993kn,Mohapatra:2002ug,Krauss:2002px,Ma:2006km,Asaka:2005an,Boehm:2006mi,Kubo:2006yx,Ma:2006fn,Hambye:2006zn,Lattanzi:2007ux,Ma:2007gq,Allahverdi:2007wt,Gu:2007ug,Sahu:2008aw,Arina:2008bb,Aoki:2008av,Ma:2008cu,Gu:2008yj,Aoki:2009vf,Gu:2010yf,Hirsch:2010ru,Esteves:2010sh,Kanemura:2011vm,Lindner:2011it,JosseMichaux:2011ba,Schmidt:2012yg,Borah:2012qr,Farzan:2012sa,Chao:2012mx,Gustafsson:2012vj,Blennow:2013pya,Law:2013saa,Hernandez:2013dta,Restrepo:2013aga,Chakraborty:2013gea,Ahriche:2014cda,Kanemura:2014rpa,Huang:2014bva,Varzielas:2015joa,Sanchez-Vega:2015qva,Fraser:2015mhb,Adhikari:2015woo,Ahriche:2016rgf,Sierra:2016qfa,Lu:2016ucn,Batell:2016zod,Ho:2016aye,Escudero:2016ksa,Bonilla:2016diq,Borah:2016zbd,Biswas:2016yan,Hierro:2016nwm,Bhattacharya:2016qsg,Chakraborty:2017dfg,Bhattacharya:2017sml,Ho:2017fte,Ghosh:2017fmr,Nanda:2017bmi,Narendra:2017uxl,Bernal:2017xat,Borah:2018gjk,Batell:2017cmf,Pospelov:2007mp,Falkowski:2009yz,Falkowski:2011xh,Cherry:2014xra,Bertoni:2014mva,Allahverdi:2016fvl,Karam:2015jta,Bhattacharya:2018ljs,Biswas:2018sib,Gehrlein:2019iwl,Hashiba:2019mzm,Dasgupta:2019lha,Samanta:2020gdw}. One interesting possibility is the so-called neutrino portal scenario where the dark sector is connected to the Standard Model (SM) through the RH neutrinos that realise the type I seesaw~\cite{Chianese:2018dsz,Chianese:2019epo,Becker:2018rve,Bian:2018mkl,Bandyopadhyay:2018qcv,Liu:2020mxj,Cosme:2020mck, Du:2020avz,Bandyopadhyay:2020qpn}. In some recent literature~\cite{Chianese:2018dsz,Chianese:2019epo}, the RH neutrino portal is achieved in a minimal extension of the LS model, including a dark scalar and a dark fermion, both charged under an additional global $U(1)_{D}$ symmetry.\footnote{Actually the model in~\cite{Chianese:2018dsz} suggested a dark $Z_2$ parity but in fact the restricted couplings considered there and here imply a larger global $U(1)_{D}$ symmetry to forbid Majorana couplings of the dark fermion.} In such a model it has been pointed out that heavy dark matter particles can be dominantly produced through the usual neutrino Yukawa interactions in the "freeze-in" scenario involving a feebly interacting massive particle (FIMP) sector~\cite{McDonald:2001vt,Hall:2009bx,Bernal:2017kxu}. Hence, we have a direct and very intriguing link between neutrinos and heavy FIMP dark matter commonly referred to as ``FIMPzilla'', since the dark matter mass may be very heavy.

On the other hand, it has been pointed out that very heavy dark matter can be also efficiently produced by gravity-mediated processes in the so-called Planckian Interacting Dark Matter (PIDM) paradigm~\cite{Garny:2015sjg,Tang:2016vch,Tang:2017hvq,Garny:2017kha,Bernal:2018qlk,Garny:2018grs,Hashiba:2018tbu}. In these effective models, the massless spin-2 graviton couples the stress-energy tensors of SM and DM particles.\footnote{The case of massive spin-2 mediators has been also examined for standard WIMP dark matter~\cite{Lee:2013bua,Lee:2014caa,Kraml:2017atm,Carrillo-Monteverde:2018phy,Kang:2020huh,Kang:2020yul,Kang:2020afi}.} Such a gravitational DM production occurring after the inflationary stage only depends on the DM mass and the reheating dynamics, which is simply parametrized by the reheating temperature $T_\mathrm{RH}$ in the instantaneous reheating scenario. In particular, there exists a peculiar relation between the DM mass and the reheating temperature to obtain the measured DM relic abundance. Very high reheating temperatures are excluded due to the gravitational overproduction of DM particles. Moreover, the gravity-mediated production in general provides constraints to the other potential interactions existing between the SM and the dark sector. For example, the impact of the gravity portal on the Higgs portal coupling in a minimal scalar dark matter model has been recently discussed in Ref.~\cite{Chianese:2020yjo}. 

In this paper, we consider dark matter in the classic type I seesaw mechanism, and investigate the interplay between the RH neutrino portal and the Planck suppressed higher dimensional operators for FIMP dark matter candidates. With this aim, we consider as a concrete example the Littlest Seesaw model augmented with a dark sector as discussed in Ref.s~\cite{Chianese:2018dsz,Chianese:2019epo}. We assume the dark fermion to be the dark matter candidate and we identify two distinct mass orderings corresponding to different allowed decay processes: dark scalar mass heavier than the RH neutrino masses (``heavier dark scalar'' mass ordering) and dark scalar mass lighter than the RH neutrino masses ( ``lighter dark scalar'' mass ordering). Hence, we exhaustively explore the parameter space of the model providing a viable dark matter candidate, and we highlight which of the two portals dominates the DM production. Hence, we characterise the impact of the gravity portal on the neutrino--dark matter relation realised through the neutrino Yukawa interactions. On the other hand, we determine the threshold values for the neutrino portal couplings below which the gravity portal is the dominant interaction and the minimal PIDM scenario of only-gravitationally interacting DM particles is achieved.

The paper is organized as follows. In Section~\ref{sec:mod}, we briefly review the general model linking the Littlest seesaw model to the dark matter. In Section~\ref{sec:DMpro}, we derive the Boltzmann equations which are required to solve the dark matter production in the freeze-in regime. In Section~\ref{sec:TRH}, we discuss the upper limit on reheating temperature imposed by the gravity-mediated production. In Section~\ref{sec:results}, we report the numerical results for two different mass orderings and the threshold values for the neutrino portal couplings. Finally, we draw our conclusions in Section~\ref{sec:con}.

%%%%%%%%%%%%%%%%%%%%%
\section{The model \label{sec:mod}}
%%%%%%%%%%%%%%%%%%%%%

The minimal model discussed here includes two right-handed neutrinos $N_{\mathrm{R}i}$ and a $U(1)_{D}$-charged dark sector (DS) consisting of a singlet complex scalar $\phi$ and a singlet Dirac fermion $\chi$. The dark global symmetry $U(1)_{D}$ assures the stability of the lightest dark particle which acts as dark matter. For the sake of concreteness, we consider the dark fermion to be the dark matter by assuming $m_\chi < m_\phi$. The irreducible representations of the new matter fields under the symmetries of the model are summarized in Tab.~\ref{tab:matter}. The full Lagrangian can be written as
\begin{equation}
\mathcal{L} = \mathcal{L}_{\rm SM} + \mathcal{L}_{\rm EH} + \mathcal{L}_{\rm DS} + \mathcal{L}_{\rm Seesaw} + \mathcal{L}_{\rm Neutrino~portal}  + \mathcal{L}_{\rm Gravity~portal} + \mathcal{L}_{\rm Higgs~portal}\,.
\label{eq:lag}
\end{equation}
\begin{table}[t!]
\centering
\begin{tabular}{|c|c|c|c|}
\hline 
& $N_{\mathrm{R}i}$ & $\phi$ & $\chi$ \\ \hline \hline 
$SU(2)_L$ & {\bf 1} & {\bf 1} & {\bf 1} \\ \hline
$U(1)_Y$ & 0 & 0 & 0 \\ \hline  \hline
$U(1)_D$ & 0 & 1 & 1 \\ \hline
\end{tabular}
\caption{\label{tab:matter}Irreducible representations of the new fields of the model under the electroweak $SU(2)_L\times U(1)_Y$ gauge symmetry and the global $U(1)_{D}$ dark symmetry. The fields $N_{\mathrm{R}i}$ are the two right-handed neutrinos, while $\phi$ and $\chi$ are a dark complex scalar and dark Dirac fermion, respectively.}
\end{table}
The first three terms are, respectively, the Standard Model Lagrangian, the Einstein-Hilbert Lagrangian and the free Lagrangian of the dark scalar and fermionic fields. The latter defines the masses of the dark scalar and the dark fermion to be $m_\phi$ and $m_\chi$, respectively. The remaining terms are
\begin{eqnarray}
\mathcal{L}_{\rm seesaw} & = & - Y_{\alpha i} \overline{L_L}_\alpha \tilde{H} N_{Ri} - \frac12 M_{Rij}\overline{N^c_{Ri}} N_{Rj} + {\rm h.c.}\,, 
\label{eq:lagNS} \\
\mathcal{L}_{\rm Neutrino~portal} & = & y_{i} \phi \, \overline{\chi}N_{Ri} + {\rm h.c.} \,, 
\label{eq:lagPortal}\\
\mathcal{L}_{\rm Gravity~portal} &=& \frac{\sqrt{8\pi}}{2\,M_\mathrm{P}} h^{\mu\nu}(T^{\rm SM}_{\mu\nu} + T^\chi_{\mu\nu} + T^\phi_{\mu\nu} + T^{N_R}_{\mu\nu})\,,
\label{eq:laggr}\\
\mathcal{L}_{\rm Higgs~portal} &= & \lambda_{\rm  \phi H}  \left| \phi \right|^2  \left| H \right|^2\,.
\label{eq:lagh}
\end{eqnarray}
The first term generates neutrino masses and mixings through the standard seesaw type-I mechanism~\cite{Minkowski:1977sc,Yanagida:1979as,GellMann:1980vs,Schechter:1980gr,Mohapatra:1979ia,Mohapatra:1980yp}. For the sake of minimality, we consider the Littlest seesaw model~\cite{King:2013iva, Bjorkeroth:2014vha, King:2015dvf,Bjorkeroth:2015ora,Bjorkeroth:2015tsa,King:2016yvg,Ballett:2016yod,King:2018fqh} where the Yukawa couplings $Y_{\alpha i}$ are completely defined by three parameters only. In particular, we have
\begin{equation}
Y = \frac{a}{v_\mathrm{SM}}  \left(\begin{array}{cc} 0 & 0 \\ 1 & 0 \\ 1 & 0\end{array}\right) + \frac{b~e^{i \frac{\eta}{2}}}{v_\mathrm{SM}}  \left(\begin{array}{cc} 0 & 1 \\ 0 & 3  \\ 0 & 1\end{array}\right)\,,
\label{eq:yuk}
\end{equation}
where $v_\mathrm{SM} = 174~\mathrm{GeV}$ is the v.e.v. of the SM Higgs field $H$. The parameters of the Littlest seesaw model given by Eq.~\eqref{eq:lagNS} are completely determined by the current neutrino oscillation data~\cite{Capozzi:2018ubv,Capozzi:2020qhw,Gariazzo:2018pei,deSalas:2020pgw,Esteban:2018azc} (taking the lightest active neutrino to be massless) and by requiring a successful leptogenesis~\cite{Davidson:2008bu}. According to the analysis reported in Ref.~\cite{King:2018fqh}, we take the following benchmark values
\begin{equation}
a = 1.42~{\rm GeV} \qquad \text{and}  \qquad b = 37.4~{\rm GeV} \qquad \text{and}  \qquad \eta = 2\pi/3 \,,
\end{equation}
defining the Yukawa couplings in Eq.~\eqref{eq:yuk} and
\begin{equation}
M_1=5.10\times 10^{10} \, \text{GeV}  \qquad \text{and} \qquad M_2=3.28\times 10^{14} \, \text{GeV} \,,
\label{eq:RHmass}
\end{equation}
for the two right-handed neutrino Majorana masses. The structure in Eq.~\eqref{eq:yuk} may be achieved in various ways, but here 
we shall simply assume these couplings and parameters as a benchmark seesaw model that describes the data.

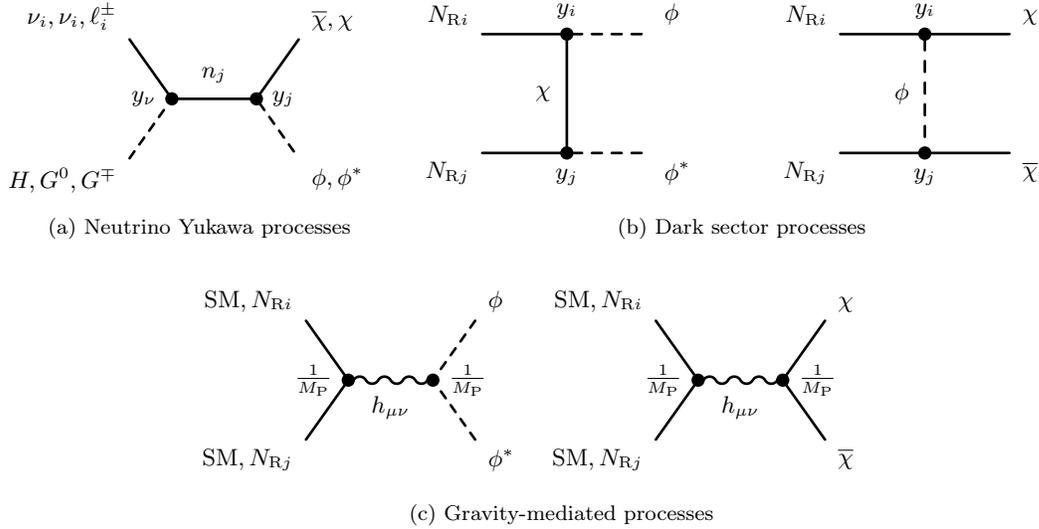
\begin{figure}[t!]
\begin{center}
\subfigure[~Neutrino Yukawa processes]{
\begin{fmffile}{NS1}
\fmfframe(23,18)(18,18){
\begin{fmfgraph*}(80,45)
\fmflabel{$\phi,\phi^*$}{o1}
\fmflabel{$\overline{\chi},\chi$}{o2}
\fmflabel{$\nu_i,\nu_i,\ell_i^\pm$}{i2}
\fmflabel{$H,G^0,G^\mp$}{i1}
\fmfv{label=$y_{j}$}{v2}
\fmfv{label=$y_\nu$}{v1}
\fmfleft{i1,i2}
\fmfright{o1,o2}
\fmf{plain}{i2,v1}
\fmf{dashes}{v2,o1}
\fmf{plain}{v2,o2}
\fmf{dashes}{i1,v1}
\fmf{plain,label=$n_j$}{v1,v2}
\fmfdotn{v}{2}
\end{fmfgraph*}}
\end{fmffile}}
%%%%%%%%%
\subfigure[~Dark sector processes]{
\begin{fmffile}{DS1}
\fmfframe(20,20)(20,20){
\begin{fmfgraph*}(80,45)
\fmflabel{$\phi^*$}{o1}
\fmflabel{$\phi$}{o2}
\fmflabel{$N_{\mathrm{R} j}$}{i1}
\fmflabel{$N_{\mathrm{R} i}$}{i2}
\fmfv{label=$y_{i}$}{v1}
\fmfv{label=$y_{j}$}{v2}
\fmfleft{i1,i2}
\fmfright{o1,o2}
\fmf{plain}{i2,v1}
\fmf{dashes}{v1,o2}
\fmf{plain}{i1,v2}
\fmf{dashes}{v2,o1}
\fmf{plain,label=$\chi$,tension=0}{v1,v2}
\fmfdotn{v}{2}
\end{fmfgraph*}
\label{fig:feynA}}
\end{fmffile}
\begin{fmffile}{DS2}
\fmfframe(20,20)(20,20){
\begin{fmfgraph*}(80,45)
\fmflabel{$\overline{\chi}$}{o1}
\fmflabel{$\chi$}{o2}
\fmflabel{$N_{\mathrm{R} j}$}{i1}
\fmflabel{$N_{\mathrm{R} i}$}{i2}
\fmfv{label=$y_{i}$}{v1}
\fmfv{label=$y_{j}$}{v2}
\fmfleft{i1,i2}
\fmfright{o1,o2}
\fmf{plain}{i2,v1}
\fmf{plain}{v1,o2}
\fmf{plain}{i1,v2}
\fmf{plain}{v2,o1}
\fmf{dashes,label=$\phi$,tension=0}{v1,v2}
\fmfdotn{v}{2}
\end{fmfgraph*}}
\end{fmffile}
\label{fig:feynB}}
%%%%%%%%%%%%%
\subfigure[~Gravity-mediated processes]{
\begin{fmffile}{G1}
\fmfframe(20,20)(20,20){
\begin{fmfgraph*}(80,45)
\fmflabel{$\phi^*$}{o1}
\fmflabel{$\phi$}{o2}
\fmflabel{${\rm SM}, N_{\mathrm{R} i}$}{i2}
\fmflabel{${\rm SM}, N_{\mathrm{R} j}$}{i1}
\fmfv{label=$\tfrac{1}{M_\mathrm{P}}$}{v1}
\fmfv{label=$\tfrac{1}{M_\mathrm{P}}$}{v2}
\fmfleft{i1,i2}
\fmfright{o1,o2}
\fmf{plain}{i2,v1}
\fmf{plain}{i1,v1}
\fmf{dashes}{o1,v2}
\fmf{dashes}{o2,v2}
\fmf{photon,label=$h_{\mu\nu}$}{v1,v2}
\fmfdotn{v}{2}
\end{fmfgraph*}}
\end{fmffile}
%%%%%%
\begin{fmffile}{G2}
\fmfframe(20,20)(20,20){
\begin{fmfgraph*}(80,45)
\fmflabel{$\overline{\chi}$}{o1}
\fmflabel{$\chi$}{o2}
\fmflabel{${\rm SM}, N_{\mathrm{R} i}$}{i2}
\fmflabel{${\rm SM}, N_{\mathrm{R} j}$}{i1}
\fmfv{label=$\tfrac{1}{M_\mathrm{P}}$}{v1}
\fmfv{label=$\tfrac{1}{M_\mathrm{P}}$}{v2}
\fmfleft{i1,i2}
\fmfright{o1,o2}
\fmf{plain}{i2,v1}
\fmf{plain}{i1,v1}
\fmf{plain}{v2,o1}
\fmf{plain}{v2,o2}
\fmf{photon,label=$h_{\mu\nu}$}{v1,v2}
\fmfdotn{v}{2}
\end{fmfgraph*}}
\end{fmffile}
\label{fig:feynC}}
%%%%%%%%%
\end{center}
\caption{\label{fig:Feyn} Processes responsible for DM production considered in this study. The neutrino Yukawa processes arise from Eq.~\eqref{eq:lagNS} with the coupling $y_\nu$ fixed by the neutrino data and leptogenesis. The two dark sector processes only derive from the neutrino portal in Eq.~\eqref{eq:lagPortal}, while the gravity-mediated ones from Eq.~\eqref{eq:laggr}. We consider the Higgs portal coupling in Eq.~\eqref{eq:lagh} to be negligible according to the results of Ref.~\cite{Chianese:2020yjo}.}
\end{figure}

The term in Eq.~\eqref{eq:lagPortal} is the right-handed neutrino portal where we allow for two different couplings $y_1$ and $y_2$ to the two right-handed neutrinos $M_1$ and $M_2$, respectively. We consider the gravity-mediated interaction among the stress-energy tensors of all the fields as reported Eq.~\eqref{eq:laggr}, where $M_\mathrm{P} = 1.2 \times 10^{19}~\mathrm{GeV}$ is the non-reduced Planck mass. Finally, the last interaction in Eq.~\eqref{eq:lagh} allowed by the symmetries of the model is the Higgs portal~\cite{Silveira:1985rk,McDonald:1993ex,Burgess:2000yq,Patt:2006fw,Cline:2012hg,Cline:2013gha,Athron:2017kgt,Chanda:2019xyl}. Such a coupling can efficiently dominate the dark matter production through the production of $\phi$ particles which successively decay into the dark matter particles $\chi$. The interplay between the gravity-mediated interaction and the Higgs portal has already been studied in Ref.~\cite{Chianese:2020yjo}. In particular, it has been reported an upper bound for the Higgs portal coupling $\lambda_{\phi H}$ such that the PIMD scenario is preserved. The DM production through the Higgs portal is proportional to the squared coupling $\lambda_{\phi H}^2$ and the ratio of dark masses $m_\chi/m_\phi$. In this work, the quantity $\lambda_{\phi H} \sqrt{m_\chi/m_\phi}$ is assumed to be at least one order of magnitude smaller than the upper bound for $\lambda_{\phi H}$ discussed in Ref.~\cite{Chianese:2020yjo}. This makes the Higgs scattering to contribute less than 1\% of the total DM relic abundance. Hence, the relevant interactions between the Standard Model and the dark sector contributing to the dark matter production are summarized in Fig.~\ref{fig:Feyn}.
%%%%%%%%%
%%%%%%%%%%%%

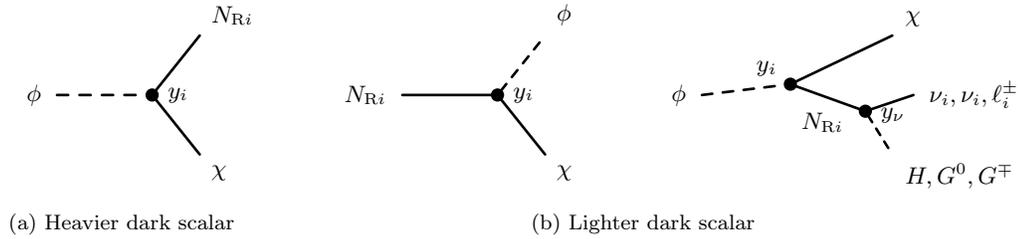
\begin{figure}[t!]
\begin{center}
\subfigure[~Heavier dark scalar]{
\begin{fmffile}{Dec1}
\fmfframe(23,18)(18,18){
\begin{fmfgraph*}(60,45)
\fmflabel{$\phi$}{i1}
\fmflabel{$\chi$}{o1}
\fmflabel{$N_{\mathrm{R} i}$}{o2}
\fmfv{label=$y_i$}{v1}
\fmfleft{i1}
\fmfright{o1,o2}
\fmf{dashes}{i1,v1}
\fmf{plain}{v1,o1}
\fmf{plain}{v1,o2}
\fmfdotn{v}{1}
\end{fmfgraph*}}
\end{fmffile}}
%%%%%%%%%
\hskip0.5cm
\subfigure[~Lighter dark scalar]{
\begin{fmffile}{Dec2}
\fmfframe(23,18)(18,18){
\begin{fmfgraph*}(60,45)
\fmflabel{$N_{\mathrm{R} i}$}{i1}
\fmflabel{$\chi$}{o1}
\fmflabel{$\phi$}{o2}
\fmfv{label=$y_i$}{v1}
\fmfleft{i1}
\fmfright{o1,o2}
\fmf{plain}{i1,v1}
\fmf{plain}{v1,o1}
\fmf{dashes}{v1,o2}
\fmfdotn{v}{1}
\end{fmfgraph*}}
\end{fmffile}
%%%%%%%%%
\begin{fmffile}{Dec3}
\fmfframe(23,18)(18,18){
\begin{fmfgraph*}(80,45)
\fmflabel{$\phi$}{i1}
\fmflabel{$\chi$}{o3}
\fmflabel{$\nu_i,\nu_i,\ell_i^\pm$}{o2}
\fmflabel{$H,G^0,G^\mp$}{o1}
\fmfv{label=$y_i$}{v1}
\fmfv{label=$y_\nu$}{v2}
\fmfleft{i1}
\fmfright{o1,o2,o3}
\fmf{dashes, tension=2}{i1,v1}
\fmf{plain,label=$N_{\mathrm{R} i}$}{v1,v2}
\fmf{plain}{v1,o3}
\fmf{dashes}{v2,o1}
\fmf{plain}{v2,o2}
\fmfdotn{v}{2}
\end{fmfgraph*}}
\end{fmffile}}
\end{center}
\caption{\label{fig:FeynDec} Decay processes which are kinematically allowed in the two different mass orderings: heavy dark scalar $m_\phi > M_i + m_\chi$ (left) and light dark scalar $M_i > m_\phi + m_\chi$ (right).}
\end{figure}

In the present paper, we aim at identifying the regions of the model parameter space where each of the processes reported in Fig.~\ref{fig:Feyn} dominates the production of dark matter in the early universe. The model has only four free parameters: the two masses $m_\chi$ and $m_\phi$, and the two dark sector couplings $y_1$ and $y_2$. However, as shown in Ref.~\cite{Chianese:2019epo}, in most of the allowed model parameter space only one of the two right-handed neutrinos at a time is efficiently coupled to the dark sector. The two right-handed neutrinos both contribute to the dark matter production only in a very narrow region around specific threshold values for the two dark sector couplings. Such threshold values mainly depend on the hierarchy between the two right-handed neutrino masses in Eq.~\eqref{eq:RHmass}. For this reason, without loss of generality we restrict the present analysis to the cases where one of the two right-handed neutrinos is effectively decoupled from the dark sector. This is achieved by simply taking the corresponding dark sector coupling to be equal to zero. For both the two scenarios, it is then possible to identify two different mass orderings:\footnote{These two mass orderings correspond respectively to ordering type A and ordering type C of the previous classification discussed in Ref.~\cite{Chianese:2019epo}. The ordering type B is instead absent in the present setting since only one of the two right-handed neutrinos at a time is assumed to be coupled to the dark sector.}
\begin{itemize}
\item {\bf Heavier dark scalar:} $m_\phi > M_i + m_\chi$;
\item {\bf Lighter dark scalar:} $M_i > m_\phi + m_\chi$;
\end{itemize}
where in all the cases the dark Dirac fermion $\chi$ is the lightest dark particle and therefore is stable. This classification of mass ordering is based on the mass of $\phi$ particles, which sets the decay processes of the scalar dark particle and the RH neutrinos. In particular, in the former case $\phi$ particles can decay into $\chi$ and a RH neutrino through two-body decay. In the latter case, the two-body decay of $\phi$ is instead kinematically suppressed while the RH neutrinos can decay into $\chi$ and $\phi$ through two-body decay. As already pointed out in Ref.s~\cite{Chianese:2018dsz,Chianese:2019epo}, in this mass ordering, the dark matter production is completely driven by such a two-body decay process of the right-handed neutrinos. We summarize in Fig.~\ref{fig:FeynDec} the decay processes kinematically allowed in the two mass orderings.
%%%%%%%%%%%%
%%%%%%%%%%%

%%%%%%%%%%%%%%%%%%%%%%
\section{Dark matter production\label{sec:DMpro}}
%%%%%%%%%%%%%%%%%%%%%%

In order to study the production of dark matter particles in the early universe, we have to solve a set of Boltzmann equations which describe how the particle number densities evolve as a function of the photon temperature $T$. In terms of the particle yields $Y_i$, the Boltzmann equations take the following general and useful forms
\begin{eqnarray}
\mathcal{H}\,T\left(1+\frac{T}{3 g^\mathfrak{s}_*\left(T\right)}\frac{{\rm d} g^\mathfrak{s}_*}{{\rm d} T}\right)^{-1}\frac{{\rm d} Y_i}{{\rm d} T} &=& \sum_{kl} \left<\Gamma_{i\rightarrow kl}\right>Y_i^{\rm eq}\left(\frac{Y_i}{Y_i^{\rm eq}}-\frac{Y_k \, Y_l}{Y_k^{\rm eq}Y_l^{\rm eq}}\right)	\nonumber \\
&& - \sum_{jk} \left<\Gamma_{j\rightarrow ik}\right>Y_j^{\rm eq}\left(\frac{Y_j}{Y_j^{\rm eq}}-\frac{Y_i \, Y_k}{Y_i^{\rm eq}Y_k^{\rm eq}}\right)
\label{eq:Boltz} \\
&&+ \mathfrak{s} \sum_{jkl} \left<\sigma_{ij\rightarrow kl}\, v_{ij}\right> Y_i^{\rm eq}Y_j^{\rm eq}\left(\frac{Y_i \, Y_j}{Y_i^{\rm eq}Y_j^{\rm eq}}-\frac{Y_k \, Y_l}{Y_k^{\rm eq}Y_l^{\rm eq}}\right) \,, \nonumber
\end{eqnarray}
The quantities $\mathcal{H}$ and $\mathfrak{s}$ denote the Hubble parameter and the entropy density of the thermal bath defined by
\begin{equation}
\mathcal{H} = \sqrt{\frac{4\pi^3{g_*}\left(T\right)}{45}}\frac{T^2}{M_{\rm P}} \qquad{\rm and}\qquad \mathfrak{s} = \frac{2\pi^2}{45} g^\mathfrak{s}_*\left(T\right) T^3 \,,
\label{eq:hubble}
\end{equation}
where $g_*$ and $g^\mathfrak{s}_*$ are the degrees of freedom of the relativistic species in the thermal bath. For temperatures higher than the electroweak scale, we have $g_* = g^\mathfrak{s}_* = 106.75$ according to the SM particle content. The right-hand side of the Boltzmann equations contains all the possible decay and scattering processes that involve the particles of species $i$. The quantity $Y^\mathrm{eq}$ denotes the yield of particles in thermal equilibrium with the photon bath given by
\begin{equation}
Y^{\rm eq}_{i} \equiv \frac{n^{\rm eq}_{i}}{\mathfrak{s}}\qquad{\rm with}\qquad n^{\rm eq}_{i} = \frac{g_i \, m_i^2 \, T}{2\pi^2} K_2\left(\frac{m_i}{T}\right)\,,
\label{eq:yeq}
\end{equation}
with $m_i$ the mass and $g_i$ the internal degrees of freedom of the particle $i$, and $K_2$ the order-2 modified Bessel function of the second kind. In the freeze-in regime~\cite{McDonald:2001vt,Hall:2009bx}, due to the feebleness of the interaction producing particles in the dark sector, we have $Y_\chi \ll Y_\chi^\mathrm{eq}$ and $Y_\phi \ll Y_\phi^\mathrm{eq}$. These conditions imply that the back-reactions depleting dark particles can be neglected, so allowing us to consider much simpler Boltzmann equations. On the other hand, all the other particles including the two right-handed neutrinos experience more efficient interactions which keep them in thermal equilibrium with the photons. Their yields follow therefore the thermal distribution for which we consider the Maxwell-Boltzmann distribution given in Eq.~\eqref{eq:yeq}. It has been indeed shown that the error committed by the use of the Maxwell-Boltzmann distribution for relativistic particles is in general compensated by other effects such as thermal mass corrections~\cite{Lebedev:2019ton}. For these reasons, to compute the final DM abundance we need to solve just two coupled Boltzmann equations for the dark scalars $\phi$ and the dark fermions $\chi$. After the dark matter yield freezes-in at the value $Y_{\rm DM,0}$, the corresponding relic abundance is given by
\begin{equation}
\Omega_{\rm DM}h^2 = \frac{2 \, \mathfrak{s}_0 \, m_\chi  \,Y_{\rm DM,0}}{\rho_{\rm crit}/h^2}\,,
\label{eq:omegaPRE}
\end{equation}
where $\mathfrak{s}_0=2891.2\,{\rm cm^3}$ and $\rho_{\rm crit}/h^2 = 1.054 \times10^{-5} {\rm GeV \, cm^{-3}}$ are the today’s entropy density and the critical density, respectively~\cite{Tanabashi:2018oca}. The factor of 2 takes into account the contribution of DM anti-particles. The predicted value has to be compared with the observed one provided by the Planck Collaboration at 68\% C.L.~\cite{Aghanim:2018eyx}:
\begin{equation}
\Omega_{\rm DM}^\mathrm{obs}h^2 = 0.120 \pm 0.001\,.
\label{eq:omegaOBS}
\end{equation}
By making this comparison, we obtain the allowed regions for the model parameter space providing a viable dark matter candidate.

Differently from the freeze-out production mechanism~\cite{Gondolo:1990dk}, in the freeze-in scenario the final DM abundance depends on the initial conditions. In the following analysis, we assume a negligible abundance of dark particles at the end of the reheating stage after inflation. In particular, we take $Y_\chi (T_\mathrm{RH}) = 0$ and $Y_\phi (T_\mathrm{RH}) = 0$ where $T_\mathrm{RH}$ denotes the reheating temperature. The reheating temperature is an additional free parameter which is allowed to range within the interval
\begin{equation}
5.10 \times 10^{10}~\mathrm{GeV} \lesssim T_\mathrm{RH} \lesssim 6.6 \times 10^{15}~\mathrm{GeV}\,.
\label{eq:TRH_range}
\end{equation}
The upper bound is set by the constraint on the Hubble rate at the end of inflation $\mathcal{H}_i$ according to the current bound on tensor modes deduced by the measurements of the CMB spectrum~\cite{Akrami:2018odb, Ade:2015lrj, Ade:2018gkx}. Under the assumption of instantaneous reheating, we have
\begin{equation}
\mathcal{H}_i (T_\mathrm{RH})= \sqrt{\frac{4\pi^3{g_*}\left(T_\mathrm{RH}\right)}{45}}\frac{T_\mathrm{RH}^2}{M_{\rm P}} \qquad \mathrm{with} \qquad  \mathcal{H}_i\lesssim 6.1 \times 10^{13}~\mathrm{GeV} \,.
\label{eq:CMBlimit}
\end{equation}
The lower bound in Eq.~\eqref{eq:TRH_range} is instead set by the requirement of a successful baryogenesis via leptogenesis. The vanilla leptogenesis scenario is indeed achieved as long as $T_\mathrm{RH} \geq M_1 = 5.10 \times 10^{10}~\mathrm{GeV}$.

In order to obtain the allowed regions of the model parameter space achieving the correct dark matter abundance, we take into account all the processes displayed in Fig.~\ref{fig:Feyn}. Differently from the neutrino Yukawa and the dark sector processes which depend on the dark sector couplings $y_1$ and $y_2$, the coupling strength of the gravity-mediated interaction is fixed by the equivalence principle. The gravitational production occurring after the inflationary stage is just controlled by the reheating dynamics, which is simply described in the instantaneous limit by the reheating temperature $T_\mathrm{RH}$. For each value of the dark particles' masses, the correct dark matter abundance is obtained in correspondence of a specific value of the reheating temperature. In other words, this implies that the gravitational production imposes an upper limit on the reheating temperature above which the dark matter is overproduced. In the next Section we therefore discuss such an upper limit on $T_\mathrm{RH}$ before solving the complete Boltzmann equations for the dark scalars and fermions.

%%%%%%%%%%%%%%
\section{Upper limit on reheating temperature from gravity-mediated production \label{sec:TRH}}
%%%%%%%%%%%%%%
In this section we only consider the effect of gravity production of the dark sector particles, switching off all the RH neutrino portal couplings.
Under the assumption of instantaneous reheating, the contributions of the gravity-mediated production to the today's yields of the dark scalars and dark fermions are respectively given by~\cite{Chianese:2020yjo}
\begin{eqnarray}
Y^\mathrm{Gravity}_{\chi, 0} & = & \int^{T_\mathrm{RH}}_0 \mathrm{d}T \, \frac{\mathfrak{s}}{\mathcal{H} \, T} \langle \sigma \, v \rangle^\mathrm{Gravity}_{\chi \chi} \left( Y^\mathrm{eq}_\chi \right)^2 \label{eq:gr_chi} \,,\\
Y^\mathrm{Gravity}_{\phi, 0} & = & \int^{T_\mathrm{RH}}_0 \mathrm{d}T \, \frac{\mathfrak{s}}{\mathcal{H} \, T} \langle \sigma \, v \rangle^\mathrm{Gravity}_{\phi \phi}  \left( Y^\mathrm{eq}_\phi \right)^2 \,, \label{eq:gr_phi}
\end{eqnarray}
where $\langle \sigma v \rangle^\mathrm{Gravity}_{\chi \chi}$ and $\langle \sigma v \rangle^\mathrm{Gravity}_{\phi \phi}$ are the total thermally averaged cross-sections related to the gravity-mediated interaction. They are given by the sum of the contributions from all the SM degrees of freedom and the ones from the two RH neutrinos:
\begin{eqnarray}
\langle \sigma \, v \rangle^\mathrm{Gravity}_{\chi \chi} & = & \langle \sigma \, v \rangle^\mathrm{Gravity}_{\mathrm{SM} \rightarrow \chi} + \sum_i \langle \sigma \, v \rangle^\mathrm{Gravity}_{N_{\mathrm{R}i} \rightarrow \chi} \,,\\
\langle \sigma \, v \rangle^\mathrm{Gravity}_{\phi \phi} & = & \langle \sigma \, v \rangle^\mathrm{Gravity}_{\mathrm{SM} \rightarrow \phi} + \sum_i \langle \sigma \, v \rangle^\mathrm{Gravity}_{N_{\mathrm{R}i} \rightarrow \phi}   \,,
\end{eqnarray}
where according to the SM matter content we have~\cite{Garny:2017kha}
\begin{eqnarray}
\langle \sigma \,  v \rangle^\mathrm{Gravity}_{\mathrm{SM} \rightarrow \chi} & = & 4 \langle \sigma\,v \rangle_{0\rightarrow1/2}  + 45 \langle \sigma  \,v \rangle_{1/2\rightarrow1/2} + 12 \langle \sigma \,v \rangle_{1\rightarrow1/2}  \,,\\
\langle \sigma \, v \rangle^\mathrm{Gravity}_{\mathrm{SM} \rightarrow \phi} & = & 4 \langle \sigma\,v \rangle_{0\rightarrow0}  + 45 \langle \sigma  \,v \rangle_{1/2\rightarrow0} + 12 \langle \sigma \,v \rangle_{1\rightarrow0}  \,, 
\end{eqnarray}
with the indices $0,1/2,1$ referring to the spin of the particles. According to the interaction in Eq.~\eqref{eq:laggr}, by neglecting the masses of SM particles we get~\cite{Garny:2017kha}
\begin{eqnarray}
\langle \sigma\,v \rangle_{0\rightarrow1/2} & = & \frac{\pi T m_\chi}{2 M_\mathrm{P}^4}  \left[ \frac15 \frac{m_\chi}{T} \left( 1 - \frac{K_1(m_\chi/T)^2}{K_2(m_\chi/T)^2}  \right) + \frac25 \frac{K_1(m_\chi/T)}{K_2(m_\chi/T)}   + \frac45 \frac{T}{m_\chi} \right]
\,, \\
\langle \sigma  \,v \rangle_{1/2\rightarrow1/2} = \langle \sigma \,v \rangle_{1\rightarrow1/2} & = & \frac{4\pi T m_\chi}{ M_\mathrm{P}^4}  \left[ \frac{2}{15} \frac{m_\chi}{T} \left( 1 - \frac{K_1(m_\chi/T)^2}{K_2(m_\chi/T)^2} \right) + \frac35 \frac{K_1(m_\chi/T)}{K_2(m_\chi/T)} + \frac65 \frac{T}{m_\chi} \right] \,, \label{eq:gr1/2_1/2} \\
\langle \sigma\,v \rangle_{0\rightarrow0} & = & \frac{\pi m_\phi^2}{8 M_\mathrm{P}^4}  \left[ \frac35 \frac{K_1(m_\phi/T)^2}{K_2(m_\phi/T)^2} + \frac25 +\frac45 \frac{T}{m_\phi} \frac{K_1(m_\phi/T)}{K_2(m_\phi/T)} + \frac85 \frac{T^2}{m_\phi^2} \right] \,, \\
\langle \sigma  \,v \rangle_{1/2\rightarrow0} = \langle \sigma \,v \rangle_{1\rightarrow0} & = & \frac{4\pi T^2}{M_\mathrm{P}^4} \left[ \frac{2}{15} \frac{m_\phi^2}{T^2}  \left( \frac{K_1(m_\phi/T)^2}{K_2(m_\phi/T)^2} -1 \right) +  \frac{6}{15} \frac{m_\phi}{T} \frac{K_1(m_\phi/T)}{K_2(m_\phi/T)} + \frac{12}{15}  \right] \,,
\end{eqnarray}
where the quantities in the square brackets asymptotically tend towards 1 for $m_\phi, m_\chi \gg T$. Differently to the SM particles, the masses of the two RH neutrinos cannot be in general neglected when compared to $m_\phi$ and $m_\chi$. When $M_i \ll m_\chi, m_\phi$, the corresponding thermally averaged cross-section to dark fermions and dark scalars are given by $\langle \sigma  \,v \rangle_{1/2\rightarrow1/2}$ and $\langle \sigma  \,v \rangle_{1/2\rightarrow0}$, respectively. Hence, the RH neutrino can be treated as massless particles and its scattering contributes as two additional degrees of freedom to the fermion degrees of freedom. When $M_i \gg m_\chi, m_\phi$, the dark particles can be treated as massless and the thermally averaged cross-sections are
\begin{eqnarray}
\langle \sigma \, v \rangle^\mathrm{Gravity}_{N_{\mathrm{R}i} \rightarrow \chi} & = & \frac{4\pi T M_i}{M_\mathrm{P}^4}  \left[ \frac{2}{15} \frac{M_i}{T} \left( 1 - \frac{K_1(M_i/T)^2}{K_2(M_i/T)^2} \right) + \frac35 \frac{K_1(M_i/T)}{K_2(M_i/T)} + \frac65 \frac{T}{M_i} \right] \,, \\
\langle \sigma \, v \rangle^\mathrm{Gravity}_{N_{\mathrm{R}i} \rightarrow \phi} & = & \frac{4\pi T^2}{M_\mathrm{P}^4} \left[ \frac{2}{15} \frac{M_i^2}{T^2}  \left( \frac{K_1(M_i/T)^2}{K_2(M_i/T)^2} -1 \right) +  \frac{6}{15} \frac{M_i}{T} \frac{K_1(M_i/T)}{K_2(M_i/T)} + \frac{12}{15}  \right] \,.
\end{eqnarray}
In order to significantly contribute to the today's yield of dark particles, it is however required that the RH neutrinos are in thermal equilibrium with the photon bath. The expressions~\eqref{eq:gr_chi} and~\eqref{eq:gr_phi} obtained from the Boltzmann equation~\eqref{eq:Boltz} indeed only holds in this case. As it will be clear later, for masses smaller than $10^{15}~\mathrm{GeV}$, the reheating temperature required to achieve the observed DM abundance is typically smaller than $10^{14}~\mathrm{GeV}$. At the highest allowed temperature, therefore the heaviest RH neutrino is far to be in thermal equilibrium and the corresponding contribution to the dark particles' yields is negligible. Regarding the lightest RH neutrino, at high temperatures $(T \gg M_1, m_\chi, m_\phi)$ the following limits apply
\begin{eqnarray}
\langle \sigma \, v \rangle^\mathrm{Gravity}_{N_{\mathrm{R}i} \rightarrow \chi} \simeq \langle \sigma \, v \rangle^\mathrm{Gravity}_{1/2 \rightarrow 1/2} & \simeq & \frac{24 \pi T^2}{5 M_\mathrm{P}^4} \,, \\
\langle \sigma \, v \rangle^\mathrm{Gravity}_{N_{\mathrm{R}i} \rightarrow \phi} \simeq \langle \sigma \, v \rangle^\mathrm{Gravity}_{1/2 \rightarrow 0} & \simeq & \frac{16 \pi T^2}{5 M_\mathrm{P}^4} \,.
\end{eqnarray}
Since the gravity production is ultraviolet dominant~\cite{Elahi:2014fsa}, the RH neutrino scattering also contributes 2 degrees of freedom if $T_\mathrm{RH} > M_1$, which is required for leptogenesis. Therefore the contribution of the lightest right-handed neutrino to the gravity production approximately adds 2 degrees of freedom to the SM fermionic degrees of freedom.
\begin{figure}[t!]
\begin{center}
\includegraphics[width=0.48\textwidth]{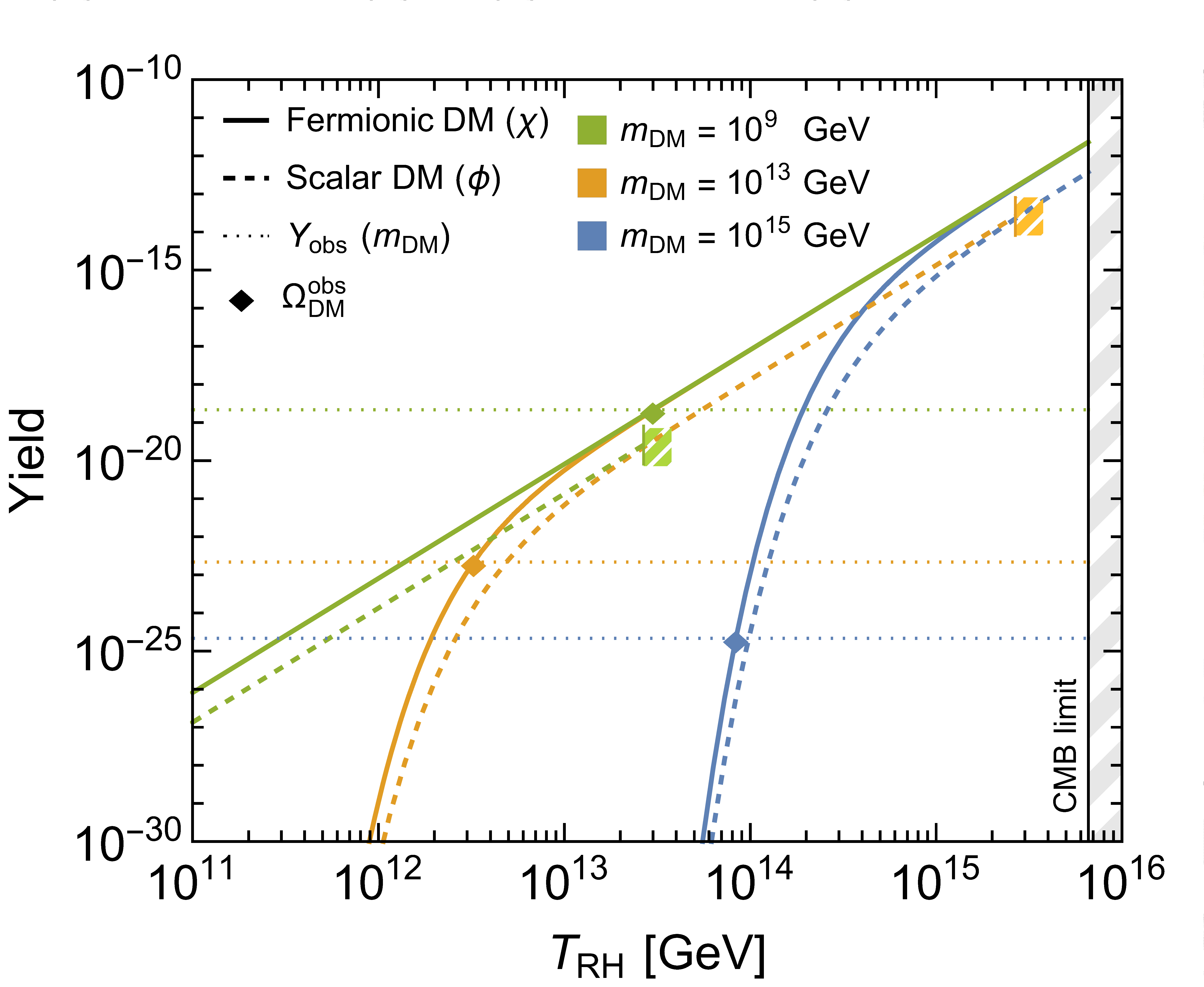}
\includegraphics[width=0.475\textwidth]{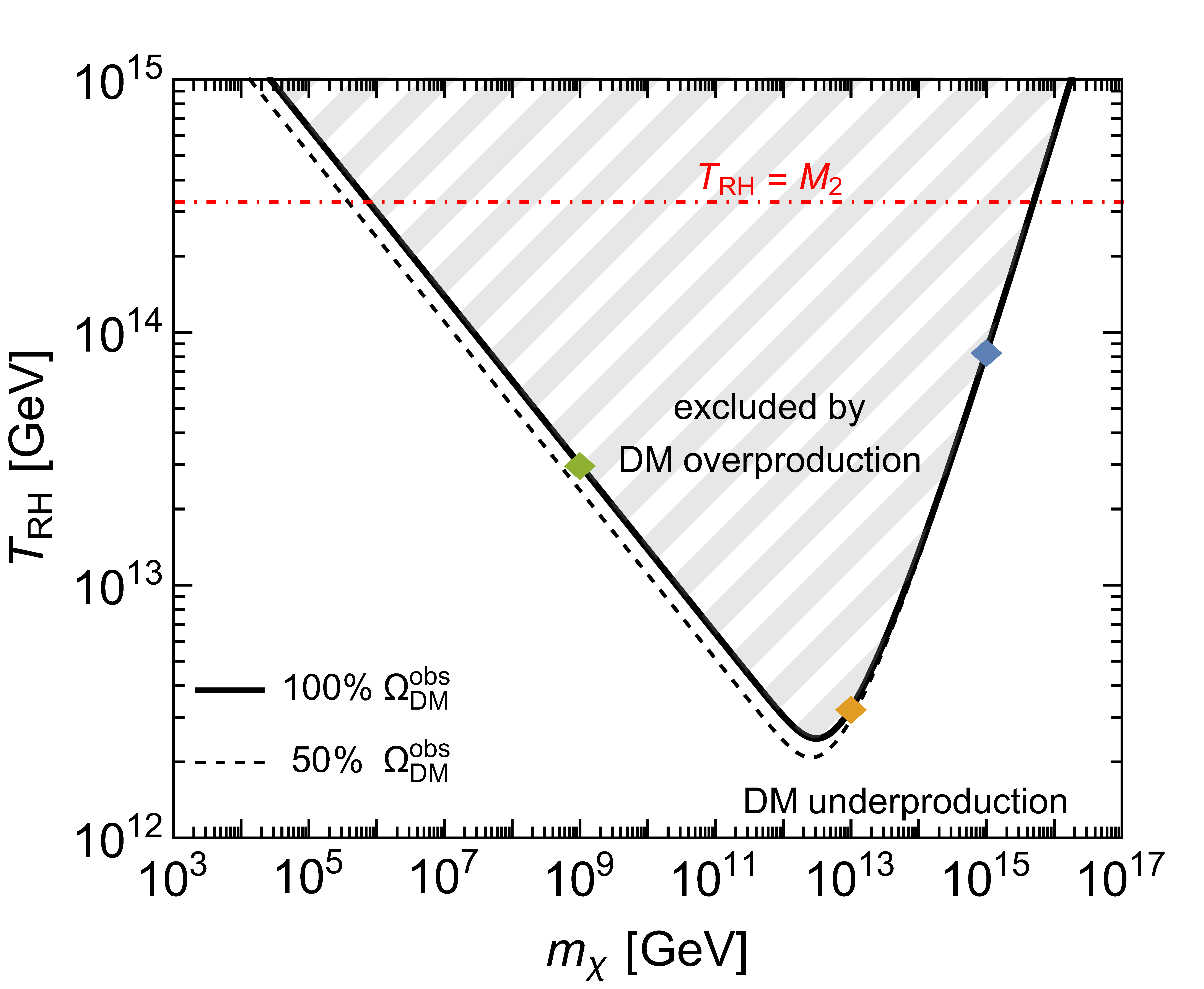}
\caption{\label{fig:Gravprod} Gravity-mediated production of dark matter in the absence of RH neutrino portal couplings.
{\it Left:} Yield as a function of reheating temperature $T_\mathrm{RH}$ for three different DM masses in the case of fermionic (solid lines) or scalar (dashed lines) DM particles. The horizontal lines correspond to the values of the yield $Y_\mathrm{obs}$ required to achieve the observed DM abundance, according to Eq.~\eqref{eq:Yobs}. The diamonds represent the correct choice of reheating temperature and DM mass providing the observed DM abundance for the fermionic case. The hatched grey region on the right is excluded by CMB measurement according to Eq.~\eqref{eq:CMBlimit}. The hatched colored regions stopping the dashed lines for scalar dark matter represent the upper bound on $T_\mathrm{RH}$ deduced by isocurvature perturbations (see Eq.~\eqref{eq:isoperturb}).
{\it Right:}  The relation between the mass $m_\chi$ of the lightest dark sector particle and the reheating temperature $T_\mathrm{RH}$ providing the 100\% (solid line) and the 50\% (dashed line) of the observed DM abundance $\Omega_\mathrm{DM}^\mathrm{obs}$ reported in Eq.~\eqref{eq:omegaOBS}. The horizontal dot-dashed red line marks the value of reheating temperature equal to the heaviest right-handed neutrino mass. The hatched region is excluded by DM overproduction, while in the white region the gravity-mediated processes leads to DM underproduction. The diamonds correspond to the ones in the left plots.}
\end{center}
\end{figure}

In left panel of Fig.~\ref{fig:Gravprod}, we report the today's yields for a fermionic dark matter (solid lines) and a scalar dark matter (dashed lines) as a function of the reheating temperature $T_\mathrm{RH}$, according to Eq.s~\eqref{eq:gr_chi} and~\eqref{eq:gr_phi}, respectively. The different colors correspond to three different values for the dark matter mass $m_\mathrm{DM}$. For all the cases, the observed DM abundance is obtained in correspondence of the intersection of the solid and dashed lines with the horizontal dotted lines. The latter indeed represent the yield
\begin{equation}
Y_\mathrm{obs}(m_\mathrm{DM}) = \frac{\rho_\mathrm{crit} \, \Omega^\mathrm{obs}_\mathrm{DM}}{2\, \mathfrak{s}_0 \, m_\mathrm{DM}} \,,
\label{eq:Yobs}
\end{equation}
obtained by inverting Eq~\eqref{eq:omegaPRE} and using the experimental value for $\Omega_\mathrm{DM}$. Therefore, the correct DM abundance for a particle of mass $m_\mathrm{DM}$ is achieved for a specific value of the reheating temperature. For example, the required values of $T_\mathrm{RH}$ for fermionic dark matter are highlighted by the three diamonds. Larger values for the reheating temperature correspond to larger values of the DM abundance and, consequently, are excluded due to DM overproduction.

The hatched grey region shows the model-independent upper bound on the reheating temperature $T_\mathrm{RH}\lesssim 6.6 \times 10^{15}~\mathrm{GeV}$ deduced by the CMB constraints on tensor modes according to Eq.~\eqref{eq:CMBlimit}. However, in the case of scalar dark matter the constraint on isocurvature perturbations typically leads to more stringent upper bounds on $T_\mathrm{RH}$ depending on the scalar dark matter mass. To suppress a large production of primordial isocurvature perturbations, one in general needs to impose $m_\mathrm{DM} \gtrsim \mathcal{H}_i$~\cite{Chung:2004nh,Nurmi:2015ema,Akrami:2018odb}. According to the instantaneous reheating scenario (see Eq.~\eqref{eq:CMBlimit}) such a constraint can be translated in an upper bound on the reheating temperature for scalar dark matter only:
\begin{equation}
m_\phi \gtrsim \mathcal{H}_i\quad \implies \quad T_\mathrm{RH} \lesssim \left[ \sqrt{\frac{45}{4\pi^3 g_*(T_\mathrm{RH})}} M_\mathrm{P} \, m_\phi \right]^{\frac12} \simeq 2.7 \times 10^{15} \left( \frac{m_\phi}{10^{13}~\mathrm{GeV}} \right)^{1/2} ~\mathrm{GeV}\,.
\label{eq:isoperturb}
\end{equation}
For the benchmark cases shown in the left panel of Fig.~\ref{fig:Gravprod}, we have $T_\mathrm{RH} \lesssim 2.7 \times (10^{13}, \, 10^{15}, \, 10^{16})~\mathrm{GeV}$ for the scalar dark matter masses $m_\mathrm{DM} = 10^{9}, \, 10^{13}, \, 10^{15}~\mathrm{GeV}$, respectively. The first two upper bounds are highlighted in the plot by the two colored hatched regions that stop the corresponding dashed lines at higher reheating temperatures. The last upper bound is instead weaker than the one imposed by CMB tensor modes. It is worth mentioning that these limits derived from isocurvature perturbations depend on the scalars potential of the model. For example, the presence of a sizeable self-interaction coupling for the scalar dark matter would relax such bounds~\cite{Markkanen:2018gcw,Tenkanen:2019aij}. Moreover, we stress that such limits do not apply to the case of fermionic dark matter.

As can be seen in the plot, the gravity-mediated production is more efficient for fermionic dark matter with respect to the scalar case. Indeed, for each value of the reheating temperature, the gravitationally-produced yield of a fermionic particle $\chi$ is always larger than the one of a scalar particle $\phi$ as long as $m_\chi \leq m_\phi$. Such a behaviour implies that in the full model, where the dark scalar is allowed to decay into the lighter dark fermion, the contribution to the DM relic from the gravity-mediated processes is dominated by the direct gravitational production of $\chi$ particles. In particular, we have
\begin{equation}
Y^\mathrm{Gravity}_{\mathrm{DM},0} = Y^\mathrm{Gravity}_{\chi,0} + Y^\mathrm{Gravity}_{\phi,0} \simeq Y^\mathrm{Gravity}_{\chi,0} \,.
\end{equation}
Remarkably, this allows us to obtain an upper limit for $T_\mathrm{RH}$ avoiding DM overproduction independently of the dark scalar mass. Such an upper limit as a function of $m_\chi$ (the dark matter mass) is shown in the right panel of Fig.~\ref{fig:Gravprod} by the black solid line. Here, the diamonds correspond to the ones in the left panel in order to highlight that the upper limit is dominated by the direct gravitational production of the lighter dark fermion. The hatched region above the solid line is therefore excluded by DM overproduction. In the region below, the gravity-mediated processes are not efficient enough leading to DM underproduction. For example, the dashed line displayed in the plot represents the relation between $m_\chi$ and $T_\mathrm{RH}$ achieving only the 50\% of the observed DM abundance. Therefore, in the region below the solid black line, other DM-interactions are required for a viable DM candidate. In the current study, these interactions arise from the right-handed neutrino portal. 

As shown in the right plot, for a fermionic dark matter mass in the range $7 \times 10^{5}~\mathrm{GeV} \lesssim m_\chi \lesssim 5 \times 10^{15}~\mathrm{GeV}$ the reheating temperature is constrained to be smaller than the mass of the heaviest right-handed neutrino $M_2 = 3.28 \times 10^{14}~\mathrm{GeV}$ highlighted by the horizontal dot-dashed red line. In the instantaneous reheating scenario, this means that the number density of $N_2$ is highly suppressed being the photon temperature smaller than $M_2$. Therefore, the upper limit on the reheating temperature provided by the gravity-mediated production implies that only the dark coupling to the lightest right-handed neutrino could provide a significant contribution to the DM abundance for $7 \times 10^{5}~\mathrm{GeV} \lesssim m_\chi \lesssim 5 \times 10^{15}~\mathrm{GeV}$. On the other hand, the coupling to the heaviest right-handed neutrino is expected to be efficient only for $m_\chi \lesssim 7 \times 10^{5}~\mathrm{GeV}$ since for $m_\chi \gtrsim 5 \times 10^{15}~\mathrm{GeV}$ the neutrino Yukawa and the dark sector processes are kinematically suppressed.

%%%%%%%%%%%%%%%%%%%%%%%%%%%
\section{Results \label{sec:results}}
%%%%%%%%%%%%%%%%%%%%%%%%%%%

In this section we discuss the allowed regions of the model parameter space, including both gravity production and RH neutrino portal couplings simultaneously, for the two mass orderings defined earlier. We obtain the threshold values for the RH neutrino portal couplings below which the dark matter is gravitationally produced.

%%%%%%%%%%%%%%%%%%%%%%%%%%%
\subsection{``Heavier dark scalar'' mass ordering \label{sec:HDS}}
%%%%%%%%%%%%%%%%%%%%%%%%%%%

Let us assume that just one of the two right-handed neutrinos effectively interacts with the dark sector through the neutrino portal coupling $y_i$. When the dark scalar is heavier than the right-handed neutrino $N_{\mathrm{R}i}$, the relevant Boltzmann equations are
\begin{eqnarray}
\mathcal{H}\,T\frac{{\rm d} Y_\phi}{{\rm d} T} & = & 
- \mathfrak{s} \left( \left<\sigma\, v\right>_{\phi\phi}^\mathrm{Gravity} + \left<\sigma\, v\right>_{\phi\phi}^\mathrm{DS} \right) \left({Y_\phi^{\rm eq}}\right)^2 
- \mathfrak{s} \left<\sigma\, v\right>^{\rm \nu-Yukawa}_{\chi\phi} Y_\phi^{\rm eq}Y_\chi^{\rm eq} 
+ \left<\Gamma_{\phi}\right>\left(Y_\phi-\frac{Y_\phi^{\rm eq}}{Y_\chi^{\rm eq}}Y_\chi\right)  \,, \label{eq:phia} \\
\mathcal{H}\,T\frac{{\rm d} Y_\chi}{{\rm d} T} & = & 
- \mathfrak{s}  \left( \left<\sigma\, v\right>_{\chi\chi}^\mathrm{Gravity} + \left<\sigma\, v\right>_{\chi\chi}^\mathrm{DS} \right)\left({Y_\chi^{\rm eq}}\right)^2 
- \mathfrak{s} \left<\sigma\, v\right>^{\rm \nu-Yukawa}_{\chi\phi} Y_\phi^{\rm eq}Y_\chi^{\rm eq}
-  \left<\Gamma_{\phi}\right>\left(Y_\phi-\frac{Y_\phi^{\rm eq}}{Y_\chi^{\rm eq}}Y_\chi\right) \,, \label{eq:chia}
\end{eqnarray}
where the thermally averaged cross-sections with superscripts ``Gravity'', ``DS'' and ``$\nu$-Yukawa'' refer to the total contributions from gravity-mediated scatterings, the dark sector processes and the neutrino Yukawa ones, respectively (see Fig.~\ref{fig:Feyn}). For the last two thermally averaged cross-sections, we use the same expressions reported in Ref.~\cite{Chianese:2019epo}. It is important to mention that the amplitudes of the DS processes are proportional to $y_i^4$, while the ones of the neutrino Yukawa scatterings are proportional to $y_i^2 y_\nu^2$, where $y_\nu = \left(U^\dagger_\nu Y\right)_{ij}/\sqrt2 \,\text{or}\, Y_{ij}$ ($U_\nu$ is the Pontecorvo–Maki–Nakagawa–Sakata matrix) depending on the type of the outgoing leptons. This different dependence on the dark sector coupling $y_i$ is the key feature that implies the switching from the dominance of the neutrino Yukawa scatterings in the dark matter production to the one of the dark sector processes as the coupling $y_i$ increases~\cite{Chianese:2018dsz, Chianese:2019epo}. In the above equations, the last terms encode the two-body decays of $\phi$ particles into $\chi$ ones and RH neutrinos. Such decays are indeed kinematically allowed in the mass ordering of heavier dark scalar where $m_\phi > M_i + m_\chi$. We have
\begin{equation}
\langle \Gamma_\phi \rangle = \frac{K_1(m_\phi / T)}{K_2(m_\phi / T)} \Gamma_{\phi \rightarrow \chi N_{\mathrm{R}i}} \,,
\end{equation}
where the partial decay width is proportional to $y_i^2$ and takes the expression
\begin{equation}
\Gamma_{\phi \rightarrow \chi N_{\mathrm{R}i}} = \frac{y_i^2\,m_\phi}{16 \pi}\left(1-\frac{m_\chi^2}{m_\phi^2}-\frac{M_i^2}{m_\phi^2}\right)\sqrt{\lambda\left(1,\frac{m_\chi^2}{m_\phi^2},\frac{M_i^2}{m_\phi^2}\right)} \simeq  \frac{y_i^2\,m_\phi}{16 \pi} \,.
\end{equation}
where $\lambda$ is the K\"allén function. Since in this mass ordering the dark scalars totally decay well before the electroweak phase transition due to the efficient two-body decay~\cite{Chianese:2018dsz, Chianese:2019epo}, the total yield of the dark matter can be described by the sum of the two Boltzmann equations:
\begin{eqnarray}
\mathcal{H}\,T \frac{{\rm d} Y_{\rm DM}}{{\rm d} T} & = & - \mathfrak{s} \left( \left<\sigma\, v\right>_{\phi\phi}^\mathrm{Gravity} + \left<\sigma\, v\right>_{\phi\phi}^\mathrm{DS} \right) \left({Y_\phi^{\rm eq}}\right)^2 - \mathfrak{s} \left( \left<\sigma\, v\right>_{\chi\chi}^\mathrm{Gravity} + \left<\sigma\, v\right>_{\chi\chi}^\mathrm{DS} \right)  \left({Y_\chi^{\rm eq}}\right)^2 \nonumber \\
&& - 2 \, \mathfrak{s} \left<\sigma\, v\right>^{\rm \nu-Yukawa}_{\chi\phi} Y_\phi^{\rm eq}Y_\chi^{\rm eq}\,. \label{eq:Beq}
\end{eqnarray}
This differential equation can be easily integrated out to provide the today's DM yield at $T=0$. Such a quantity is given by the sum of different contributions related to the dark sector couplings:
\begin{equation}
Y_{\rm DM,0} = Y_{\chi,0} + Y_{\phi,0}= Y^{\rm Gravity} + Y^{\rm DS}  + 2\,Y^{\rm \nu-Yukawa}\,,
\label{eq:DMyield}
\end{equation}
where $Y^{\rm Gravity}$ is the gravitational contribution given by the sum of the two quantities in Eq.s~\eqref{eq:gr_chi} and~\eqref{eq:gr_phi} (see previous Section), and
\begin{eqnarray}
Y^{\rm DS}&  = & \int_{0}^{T_{\rm RH}} {\rm d}T \, \frac{\mathfrak{s}}{\mathcal{H}\,T}
\left[ \left<\sigma\, v\right>_{\phi\phi}^\mathrm{DS}\left({Y_\phi^{\rm eq}} \right)^2 
+ \left<\sigma\, v\right>_{\chi\chi}^\mathrm{DS} \left({Y_\chi^{\rm eq}} \right)^2 \right] \,, \\
%%%%
Y^{\rm \nu-Yukawa} & = & \int_{0}^{T_{\rm RH}} {\rm d}T \, \frac{\mathfrak{s}}{\mathcal{H}\,T}
\left[ \left<\sigma\, v\right>^{\rm \nu-Yukawa}_{\chi\phi} Y_\phi^{\rm eq}Y_\chi^{\rm eq} \right] \,.
\end{eqnarray}
%%%%%%%%%%%%%%%%%%%%%%%%%%%
\begin{figure}[t!]
\begin{center}
\vspace{-15pt}
\includegraphics[width=0.48\textwidth]{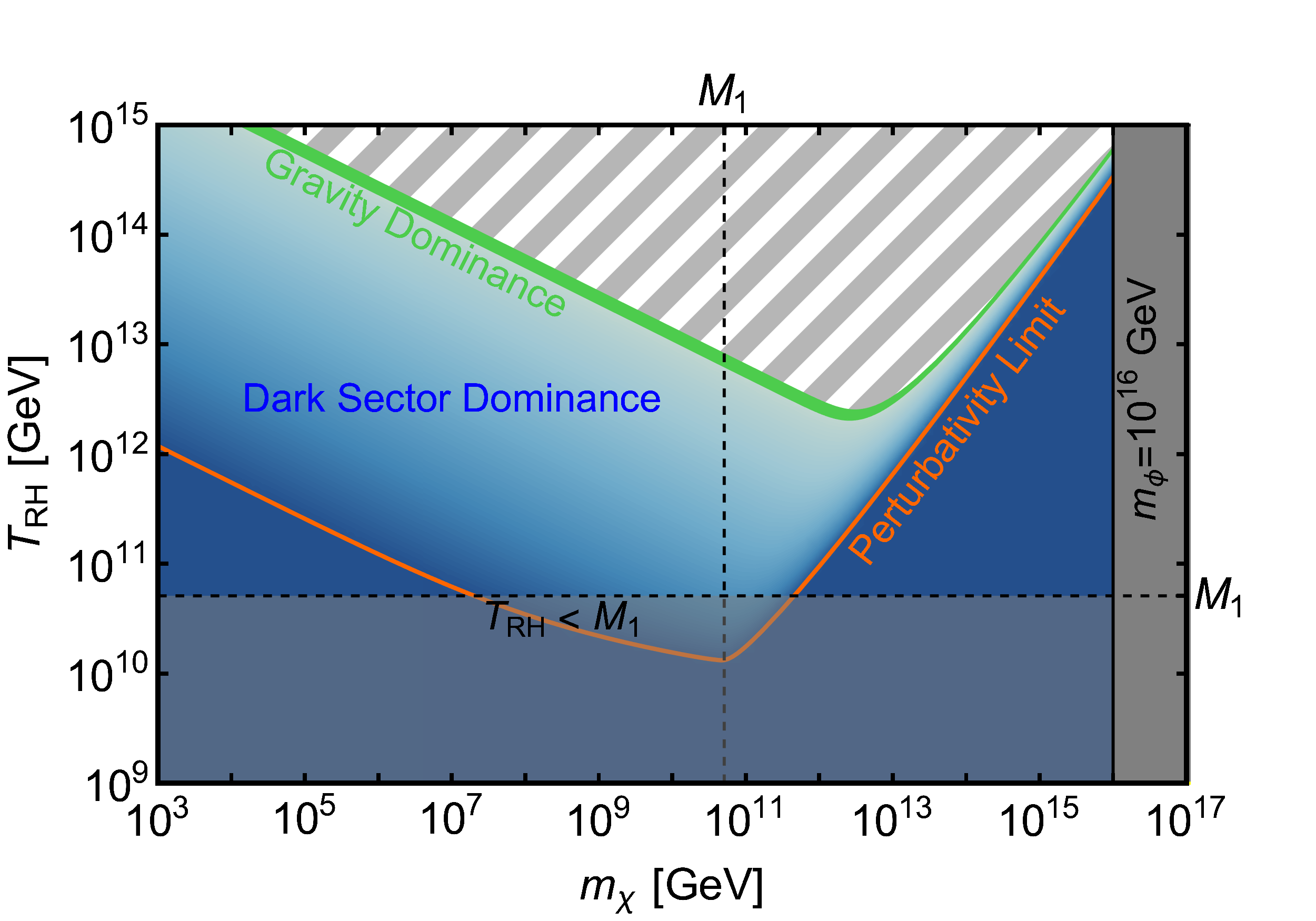}
\includegraphics[width=0.48\textwidth]{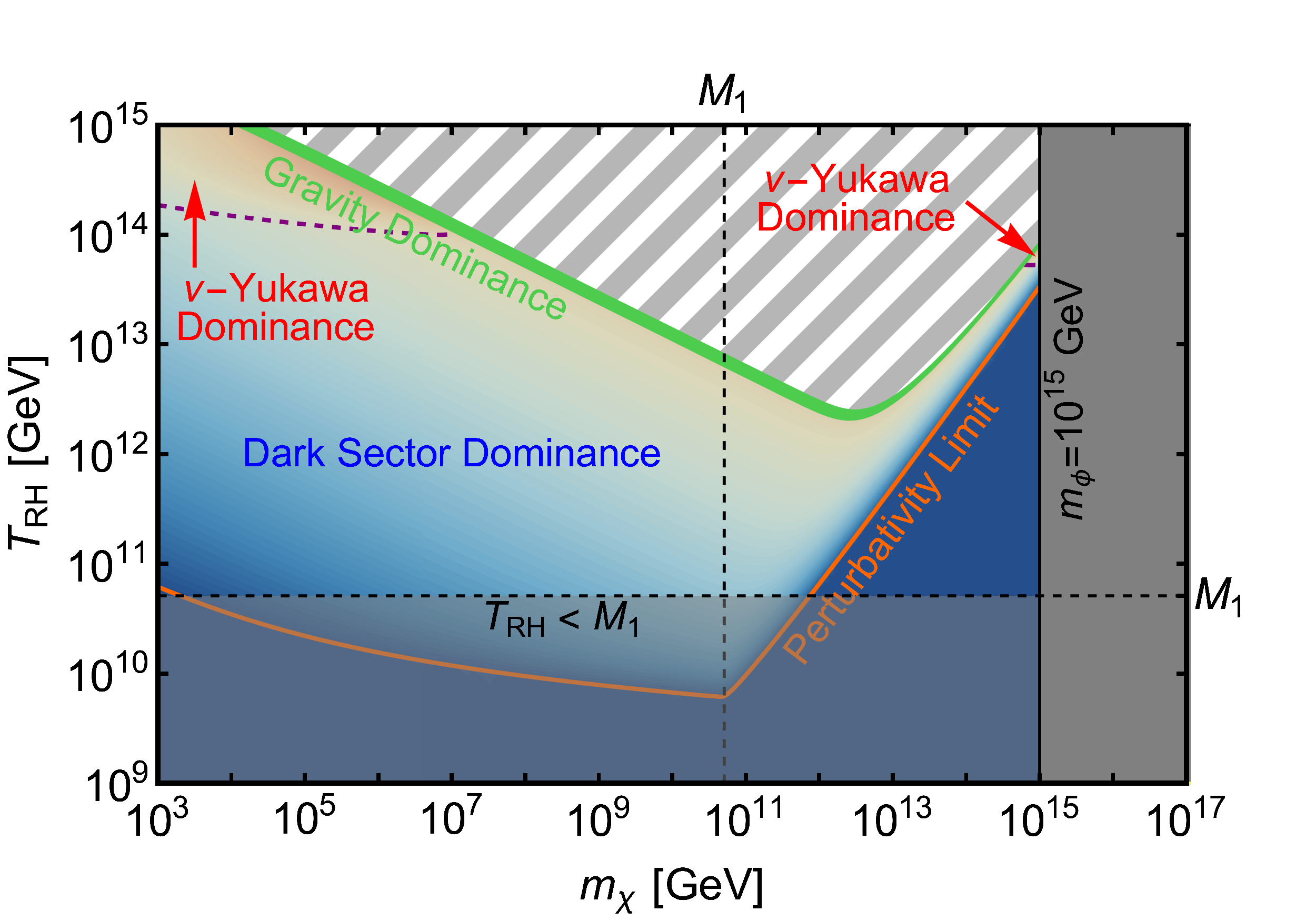}\\
\includegraphics[width=0.48\textwidth]{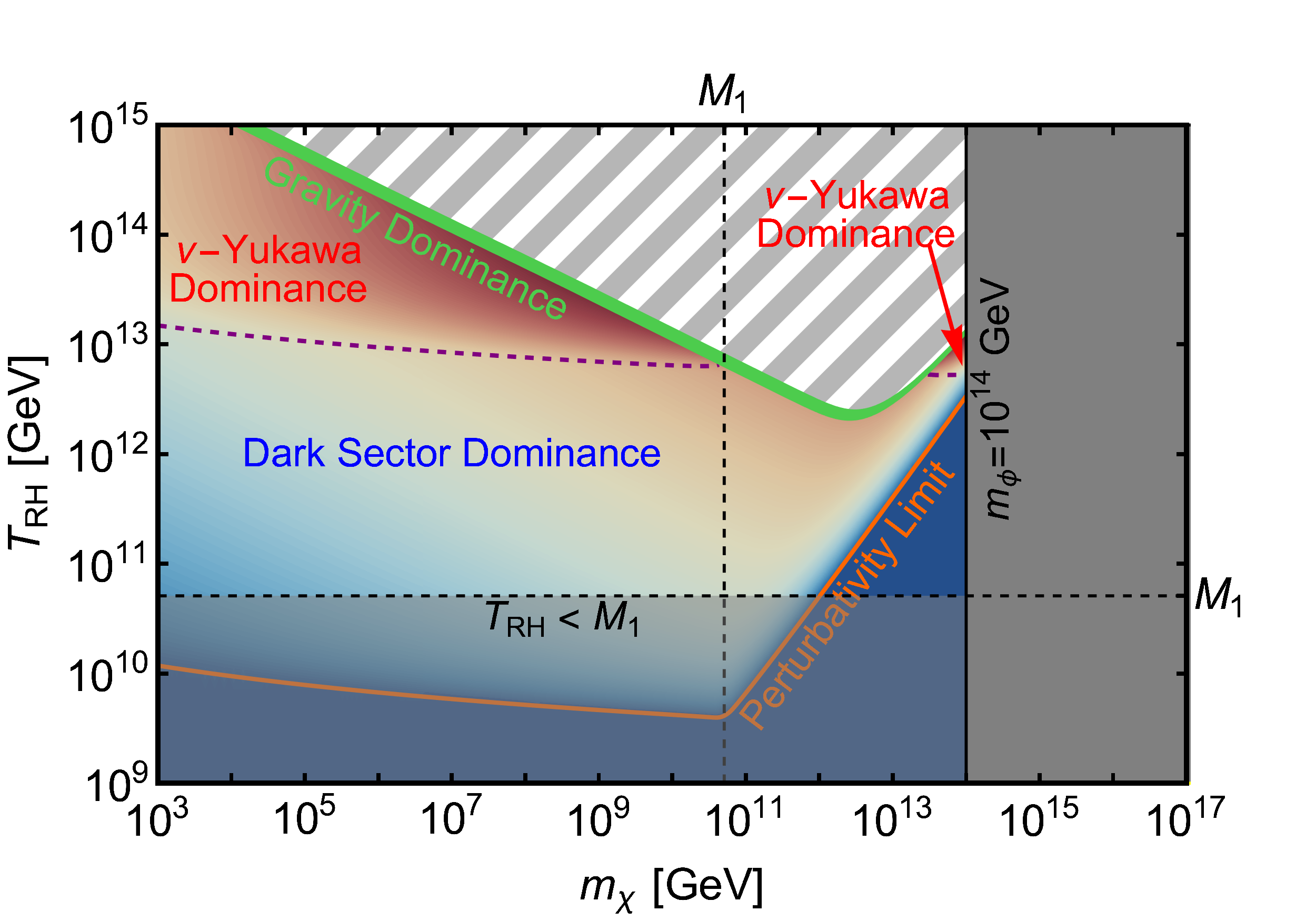}
\includegraphics[width=0.48\textwidth]{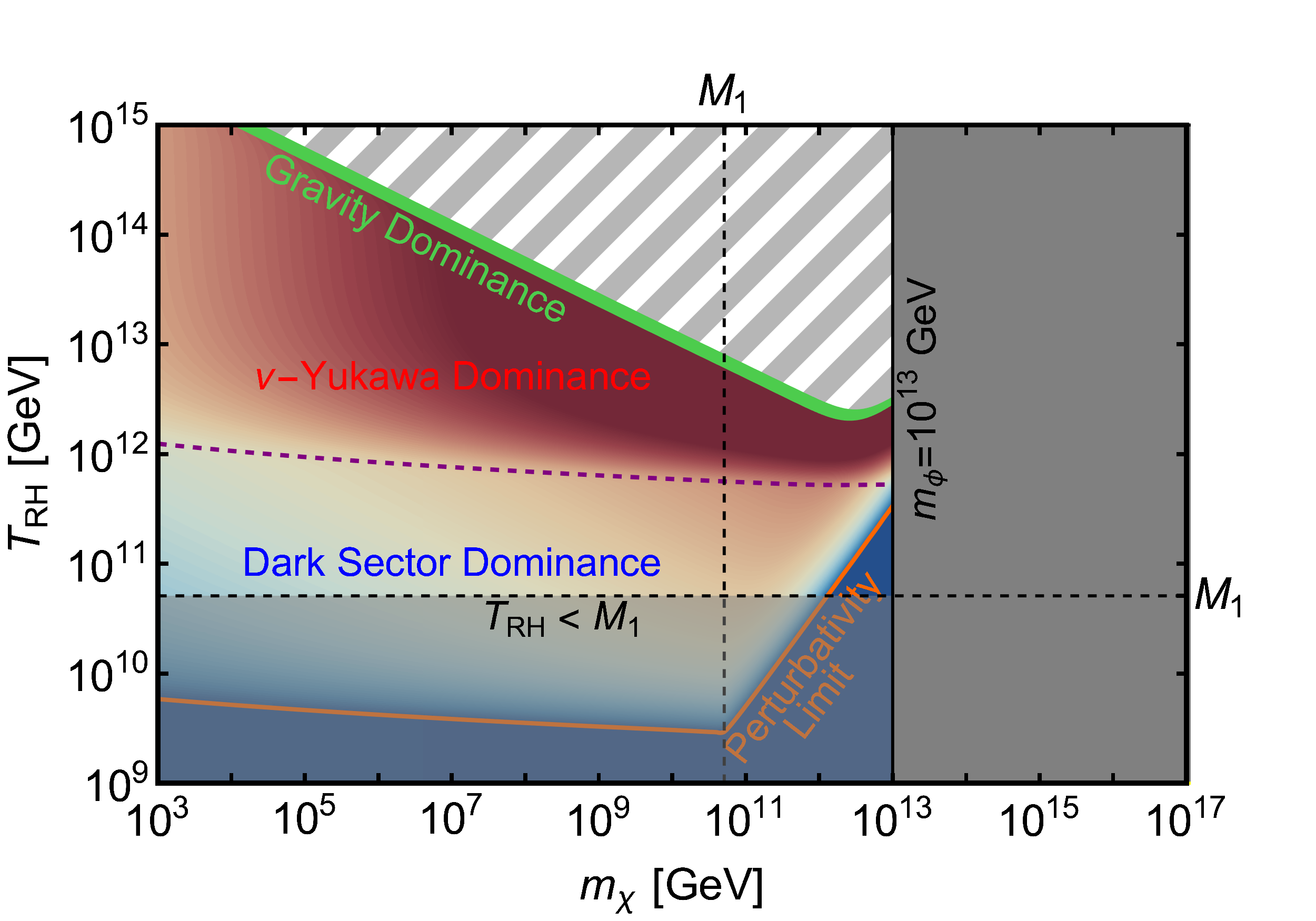}\\
\includegraphics[width=0.48\textwidth]{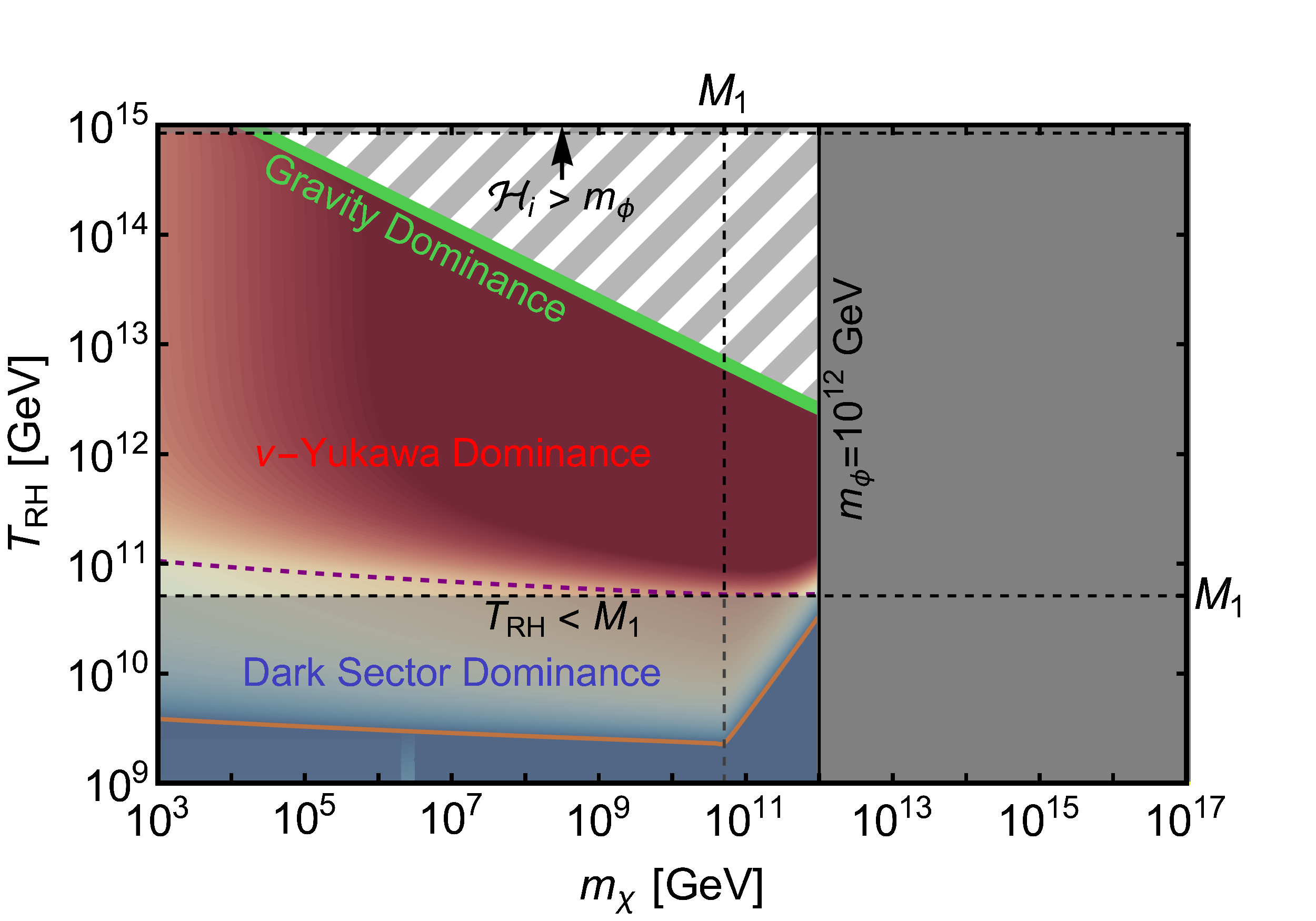}
\includegraphics[width=0.48\textwidth]{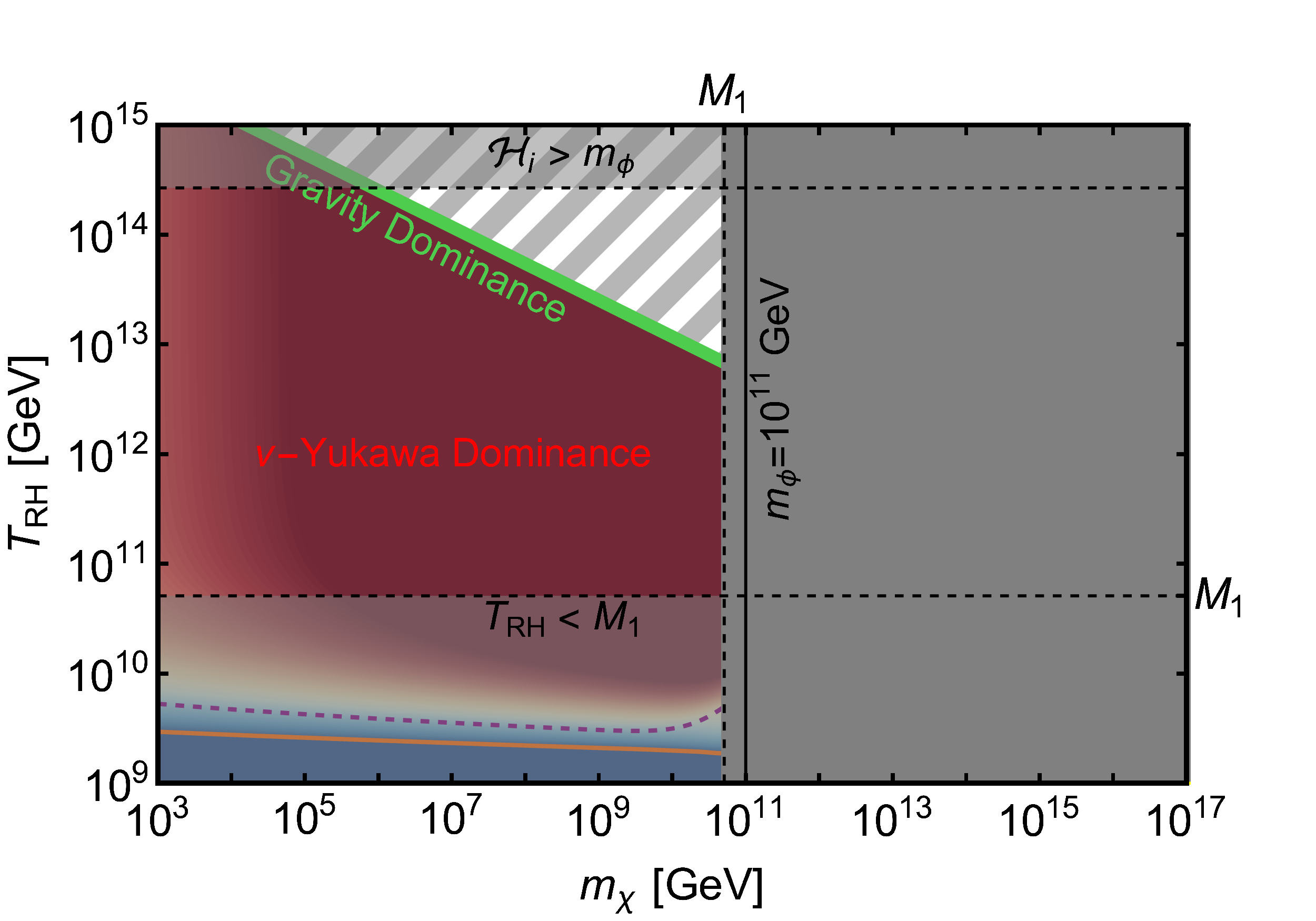}\\
\includegraphics[width=0.40\textwidth]{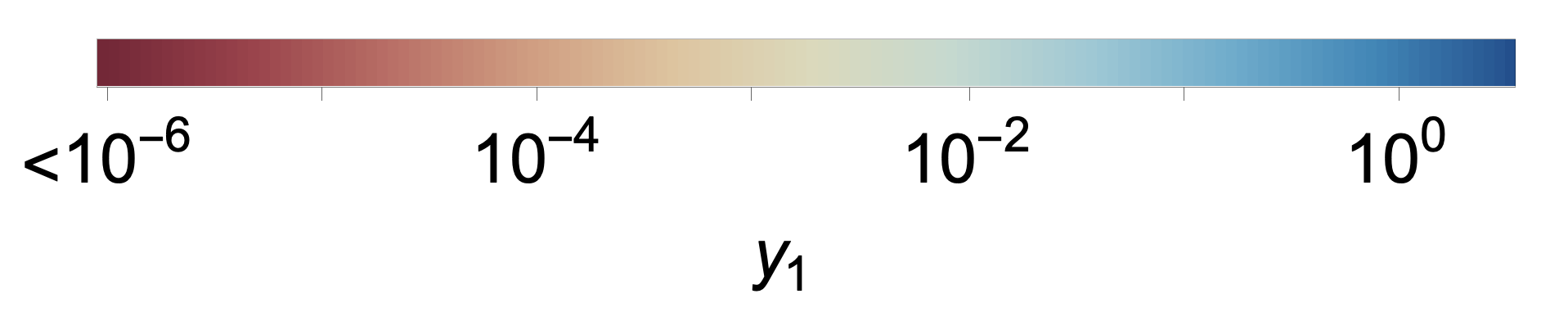}
\caption{\label{fig:HeavierPhi_1} Dark sector coupling $y_1$ to the lightest right-handed neutrino $N_{\mathrm{R}1}$ in the plane $m_\chi$--$T_\mathrm{RH}$ achieving the observed dark matter abundance in the mass ordering $m_\phi > M_1 + m_\chi$ (heavier dark scalar). The reddish (blueish) colors refer to small (large) dark sector coupling $y_1$. The different panels correspond to different values of the dark scalar mass $m_\phi$ from $10^{16}~\mathrm{GeV}$ (top-left panel) to $10^{11}~\mathrm{GeV}$ (bottom-right panel). In all the plots, the dark matter production is dominated by gravity-mediated processes in the green bands (see Fig.~\ref{fig:Gravprod}), by neutrino Yukawa processes between the green bands and the purple dashed lines (if present), and by the dark sector processes down to the orange solid lines representing the perturbativity constraint on the coupling ($y_1 \leq \sqrt{4 \pi}$). The regions excluded are: the grey area on the right due to the mass ordering, the hatched region on the top due to DM overproduction (see Fig.~\ref{fig:Gravprod}), the shaded region on the bottom due to leptogenesis ($T_\mathrm{RH}>M_1$), and the shaded area on the top (if present) due to limit on isocurvature perturbation implying $m_\phi \gtrsim \mathcal{H}_i$.}
\vspace{-15pt}
\end{center}
\end{figure}
%%%%%%%%%%%%%%%%%%%%%%%%%%%
Depending on the free parameters of the model (the two dark masses, the dark sector couplings and the reheating temperature), we expect that only one of the three terms in Eq.~\eqref{eq:DMyield} at a time dominates the dark matter production. Hence, we aim at highlighting the dominant interaction driving the dark matter production in the model parameter space where the observed DM abundance is achieved.

In Fig.~\ref{fig:HeavierPhi_1}, we report the results concerning the scenario where only the lightest right-handed neutrino is coupled to the dark sector ($y_1 \neq 0$ and $y_2= 0$). In the plots, it is shown the dark sector coupling $y_1$ required to achieve the observed DM relic abundance as a function of the reheating temperature $T_\mathrm{RH}$ and the two dark masses. The different plots correspond to different values of the dark scalar mass $m_\phi$ from $10^{16}~\mathrm{GeV}$ (top-left panel) to $10^{11}~\mathrm{GeV}$ (bottom-right panel). The latter is the lowest possible value for $m_\phi$ according to the mass ordering. The required dark sector coupling is color-coded. The reddish (blueish) colors correspond to small (large) dark sector couplings. In particular, the darkest red color represents a coupling smaller than $10^6$, while the darkest blue color means a coupling larger than $\sqrt{4\pi}$ considered as the threshold for the perturbative calculations.\footnote{Such large values for the dark sector are still not able to establish the thermal equilibrium between the photon bath and the dark sector due to the Boltzmann exponential suppression occurring for $m \gg T$. This justifies the assumption of freeze-in production in all the model parameter space explored in this paper.} The regions excluded by the requirement of perturbativity are bounded from above by the orange solid lines referring to $y_1 = \sqrt{4\pi}$. Other regions excluded are the grey regions on the right according to the mass ordering $m_\phi > m_\chi + M_1$ and the hatched regions on the top due to DM overproduction through the gravity-mediated processes (see Fig.~\ref{fig:Gravprod} and related discussion). Moreover, the shaded regions on the bottom are excluded when requiring a successful vanilla leptogenesis ($T_\mathrm{RH} > M_1$), while the ones on the top in the lower two panels are disallowed by the growing of isocurvature perturbations according to the value of the dark scalar mass (see Eq.~\eqref{eq:isoperturb}).

In all the plots we highlight the regions where each of the different processes in Fig.~\ref{fig:Feyn} dominates the dark matter production:
\begin{itemize}
\item {\bf Gravity dominance.} The green band is where the gravity-mediated interaction provides the main contribution to the today's DM yield. In particular, it is delimited from above by requiring that the 100\% of the total DM abundance is accounted for by the gravitational production (solid black line in Fig.~\ref{fig:Gravprod}) and from below by requiring only a 50\% contribution from gravity (dashed black line in Fig.~\ref{fig:Gravprod}). Hence, in this region the dark matter is gravitationally produced and can be regarded as a Planckian Interacting Dark Matter (PIDM) candidate. As previously discussed, the green region of gravity dominance is independent of the dark scalar mass and therefore it is equal in all the plots of Fig.~\ref{fig:HeavierPhi_1}.
\item {\bf Neutrino Yukawa dominance.} The region (if present) between the green band and the dashed purple line is where the usual neutrino Yukawa scatterings dominate the dark matter production. As pointed out in the previous works~\cite{Chianese:2018dsz, Chianese:2019epo}, this is the most interesting scenario where a direct link between neutrino physics and dark matter production exists. Indeed, the neutrino Yukawa couplings that are completely determined by neutrino oscillations data and leptogenesis are also responsible for producing the dark matter in the early universe. Remarkably, the neutrino-dark matter relation is preserved when taking into account the gravitational production.
\item {\bf Dark sector dominance.} The dark sector processes drive the dark matter production for larger values of the dark sector couplings along with lower reheating temperatures. The regions where this occurs are bounded from above by the lowest between the green band and the dashed purple lines and from below by the orange solid line (perturbativity limit).
%%%%%%%%%%%%%%%%%%%%
\end{itemize}
\begin{figure}[t!]
\begin{center}
\includegraphics[width=0.48\textwidth]{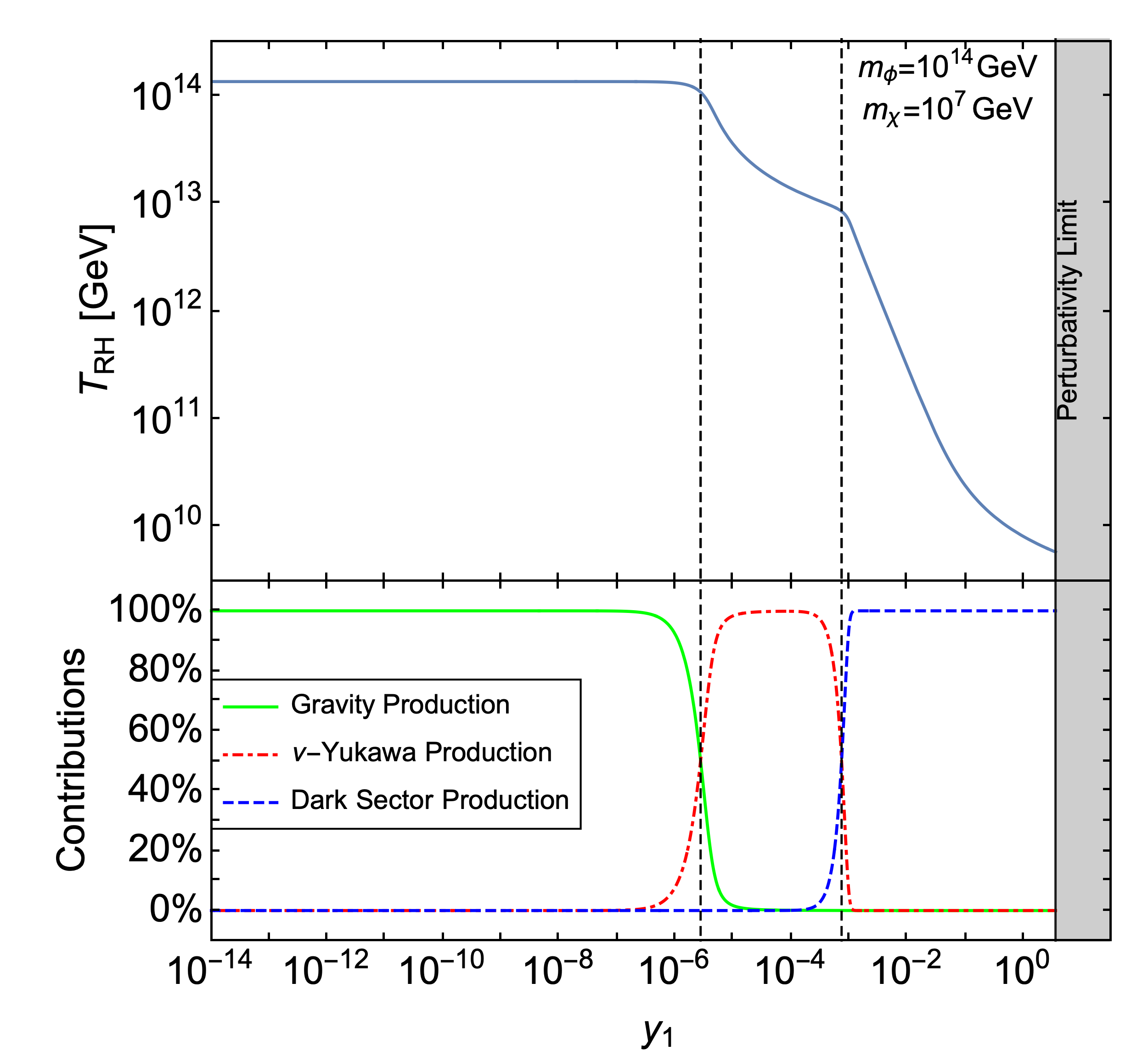}
\includegraphics[width=0.48\textwidth]{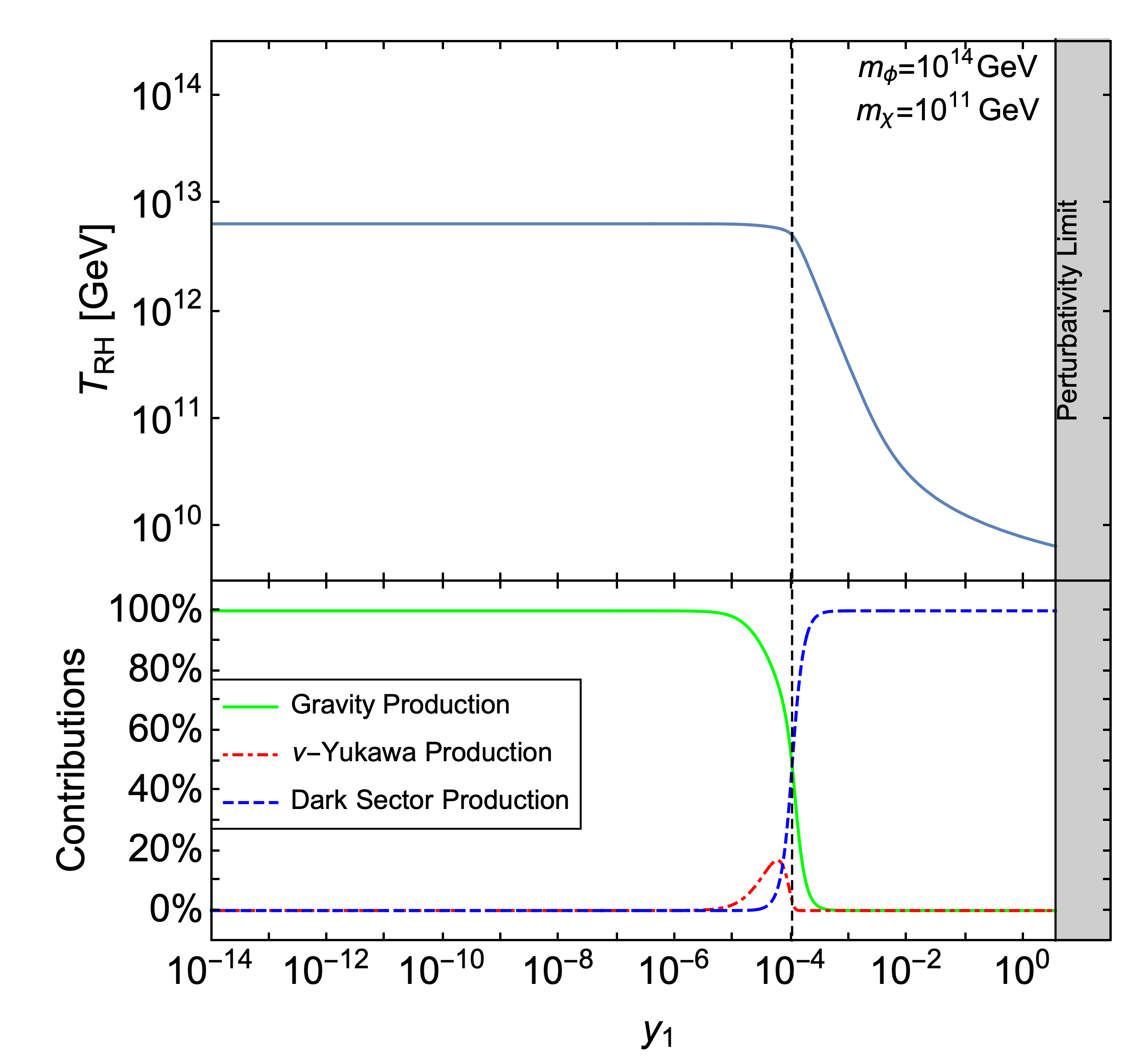}
\caption{\label{fig:bench}Relation between the reheating temperature $T_\mathrm{RH}$ and the dark sector coupling $y_1$ achieving the observed DM abundance for two benchmark cases: $m_\chi = 10^{7}~\mathrm{GeV}$ and $m_\phi = 10^{14}~\mathrm{GeV}$ (left plot); $m_\chi = 10^{11}~\mathrm{GeV}$ and $m_\phi = 10^{14}~\mathrm{GeV}$ (right plot). The lower panels show the relative contributions to the DM abundance of the three different processes of the model: gravity-mediated production (green solid lines); neutrino Yukawa production (red dot-dashed lines); dark sector production (blue dashed lines). The vertical black dashed lines highlight the coupling at which the dominant process producing more than 50\% of the total dark matter abundance changes.}
\end{center}
\end{figure}
%%%%%%%%%%%%%%%%%%%%
\begin{figure}[t!]
\begin{center}
\includegraphics[width=0.48\textwidth]{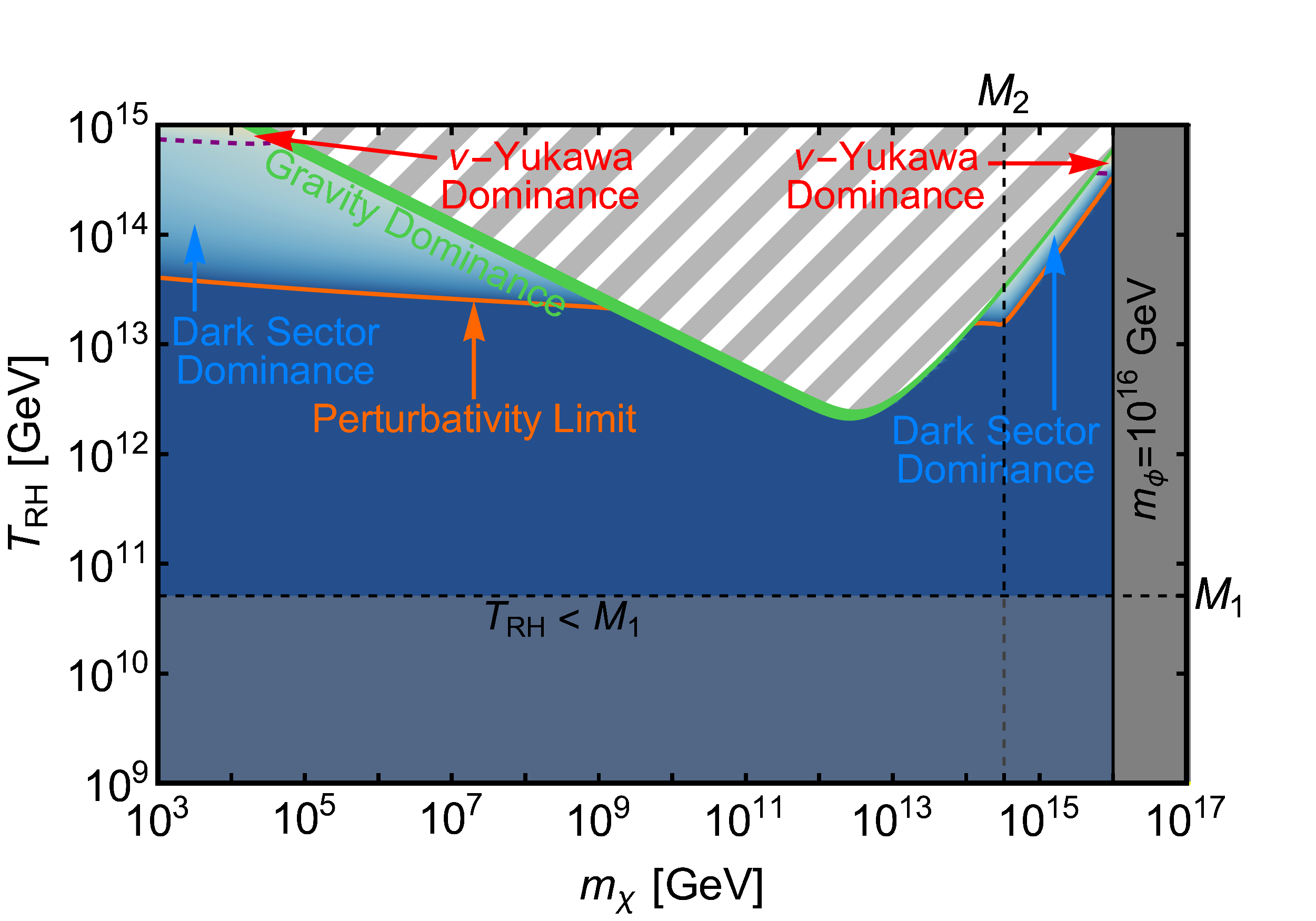}
\includegraphics[width=0.48\textwidth]{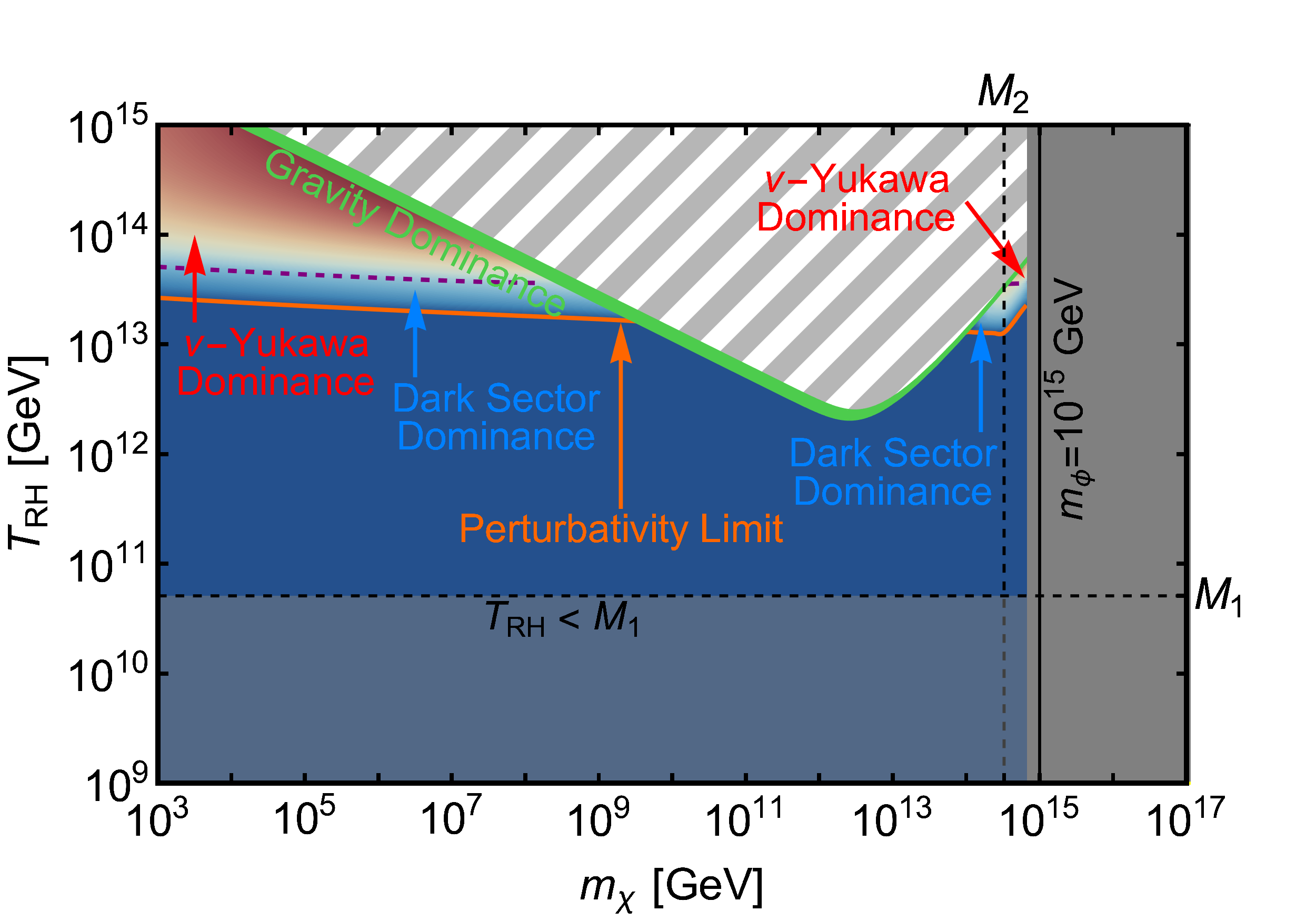}\\
\includegraphics[width=0.40\textwidth]{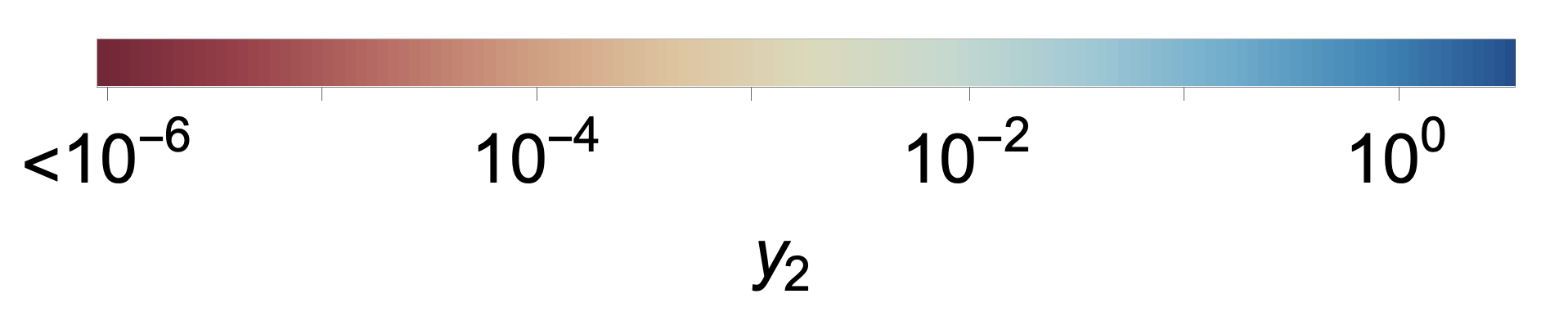}
\caption{\label{fig:HeavierPhi_2} Dark sector coupling $y_2$ to the heaviest right-handed neutrino $N_{\mathrm{R}2}$ in the plane $m_\chi$--$T_\mathrm{RH}$ achieving the observed dark matter abundance in the mass ordering $m_\phi > M_2 + m_\chi$ (heavier dark scalar). The description of the plots is the same as the one in Fig.~\ref{fig:HeavierPhi_1}.}
\end{center}
\end{figure}
%%%%%%%%%%%%%%%%%%%%
For given values of the dark masses, the dominance of a specific class of processes is determined by the relation between the dark sector coupling and the reheating temperature achieving the observed DM abundance. In general, the lower the reheating temperature, the larger the dark sector coupling required to compensate the Boltzmann suppression in the thermal particle distributions occurring for $m \gg T$. Such a behaviour is further explained in Fig.~\ref{fig:bench} where we focus on two benchmark cases: $m_\chi = 10^{7}~\mathrm{GeV}$ and $m_\phi = 10^{14}~\mathrm{GeV}$ (left plot); $m_\chi = 10^{11}~\mathrm{GeV}$ and $m_\phi = 10^{14}~\mathrm{GeV}$ (right plot). In the plots, the upper panels show the aforementioned relation between the dark coupling $y_1$ and the reheating temperature, while the lower panels highlight the relative contributions of the three different processes to the DM relic abundance. In all the cases, for very small values of the dark coupling, the gravity-mediated processes (green solid lines) completely dominate the dark matter production providing the 100\% of the observed $\Omega_\mathrm{DM}h^2$. In this regime, the reheating temperature is fixed to set the required value for the Hubble rate at the end of inflation $\mathcal{H}_i$. As the coupling increases, the reheating temperature has to decreases accordingly, so suppressing the gravitational production. In the left plot, the dominance of neutrino Yukawa processes (red dot-dashed lines) can occur since the reheating temperature is allowed to be of the same order of $m_\phi$. Then, as the coupling further increases, the dark matter production becomes fully dominated by the dark sector processes (blue dashed lines). The couplings at which the dominance changes are shown by the vertical black lines. In the right plot, the neutrino Yukawa processes instead are kinematically suppressed being the reheating temperature always smaller than the dark scalar mass. Hence, they can only provide a contribution of $\sim 10\%$ to the DM relic density and never dominate the DM production. In this benchmark case, the dominance of gravity-mediated processes directly gives way to the one of the dark sector scatterings.

In Fig.~\ref{fig:HeavierPhi_2}, we report the results obtained when only the heaviest right-handed neutrino is coupled to the dark sector ($y_1 = 0$ and $y_2 \neq 0$). As before, we show the dark sector coupling $y_2$ required to achieve the observed DM relic abundance as a function of the reheating temperature $T_\mathrm{RH}$ and the two dark masses. Differently from Fig.~\ref{fig:HeavierPhi_1}, here we consider only two benchmark values for the dark scalar mass, $m_\phi = 10^{16}~\mathrm{GeV}$ and $m_\phi = 10^{15}~\mathrm{GeV}$, since smaller values are incompatible with the requirement of $m_\phi > m_\chi + M_2$. In this scenario, larger dark sector couplings are required to overcome the gravitational production being the other processes kinematically suppressed for $T_\mathrm{RH} \sim M_2 = 3.28 \times 10^{14}~\mathrm{GeV}$. This behaviour is in agreement with the results obtained in the previous analysis~ \cite{Chianese:2019epo}. This implies that the perturbative limit on the dark sector coupling $y_2$ is reached at higher reheating temperatures with respect to the coupling $y_1$. For this reason, there exists only a small region for heavy fermionic dark matter ($m_\chi \gtrsim 10^{14}~\mathrm{GeV}$) where the dark matter particles are not gravitationally produced. On the other hand, the coupling to the heaviest right-handed neutrino prefers lighter masses for the fermionic dark matter candidate ($m_\chi \lesssim 10^{10}~\mathrm{GeV}$) where the dominance of neutrino Yukawa processes can occur at high reheating temperature.

%%%%%%%%%%%%%%%%%%%%%%%%%%%
\subsection{``Lighter dark scalar'' mass ordering \label{sec:LDS}}
%%%%%%%%%%%%%%%%%%%%%%%%%%%

When the dark scalar is lighter than the right-handed neutrino $i$ effectively coupled to the dark sector, the dominant process competing in the dark matter production with the gravity-mediated scatterings is the two-body decay $N_{\mathrm{R}i} \rightarrow \phi + \chi$. Such a process is indeed proportional to $y_i^2$ and, consequently, its rate is larger than the ones of the neutrino Yukawa and dark sector scatterings. Assuming that the dark scalar totally decay into the lighter dark fermion, the Boltzmann equation for the total DM yield is now simply given by
\begin{equation}
\mathcal{H}\,T\frac{{\rm d} Y_{\rm DM}}{{\rm d} T} =
- \mathfrak{s} \left<\sigma\, v\right>_{\phi\phi}^\mathrm{Gravity} \left({Y_\phi^{\rm eq}}\right)^2 
- \mathfrak{s} \left<\sigma\, v\right>^\mathrm{Gravity}_{\chi\chi} \left(Y_\chi^{\rm eq}\right)^2 
- 2\left<\Gamma_{N_{\mathrm{R}i}}\right>Y_{N_{\mathrm{R}i}}^{\rm eq} \,,
\end{equation}
where $\left<\Gamma_{N_{\mathrm{R}i}}\right>$ is the thermally average of the partial decay width
\begin{equation}
\Gamma_{N_{\mathrm{R}i} \rightarrow \phi \chi} = \frac{y_i^2\,M_1}{32 \pi}\left(1+\frac{m_\chi^2}{M_i^2}-\frac{m_\phi^2}{M_i^2}\right)\sqrt{\lambda\left(1,\frac{m_\chi^2}{M_i^2},\frac{m_\phi^2}{M_i^2}\right)} \simeq  \frac{y_i^2\,M_i}{32 \pi} \,.
\label{eq:decn1}
\end{equation}
The last equality in the above expression holds in the limit of $M_i \gg m_\phi, m_\chi$. For values of the reheating temperature slightly smaller than the upper limit corresponding to the gravitational DM overproduction discussed in Fig.~\ref{fig:Gravprod} and larger than $M_i$, the correct DM abundance is achieved when
\begin{equation}
y_1 \simeq  8.76 \times 10^{-11} \left(\frac{M_1}{5.10 \times 10^{10}~\mathrm{GeV}}\right)^{1/2} \left( \frac{10^7~\mathrm{GeV}}{m_\chi}\right)^{1/2} \qquad \text{with $y_2 =0$}\,,
\label{eq:lighter_coupling}
\end{equation}
or 
\begin{equation}
y_2  \simeq  7.02\times 10^{-9} \left(\frac{M_2}{3.28 \times 10^{14}~\mathrm{GeV}}\right)^{1/2} \left( \frac{10^7~\mathrm{GeV}}{m_\chi}\right)^{1/2} \qquad \text{with $y_1 =0$}\,,
\label{eq:heavier_coupling}
\end{equation}
when taking the limit $M_i \gg m_\phi, m_\chi$ in the two scenarios considered. These values also represent a threshold for the corresponding dark sector coupling below which the gravity portal dominates over the neutrino portal.

The final DM abundance is obtained by the three-body decays of the dark scalars, whose decay width is
\begin{equation}
\Gamma_\phi^{\rm 3-body} \simeq \frac{y_{i}^2\tilde{y}_\nu^2}{1536 \pi^3}\frac{m_\phi^3}{M_{i}^2}\left( 1 + \frac34\frac{m_\phi^2}{M_{i}^2} \right)\,,
\label{eq:dec3}
\end{equation}
where $\tilde{y}_\nu^2$ is the effective squared Yukawa coupling. The decays become efficient when their rate $\left< \Gamma_\phi^{\rm 3-body} \right>$ overcomes the expansion rate of the universe set by Hubble parameter $\mathcal{H}$. Differently from the previous case, the three-body decay processes are further suppressed by the presence of the mediator and the additional insertion of the Yukawa interaction given by $\tilde{y}_\nu^2$. This implies that the decay of the dark scalars in general occurs at a later time. For the neutrino portal with the lightest right-handed neutrino, it occurs when the quantity
\begin{eqnarray}
\frac{\left< \Gamma_\phi^{\rm 3-body} \right>}{\mathcal{H}} \simeq
\begin{dcases}
\left( \frac{y_1}{6.04 \times 10^{-13}} \right)^2 \left( \frac{5.10 \times 10^{10}~\mathrm{GeV}}{M_1} \right)^2 \left( \frac{m_\phi}{10^{10}~\mathrm{GeV}} \right)^3 \left( \frac{1~\mathrm{MeV}}{T} \right)^2 & \quad \mathrm{for}~ m_\phi \gg T \\
\left( \frac{y_1}{2.70\times 10^{-10}} \right)^2 \left( \frac{5.10 \times 10^{10}~\mathrm{GeV}}{M_1} \right)^2 \left( \frac{m_\phi}{10^{10}~\mathrm{GeV}} \right)^4 \left( \frac{10^3~\mathrm{GeV}}{T} \right)^3 & \quad \mathrm{for}~ m_\phi \ll T
\end{dcases} \,.
\label{eq:3b_rate1}
\end{eqnarray}
is larger than unity. For example, using the upper expression this means that for $m_\phi = 10^{10}~\mathrm{GeV}$ the decay of the dark scalars occurs before Big Bang Nucleosynthesis (BBN) ($T \gtrsim 1~\mathrm{MeV}$) if the dark sector coupling is larger than $6.04 \times 10^{-13}$. On the other hand, the dark scalars would quickly decay at $T \gtrsim m_\phi$ only if the dark sector coupling is larger than $2.70\times 10^{-10}$ according to the lower expression in Eq.~\eqref{eq:3b_rate1}. However, such a large coupling would overproduce dark matter particles according to Eq.~\eqref{eq:lighter_coupling} when requiring $m_\chi < m_\phi$. Hence, by considering the viable case $m_\phi \gg T$ and plugging Eq.~\eqref{eq:lighter_coupling} into Eq.~\eqref{eq:3b_rate1}, we get the following estimate for the temperature $T_\phi$ at which the three-body scalar decay occurs:
\begin{equation}
\left< \Gamma_\phi^{\rm 3-body} \right> (T_\phi) = \mathcal{H}(T_\phi) \quad \implies \quad T_\phi = 0.14~\mathrm{GeV} \left( \frac{5.10 \times 10^{10}~\mathrm{GeV}}{M_1} \right)^{1/2} \left( \frac{10^7~\mathrm{GeV}}{m_\chi} \right)^{1/2} \left( \frac{m_\phi}{10^{10}~\mathrm{GeV}} \right)^{3/2} \,.
\end{equation}
Then, requiring the dark scalar decay to occur before BBN provides the following lower bound on the dark scalar mass
\begin{equation}
T_\phi \gtrsim 1~\mathrm{MeV}  \quad \implies \quad m_\phi \gtrsim 3.62 \times 10^{8}~\mathrm{GeV} \left(\frac{M_1}{5.10 \times 10^{10}~\mathrm{GeV}}\right)^{1/3}\left( \frac{m_\chi}{10^{7}~\mathrm{GeV}} \right)^{1/3} \,,
\label{eq:lim_mf_1}
\end{equation}
which is valid as long as $m_\phi < M_1 = 5.10 \times 10^{10}~\mathrm{GeV}$.

In the case of the neutrino portal with the heaviest right-handed neutrino, we instead have 
\begin{eqnarray}
\frac{\left< \Gamma_\phi^{\rm 3-body} \right>}{\mathcal{H}} \simeq
\begin{dcases}
\left( \frac{y_2}{6.03 \times 10^{-11}} \right)^2 \left( \frac{3.28 \times 10^{14}~\mathrm{GeV}}{M_2} \right)^2 \left( \frac{m_\phi}{10^{10}~\mathrm{GeV}} \right)^3 \left( \frac{1~\mathrm{MeV}}{T} \right)^2 & \quad \mathrm{for}~ m_\phi \gg T \\
\left( \frac{y_2}{2.70 \times 10^{-8}} \right)^2 \left( \frac{3.28 \times 10^{14}~\mathrm{GeV}}{M_2} \right)^2 \left( \frac{m_\phi}{10^{10}~\mathrm{GeV}} \right)^4 \left( \frac{10^3~\mathrm{GeV}}{T} \right)^3 & \quad \mathrm{for}~ m_\phi \ll T
\end{dcases} \,,
\label{eq:3b_rate2}
\end{eqnarray}
and the temperature $T_\phi$ at which the three-body scalar decay occurs is 
\begin{equation}
\left< \Gamma_\phi^{\rm 3-body} \right> (T_\phi) =\mathcal{H}(T_\phi) \quad \implies \quad T_\phi = 0.12~\mathrm{GeV} \left( \frac{3.28 \times 10^{14}~\mathrm{GeV}}{M_2} \right)^{1/2} \left( \frac{10^7~\mathrm{GeV}}{m_\chi} \right)^{1/2} \left( \frac{m_\phi}{10^{10}~\mathrm{GeV}} \right)^{3/2} \,.
\end{equation}
The lower bound on the dark scalar mass allowing the dark scalar to decay before BBN is 
\begin{equation}
T_\phi \gtrsim 1~\mathrm{MeV}  \quad \implies \quad m_\phi \gtrsim 4.19 \times 10^{8}~\mathrm{GeV} \left(\frac{M_2}{3.28 \times 10^{14}~\mathrm{GeV}}\right)^{1/3}\left( \frac{m_\chi}{10^{7}~\mathrm{GeV}} \right)^{1/3} \,.
\label{eq:lim_mf_2}
\end{equation}
When fixing the right-handed neutrino masses to their benchmark values, such a lower bound is therefore almost independent of which of the two right-handed neutrinos is efficiently coupled to the dark sector. Moreover, it mildly depends on the dark matter mass $m_\chi$. However, as can been seen from Eq.s~\eqref{eq:3b_rate1} and~\eqref{eq:3b_rate2}, the coupling $y_2$ achieving the correct dark matter abundance is required to be larger with respect to the coupling $y_1$. This is in agreement with the results obtained in the ``heavier dark scalar'' mass ordering.

%%%%%%%%%%%%%%%%%%%%%%%%%%%
\subsection{Threshold values for neutrino portal couplings \label{sec:limits}}
%%%%%%%%%%%%%%%%%%%%%%%%%%%
\begin{figure}[t!]
\begin{center}
\includegraphics[width=0.48\textwidth]{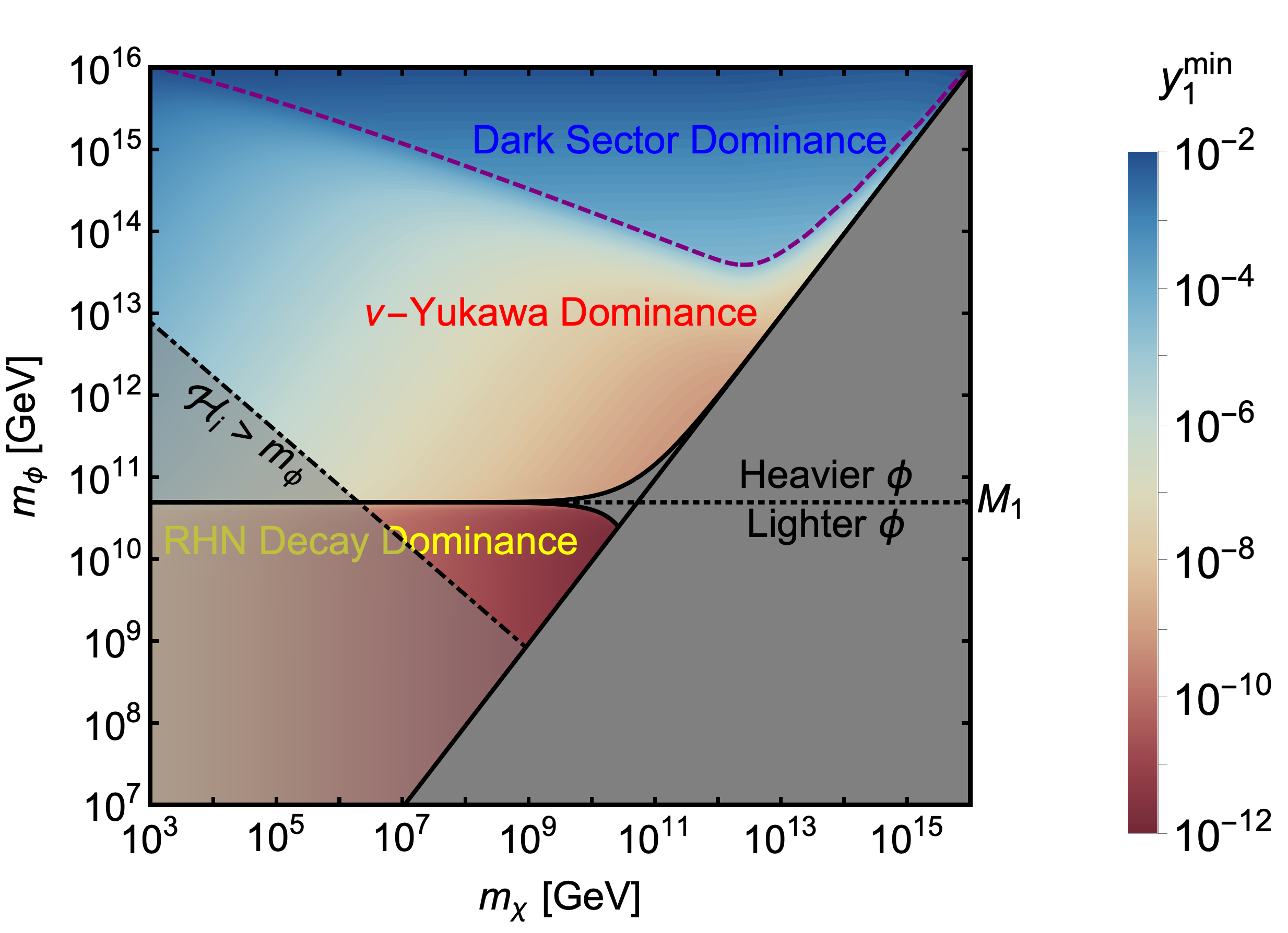}
\includegraphics[width=0.48\textwidth]{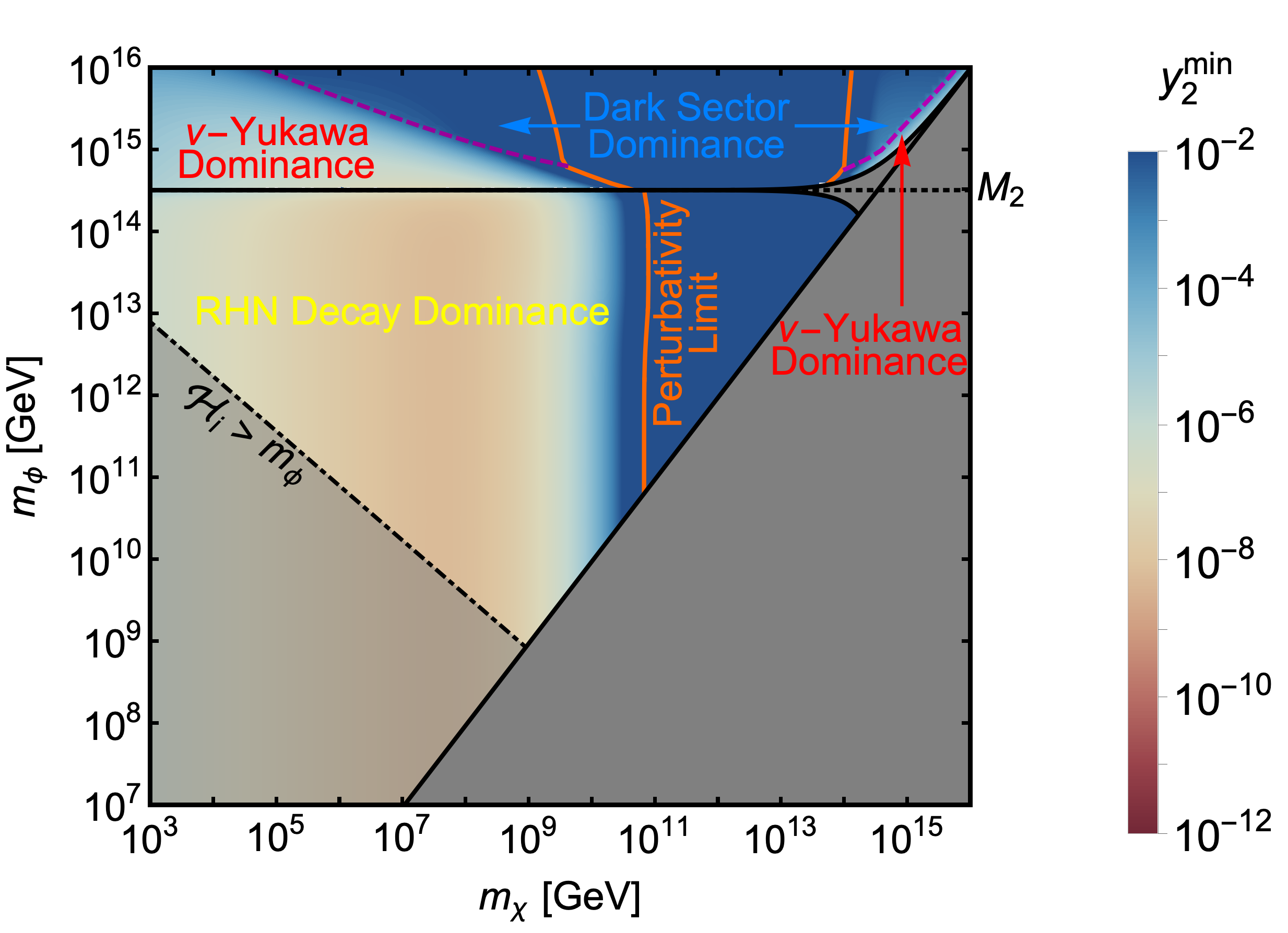}
\caption{\label{fig:lower_limit} Threshold values for the two dark sector couplings above which the gravity-mediated production is sub-dominant (a contribution to $\Omega_\mathrm{DM} h^2$ smaller than 50\%) as a function of the dark masses. The ``heavier dark scalar'' (``lighter dark scalar'') mass ordering corresponds to the region above (below) the horizontal dotted line. The purple dashed lines delimit the region where the dark sector and the neutrino Yukawa dominances occur above the dotted line. In the ``lighter $\phi$'' mass ordering the RHN decay dominance occurs. The regions excluded are: between the orange lines in the left plot due to perturbativity; the grey regions due to the kinetic constraints of the two mass orderings and the requirement $m_\chi < m_\phi$; the shaded region below the dot-dashed black line due to isocurvature perturbations ($\mathcal{H}_i > m_\phi$).}
\end{center}
\end{figure}

In order to further study the interplay between the gravity and neutrino portals, we estimate the threshold values $y_1^\mathrm{min}$ and $y_2^\mathrm{min}$ for the two dark sector couplings above which the gravity-mediated interaction does not provide the dominant contribution to the dark matter abundance. For given values of the two dark masses, these threshold values are obtained in correspondence of a 50\% contribution from the gravity-mediated processes to the dark matter relic density. Therefore, in the ``heavier dark scalar'' mass ordering, for given values of the two dark masses, the reheating temperature is uniquely fixed by the lower contour of the green band showing the gravity dominance in the plots of Fig.s~\ref{fig:HeavierPhi_1} and~\ref{fig:HeavierPhi_2}. For lower values of the reheating temperature, the gravitational production is suppressed and dark sector couplings larger than $y_1^\mathrm{min}$ and $y_2^\mathrm{min}$ are indeed required to achieve the correct dark matter abundance. In the ``lighter dark scalar'' mass ordering, the reheating temperature is instead fixed by following the dashed black line in Fig.~\ref{fig:Gravprod}. In this mass ordering, the required dark sector coupling is independent of the reheating temperature which only sets the gravitational production.

The threshold values $y_1^\mathrm{min}$ and $y_2^\mathrm{min}$ obtained in this way are reported in Fig.~\ref{fig:lower_limit}. In the two plots, the parameter space is divided in the two different mass orderings (heavier and lighter $\phi$) with respect to the masses of the right-handed neutrinos (horizontal dotted black lines). Moreover, we highlight the regions where the different processes induced by the neutrino portal dominate the dark matter production:
\begin{itemize}
\item {\bf Dark sector dominance.} In the ``heavier dark scalar'' mass ordering, the dark sector scatterings dominate the dark matter production in the region above the purple dashed line and below the orange lines (if present).
\item {\bf Neutrino Yukawa dominance.} In the ``heavier dark scalar'' mass ordering, the neutrino Yukawa dominance occurs between the horizontal black dotted line and the purple dashed line.
\item {\bf RHN decay dominance.} In the ``lighter dark scalar'' mass ordering (below the horizontal black dotted line), the main contribution to the dark matter relic density comes from the two-body decays of right-handed neutrinos.
\end{itemize}
Remarkably, the dominance of a class of processes is set by the strength of the neutrino portal coupling. For example, looking at the right plot, very small dark sector couplings (reddish colors) generally correspond to the RHN decay dominance, intermediate couplings to the neutrino Yukawa dominance, and large couplings approaching the perturbativity limit (bluish colors) to the dark sector dominance (the darkest blue region between the orange lines in the left plots is excluded by the perturbativity constraint). Moreover, we note once again that the dark sector coupling $y_2$ to the second right-handed neutrino has to take larger values when compared to $y_1$ in order to obtain the correct dark matter abundance. In the plots, the grey regions are forbidden in agreement with the requirement of a fermionic dark matter ($m_\chi < m_\phi$) and the kinetic constraints of the two mass orderings (small triangles displayed for $m_\phi$ of the order of the right-handed neutrino mass). The shaded regions below the dot-dashed black line (labeled by $\mathcal{H}_i > m_\phi$) are instead excluded by the constraints on the isocurvature perturbations. As can be seen in the figure, these last constraints disfavour the RHN decay dominance for the lightest right-handed neutrino which is allowed only in a small region of the model parameter space. On the other hand, for the heaviest right-handed neutrino the RHN decay dominance occurs in a large part of the parameter space so favouring lighter dark matter masses.

%%%%%%%%%%%%%%%%%%%%
\section{Conclusions \label{sec:con}}
%%%%%%%%%%%%%%%%%%%%

In the present paper we have studied a minimal and realistic model accounting for neutrino oscillation data, the baryon asymmetry of the universe, and a viable dark matter candidate. The model is based on the classic type I seesaw mechanism, augmented by a dark sector which may be accessed either through gravitational couplings or via the right-handed neutrino portal. The minimal type I seesaw model considered involves two right-handed neutrinos according to the Littlest seesaw model~\cite{King:2013iva, Bjorkeroth:2014vha, King:2015dvf,Bjorkeroth:2015ora,Bjorkeroth:2015tsa,King:2016yvg,Ballett:2016yod,King:2018fqh}. Accounting for the observed neutrino data and requiring a successful leptogenesis completely fix the neutrino Yukawa sector of the model~\cite{King:2018fqh}. The dark sector is charged under a global $U(1)_D$ symmetry and consists of a dark fermion and a dark scalar with the former assumed to be the lighter dark particle. In this framework, we have investigated the production of dark matter particles in the early universe and explored the model parameter space which achieves the observed dark matter abundance. In particular, we have analyzed in detail the interplay between the RH neutrino and gravity portals in the dark matter production. The neutrino portal induces two distinct processes: the neutrino Yukawa scatterings where the right-handed neutrinos act as mediators, and the dark sector scatterings where they annihilate into dark particles. Moreover, when the right-handed neutrinos are heavier than the dark particles, the dark matter $\chi$ can be directly produced by the two-body decays $N_\mathrm{R} \rightarrow \chi + \phi$ and the subsequent three-body decays of $\phi$ particles.

Firstly, we have shown that, ignoring the RH neutrino portal couplings, the gravity-mediated interaction provides an upper limit on the reheating temperature as a function of the dark matter mass $m_\chi$ (see Fig.~\ref{fig:Gravprod}). In particular, for $7 \times 10^{5}~\mathrm{GeV} \lesssim m_\chi \lesssim 5 \times 10^{15}~\mathrm{GeV}$ the reheating temperature $T_\mathrm{RH}$ has to be smaller than the second right-handed neutrino mass ($M_2 = 3.28 \times 10^{14}~\mathrm{GeV}$) in order to avoid dark matter's overproduction through the gravitational interaction. In this mass range, the dark matter production through the neutrino portal with the second right-handed neutrino is therefore highly suppressed in the instantaneous reheating scenario. On the other hand, the second right-handed neutrino is favourably allowed to drive the dark matter production for smaller dark matter masses. The same suppression does not occur for the first right-handed neutrino being its mass ($M_1 = 5.10 \times 10^{10}~\mathrm{GeV}$) always lighter than the upper bound on the reheating temperature. 

Including both gravity and the RH neutrino portal couplings, in Fig.s~\ref{fig:HeavierPhi_1} and~\ref{fig:HeavierPhi_2} we have then respectively reported the values for the RH neutrino portal couplings $y_1$ and $y_2$ required to achieve the correct dark matter abundance when the dark scalars are heavier than the right-handed neutrinos (``heavier dark scalar'' mass ordering). In these plots, we have highlighted the regions of the parameter space where the three different processes (gravity-mediated, neutrino Yukawa and dark sector scatterings) dominate the dark matter production. We have demonstrated that larger values for the coupling $y_2$ are required to achieve a viable dark matter candidate with respect to the coupling $y_1$. This result has two important implications. Firstly, the neutrino portal with the second right-handed neutrino is not able to overcome the gravitational production in the mass range $1.37 \times 10^{9}~\mathrm{GeV} \lesssim m_\chi \lesssim 1.26 \times 10^{14}~\mathrm{GeV}$ due to the perturbativity constraint. In the absence of an efficient coupling $y_1$ with the first right-handed neutrino, such a dark matter candidate can be truly regarded as a Planckian Interacting Dark Matter candidate~\cite{Garny:2015sjg,Tang:2016vch,Tang:2017hvq,Garny:2017kha,Bernal:2018qlk,Garny:2018grs,Hashiba:2018tbu}. Secondly, the neutrino Yukawa processes mediated by the first right-handed neutrino (non-zero $y_1$ coupling) dominate the dark matter production in a large part of the model parameter space. In particular, for a dark scalar mass $m_\phi$ of the same order of $M_1$ (bottom panels in Fig.~\ref{fig:HeavierPhi_1}), the neutrino Yukawa dominance can easily occur with values of the coupling $y_1$ smaller than $10^6$. In this case, the dark sector dominance is instead excluded by the requirement of successful leptogenesis ($T_\mathrm{RH} > M_1$). Remarkably, we have therefore shown that the neutrino Yukawa dominance providing a direct link between dark matter and neutrinos can be easily achieved despite the presence of gravity-mediated processes. Although more constrained, the neutrino Yukawa dominance can also occur for smaller dark matter masses ($m_\chi < \mathcal{O}(10^9~\mathrm{GeV})$) with the second right-handed neutrino acting as a mediator as displayed in Fig.~\ref{fig:HeavierPhi_2}.

When the dark scalars are lighter than the right-handed neutrinos (``lighter dark scalar'' mass ordering), we have derived the analytical expressions in Eq.s~\eqref{eq:lighter_coupling} and~\eqref{eq:heavier_coupling} for the two neutrino portal couplings $y_1$ and $y_2$, respectively, providing the correct dark matter abundance. In particular, they are independent of the reheating temperature as long as it is larger than the RH neutrino masses. Moreover, we have shown that in order to avoid late dark scalar decays during Big Bang Nucleosynthesis, the dark scalar mass has to be larger than $\sim 4 \times 10^{8}~\mathrm{GeV}$ for a dark matter mass of $10^7~\mathrm{GeV}$ (see Eq.s~\eqref{eq:lim_mf_1} and~\eqref{eq:lim_mf_2}).

Finally, we have discussed in Fig.~\ref{fig:lower_limit} the threshold values for the RH neutrino portal couplings above which the gravity-mediated production is sub-dominant. For given values of the two dark masses, a Planckian Interacting Dark Matter candidate is achieved if the neutrino portal couplings are smaller than $y_1^\mathrm{min}$ and $y_2^\mathrm{min}$. Moreover, in these plots we have clearly shown that there exist distinct regions where the different processes induced by the neutrino portal separately dominate the dark matter production. In particular, very small dark sector couplings ($y_1 = \mathcal{O}(10^{-10})$ and $y_2 = \mathcal{O}(10^{-8})$) favour the dominance of the two-body decay of right-handed neutrinos in the ``lighter dark scalar'' mass ordering. However, such a dominance is mostly constrained by the limit on isocurvature perturbations  in the case of a coupling to the lightest right-handed neutrino, while it is preferred in the case of a non-zero $y_2$ coupling. In the ``heavier dark scalar'' mass ordering, the neutrino Yukawa dominance is favoured by intermediate couplings ($y_1 = \mathcal{O}(10^{-6})$ and $y_2 = \mathcal{O}(10^{-4})$) while the dark sector dominance generally requires large couplings ($y_1 = \mathcal{O}(10^{-2})$ and $y_2 = \mathcal{O}(1)$). Such mass ordering is highly constrained in the case of a coupling only to the heaviest right-handed neutrino due to the upper limit on the reheating temperature of the universe imposed by the gravitational production.

In conclusion, the classic type I seesaw mechanism with very heavy right-handed neutrinos, remains an attractive and plausible explanation of neutrino mass and mixing, which can accommodate not only leptogenesis, but also dark matter, via a simple dark sector extension, which may be accessed either gravitationally or via the RH neutrino portal, or a delicate interplay of both mechanisms, resulting in FIMP dark matter produced via the ``freeze-in'' mechanism.

%%%%%%%%%%%%%%%%%%%%
\section*{Acknowledgments}
BF acknowledges the Chinese Scholarship Council (CSC) Grant No.\ 201809210011 under agreements [2018]3101 and [2019]536. SFK acknowledges the STFC Consolidated Grant ST/L000296/1 and the European Union's Horizon 2020 Research and Innovation programme under Marie Sk\l {}odowska-Curie grant agreements Elusives ITN No.\ 674896 and InvisiblesPlus RISE No.\ 690575.

%%%%%%%%%%%%%%%%%%%%
\bibliography{DarkMatter}

\begin{thebibliography}{143}%
\makeatletter
\providecommand \@ifxundefined [1]{%
 \@ifx{#1\undefined}
}%
\providecommand \@ifnum [1]{%
 \ifnum #1\expandafter \@firstoftwo
 \else \expandafter \@secondoftwo
 \fi
}%
\providecommand \@ifx [1]{%
 \ifx #1\expandafter \@firstoftwo
 \else \expandafter \@secondoftwo
 \fi
}%
\providecommand \natexlab [1]{#1}%
\providecommand \enquote  [1]{``#1''}%
\providecommand \bibnamefont  [1]{#1}%
\providecommand \bibfnamefont [1]{#1}%
\providecommand \citenamefont [1]{#1}%
\providecommand \href@noop [0]{\@secondoftwo}%
\providecommand \href [0]{\begingroup \@sanitize@url \@href}%
\providecommand \@href[1]{\@@startlink{#1}\@@href}%
\providecommand \@@href[1]{\endgroup#1\@@endlink}%
\providecommand \@sanitize@url [0]{\catcode `\\12\catcode `\$12\catcode
  `\&12\catcode `\#12\catcode `\^12\catcode `\_12\catcode `\%12\relax}%
\providecommand \@@startlink[1]{}%
\providecommand \@@endlink[0]{}%
\providecommand \url  [0]{\begingroup\@sanitize@url \@url }%
\providecommand \@url [1]{\endgroup\@href {#1}{\urlprefix }}%
\providecommand \urlprefix  [0]{URL }%
\providecommand \Eprint [0]{\href }%
\providecommand \doibase [0]{http://dx.doi.org/}%
\providecommand \selectlanguage [0]{\@gobble}%
\providecommand \bibinfo  [0]{\@secondoftwo}%
\providecommand \bibfield  [0]{\@secondoftwo}%
\providecommand \translation [1]{[#1]}%
\providecommand \BibitemOpen [0]{}%
\providecommand \bibitemStop [0]{}%
\providecommand \bibitemNoStop [0]{.\EOS\space}%
\providecommand \EOS [0]{\spacefactor3000\relax}%
\providecommand \BibitemShut  [1]{\csname bibitem#1\endcsname}%
\let\auto@bib@innerbib\@empty
%</preamble>
\bibitem [{\citenamefont {{Ohlsson}}(2016)}]{2016NuPhB.908....1O}%
  \BibitemOpen
  \bibfield  {author} {\bibinfo {author} {\bibfnamefont {T.}~\bibnamefont
  {{Ohlsson}}},\ }\href {\doibase 10.1016/j.nuclphysb.2016.04.024} {\bibfield
  {journal} {\bibinfo  {journal} {Nuclear Physics B}\ }\textbf {\bibinfo
  {volume} {908}},\ \bibinfo {pages} {1} (\bibinfo {year} {2016})}\BibitemShut
  {NoStop}%
\bibitem [{\citenamefont {Minkowski}(1977)}]{Minkowski:1977sc}%
  \BibitemOpen
  \bibfield  {author} {\bibinfo {author} {\bibfnamefont {P.}~\bibnamefont
  {Minkowski}},\ }\href {\doibase 10.1016/0370-2693(77)90435-X} {\bibfield
  {journal} {\bibinfo  {journal} {Phys. Lett.}\ }\textbf {\bibinfo {volume}
  {67B}},\ \bibinfo {pages} {421} (\bibinfo {year} {1977})}\BibitemShut
  {NoStop}%
%%CITATION = PHLTA,67B,421;%%
\bibitem [{\citenamefont {Yanagida}(1979)}]{Yanagida:1979as}%
  \BibitemOpen
  \bibfield  {author} {\bibinfo {author} {\bibfnamefont {T.}~\bibnamefont
  {Yanagida}},\ }\bibfield  {booktitle} {\emph {\bibinfo {booktitle}
  {{Proceedings: Workshop on the Unified Theories and the Baryon Number in the
  Universe: Tsukuba, Japan, February 13-14, 1979}}},\ }\href@noop {} {\bibfield
   {journal} {\bibinfo  {journal} {Conf. Proc.}\ }\textbf {\bibinfo {volume}
  {C7902131}},\ \bibinfo {pages} {95} (\bibinfo {year} {1979})}\BibitemShut
  {NoStop}%
%%CITATION = CONFP,C7902131,95;%%
\bibitem [{\citenamefont {Gell-Mann}\ \emph {et~al.}(1979)\citenamefont
  {Gell-Mann}, \citenamefont {Ramond},\ and\ \citenamefont
  {Slansky}}]{GellMann:1980vs}%
  \BibitemOpen
  \bibfield  {author} {\bibinfo {author} {\bibfnamefont {M.}~\bibnamefont
  {Gell-Mann}}, \bibinfo {author} {\bibfnamefont {P.}~\bibnamefont {Ramond}}, \
  and\ \bibinfo {author} {\bibfnamefont {R.}~\bibnamefont {Slansky}},\
  }\bibfield  {booktitle} {\emph {\bibinfo {booktitle} {{Supergravity Workshop
  Stony Brook, New York, September 27-28, 1979}}},\ }\href@noop {} {\bibfield
  {journal} {\bibinfo  {journal} {Conf. Proc.}\ }\textbf {\bibinfo {volume}
  {C790927}},\ \bibinfo {pages} {315} (\bibinfo {year} {1979})},\ \Eprint
  {http://arxiv.org/abs/1306.4669} {arXiv:1306.4669 [hep-th]} \BibitemShut
  {NoStop}%
%%CITATION = ARXIV:1306.4669;%%
\bibitem [{\citenamefont {Schechter}\ and\ \citenamefont
  {Valle}(1980)}]{Schechter:1980gr}%
  \BibitemOpen
  \bibfield  {author} {\bibinfo {author} {\bibfnamefont {J.}~\bibnamefont
  {Schechter}}\ and\ \bibinfo {author} {\bibfnamefont {J.~W.~F.}\ \bibnamefont
  {Valle}},\ }\href {\doibase 10.1103/PhysRevD.22.2227} {\bibfield  {journal}
  {\bibinfo  {journal} {Phys. Rev.}\ }\textbf {\bibinfo {volume} {D22}},\
  \bibinfo {pages} {2227} (\bibinfo {year} {1980})}\BibitemShut {NoStop}%
%%CITATION = PHRVA,D22,2227;%%
\bibitem [{\citenamefont {Mohapatra}\ and\ \citenamefont
  {Senjanovic}(1980)}]{Mohapatra:1979ia}%
  \BibitemOpen
  \bibfield  {author} {\bibinfo {author} {\bibfnamefont {R.~N.}\ \bibnamefont
  {Mohapatra}}\ and\ \bibinfo {author} {\bibfnamefont {G.}~\bibnamefont
  {Senjanovic}},\ }\href {\doibase 10.1103/PhysRevLett.44.912} {\bibfield
  {journal} {\bibinfo  {journal} {Phys. Rev. Lett.}\ }\textbf {\bibinfo
  {volume} {44}},\ \bibinfo {pages} {912} (\bibinfo {year} {1980})},\ \bibinfo
  {note} {[,231(1979)]}\BibitemShut {NoStop}%
%%CITATION = PRLTA,44,912;%%
\bibitem [{\citenamefont {Mohapatra}\ and\ \citenamefont
  {Senjanovic}(1981)}]{Mohapatra:1980yp}%
  \BibitemOpen
  \bibfield  {author} {\bibinfo {author} {\bibfnamefont {R.~N.}\ \bibnamefont
  {Mohapatra}}\ and\ \bibinfo {author} {\bibfnamefont {G.}~\bibnamefont
  {Senjanovic}},\ }\href {\doibase 10.1103/PhysRevD.23.165} {\bibfield
  {journal} {\bibinfo  {journal} {Phys. Rev.}\ }\textbf {\bibinfo {volume}
  {D23}},\ \bibinfo {pages} {165} (\bibinfo {year} {1981})}\BibitemShut
  {NoStop}%
%%CITATION = PHRVA,D23,165;%%
\bibitem [{\citenamefont {King}(2013)}]{King:2013iva}%
  \BibitemOpen
  \bibfield  {author} {\bibinfo {author} {\bibfnamefont {S.~F.}\ \bibnamefont
  {King}},\ }\href {\doibase 10.1007/JHEP07(2013)137} {\bibfield  {journal}
  {\bibinfo  {journal} {JHEP}\ }\textbf {\bibinfo {volume} {07}},\ \bibinfo
  {pages} {137} (\bibinfo {year} {2013})},\ \Eprint
  {http://arxiv.org/abs/1304.6264} {arXiv:1304.6264 [hep-ph]} \BibitemShut
  {NoStop}%
%%CITATION = ARXIV:1304.6264;%%
\bibitem [{\citenamefont {Björkeroth}\ and\ \citenamefont
  {King}(2015)}]{Bjorkeroth:2014vha}%
  \BibitemOpen
  \bibfield  {author} {\bibinfo {author} {\bibfnamefont {F.}~\bibnamefont
  {Björkeroth}}\ and\ \bibinfo {author} {\bibfnamefont {S.~F.}\ \bibnamefont
  {King}},\ }\href {\doibase 10.1088/0954-3899/42/12/125002} {\bibfield
  {journal} {\bibinfo  {journal} {J. Phys.}\ }\textbf {\bibinfo {volume}
  {G42}},\ \bibinfo {pages} {125002} (\bibinfo {year} {2015})},\ \Eprint
  {http://arxiv.org/abs/1412.6996} {arXiv:1412.6996 [hep-ph]} \BibitemShut
  {NoStop}%
%%CITATION = ARXIV:1412.6996;%%
\bibitem [{\citenamefont {King}(2016)}]{King:2015dvf}%
  \BibitemOpen
  \bibfield  {author} {\bibinfo {author} {\bibfnamefont {S.~F.}\ \bibnamefont
  {King}},\ }\href {\doibase 10.1007/JHEP02(2016)085} {\bibfield  {journal}
  {\bibinfo  {journal} {JHEP}\ }\textbf {\bibinfo {volume} {02}},\ \bibinfo
  {pages} {085} (\bibinfo {year} {2016})},\ \Eprint
  {http://arxiv.org/abs/1512.07531} {arXiv:1512.07531 [hep-ph]} \BibitemShut
  {NoStop}%
%%CITATION = ARXIV:1512.07531;%%
\bibitem [{\citenamefont {Björkeroth}\ \emph
  {et~al.}(2015{\natexlab{a}})\citenamefont {Björkeroth}, \citenamefont
  {de~Anda}, \citenamefont {de~Medeiros~Varzielas},\ and\ \citenamefont
  {King}}]{Bjorkeroth:2015ora}%
  \BibitemOpen
  \bibfield  {author} {\bibinfo {author} {\bibfnamefont {F.}~\bibnamefont
  {Björkeroth}}, \bibinfo {author} {\bibfnamefont {F.~J.}\ \bibnamefont
  {de~Anda}}, \bibinfo {author} {\bibfnamefont {I.}~\bibnamefont
  {de~Medeiros~Varzielas}}, \ and\ \bibinfo {author} {\bibfnamefont {S.~F.}\
  \bibnamefont {King}},\ }\href {\doibase 10.1007/JHEP06(2015)141} {\bibfield
  {journal} {\bibinfo  {journal} {JHEP}\ }\textbf {\bibinfo {volume} {06}},\
  \bibinfo {pages} {141} (\bibinfo {year} {2015}{\natexlab{a}})},\ \Eprint
  {http://arxiv.org/abs/1503.03306} {arXiv:1503.03306 [hep-ph]} \BibitemShut
  {NoStop}%
%%CITATION = ARXIV:1503.03306;%%
\bibitem [{\citenamefont {Björkeroth}\ \emph
  {et~al.}(2015{\natexlab{b}})\citenamefont {Björkeroth}, \citenamefont
  {de~Anda}, \citenamefont {de~Medeiros~Varzielas},\ and\ \citenamefont
  {King}}]{Bjorkeroth:2015tsa}%
  \BibitemOpen
  \bibfield  {author} {\bibinfo {author} {\bibfnamefont {F.}~\bibnamefont
  {Björkeroth}}, \bibinfo {author} {\bibfnamefont {F.~J.}\ \bibnamefont
  {de~Anda}}, \bibinfo {author} {\bibfnamefont {I.}~\bibnamefont
  {de~Medeiros~Varzielas}}, \ and\ \bibinfo {author} {\bibfnamefont {S.~F.}\
  \bibnamefont {King}},\ }\href {\doibase 10.1007/JHEP10(2015)104} {\bibfield
  {journal} {\bibinfo  {journal} {JHEP}\ }\textbf {\bibinfo {volume} {10}},\
  \bibinfo {pages} {104} (\bibinfo {year} {2015}{\natexlab{b}})},\ \Eprint
  {http://arxiv.org/abs/1505.05504} {arXiv:1505.05504 [hep-ph]} \BibitemShut
  {NoStop}%
%%CITATION = ARXIV:1505.05504;%%
\bibitem [{\citenamefont {King}\ and\ \citenamefont
  {Luhn}(2016)}]{King:2016yvg}%
  \BibitemOpen
  \bibfield  {author} {\bibinfo {author} {\bibfnamefont {S.~F.}\ \bibnamefont
  {King}}\ and\ \bibinfo {author} {\bibfnamefont {C.}~\bibnamefont {Luhn}},\
  }\href {\doibase 10.1007/JHEP09(2016)023} {\bibfield  {journal} {\bibinfo
  {journal} {JHEP}\ }\textbf {\bibinfo {volume} {09}},\ \bibinfo {pages} {023}
  (\bibinfo {year} {2016})},\ \Eprint {http://arxiv.org/abs/1607.05276}
  {arXiv:1607.05276 [hep-ph]} \BibitemShut {NoStop}%
%%CITATION = ARXIV:1607.05276;%%
\bibitem [{\citenamefont {Ballett}\ \emph {et~al.}(2017)\citenamefont
  {Ballett}, \citenamefont {King}, \citenamefont {Pascoli}, \citenamefont
  {Prouse},\ and\ \citenamefont {Wang}}]{Ballett:2016yod}%
  \BibitemOpen
  \bibfield  {author} {\bibinfo {author} {\bibfnamefont {P.}~\bibnamefont
  {Ballett}}, \bibinfo {author} {\bibfnamefont {S.~F.}\ \bibnamefont {King}},
  \bibinfo {author} {\bibfnamefont {S.}~\bibnamefont {Pascoli}}, \bibinfo
  {author} {\bibfnamefont {N.~W.}\ \bibnamefont {Prouse}}, \ and\ \bibinfo
  {author} {\bibfnamefont {T.}~\bibnamefont {Wang}},\ }\href {\doibase
  10.1007/JHEP03(2017)110} {\bibfield  {journal} {\bibinfo  {journal} {JHEP}\
  }\textbf {\bibinfo {volume} {03}},\ \bibinfo {pages} {110} (\bibinfo {year}
  {2017})},\ \Eprint {http://arxiv.org/abs/1612.01999} {arXiv:1612.01999
  [hep-ph]} \BibitemShut {NoStop}%
%%CITATION = ARXIV:1612.01999;%%
\bibitem [{\citenamefont {King}\ \emph {et~al.}(2018)\citenamefont {King},
  \citenamefont {Molina~Sedgwick},\ and\ \citenamefont
  {Rowley}}]{King:2018fqh}%
  \BibitemOpen
  \bibfield  {author} {\bibinfo {author} {\bibfnamefont {S.~F.}\ \bibnamefont
  {King}}, \bibinfo {author} {\bibfnamefont {S.}~\bibnamefont
  {Molina~Sedgwick}}, \ and\ \bibinfo {author} {\bibfnamefont {S.~J.}\
  \bibnamefont {Rowley}},\ }\href {\doibase 10.1007/JHEP10(2018)184} {\bibfield
   {journal} {\bibinfo  {journal} {JHEP}\ }\textbf {\bibinfo {volume} {10}},\
  \bibinfo {pages} {184} (\bibinfo {year} {2018})},\ \Eprint
  {http://arxiv.org/abs/1808.01005} {arXiv:1808.01005 [hep-ph]} \BibitemShut
  {NoStop}%
%%CITATION = ARXIV:1808.01005;%%
\bibitem [{\citenamefont {Aghanim}\ \emph {et~al.}(2018)\citenamefont {Aghanim}
  \emph {et~al.}}]{Aghanim:2018eyx}%
  \BibitemOpen
  \bibfield  {author} {\bibinfo {author} {\bibfnamefont {N.}~\bibnamefont
  {Aghanim}} \emph {et~al.} (\bibinfo {collaboration} {Planck}),\ }\href@noop
  {} {\  (\bibinfo {year} {2018})},\ \Eprint {http://arxiv.org/abs/1807.06209}
  {arXiv:1807.06209 [astro-ph.CO]} \BibitemShut {NoStop}%
\bibitem [{\citenamefont {Caldwell}\ and\ \citenamefont
  {Mohapatra}(1993)}]{Caldwell:1993kn}%
  \BibitemOpen
  \bibfield  {author} {\bibinfo {author} {\bibfnamefont {D.~O.}\ \bibnamefont
  {Caldwell}}\ and\ \bibinfo {author} {\bibfnamefont {R.~N.}\ \bibnamefont
  {Mohapatra}},\ }\href {\doibase 10.1103/PhysRevD.48.3259} {\bibfield
  {journal} {\bibinfo  {journal} {Phys. Rev.}\ }\textbf {\bibinfo {volume}
  {D48}},\ \bibinfo {pages} {3259} (\bibinfo {year} {1993})},\ \bibinfo {note}
  {[,603(1993)]}\BibitemShut {NoStop}%
%%CITATION = PHRVA,D48,3259;%%
\bibitem [{\citenamefont {Mohapatra}\ and\ \citenamefont
  {Perez-Lorenzana}(2003)}]{Mohapatra:2002ug}%
  \BibitemOpen
  \bibfield  {author} {\bibinfo {author} {\bibfnamefont {R.~N.}\ \bibnamefont
  {Mohapatra}}\ and\ \bibinfo {author} {\bibfnamefont {A.}~\bibnamefont
  {Perez-Lorenzana}},\ }\href {\doibase 10.1103/PhysRevD.67.075015} {\bibfield
  {journal} {\bibinfo  {journal} {Phys. Rev.}\ }\textbf {\bibinfo {volume}
  {D67}},\ \bibinfo {pages} {075015} (\bibinfo {year} {2003})},\ \Eprint
  {http://arxiv.org/abs/hep-ph/0212254} {arXiv:hep-ph/0212254 [hep-ph]}
  \BibitemShut {NoStop}%
%%CITATION = HEP-PH/0212254;%%
\bibitem [{\citenamefont {Krauss}\ \emph {et~al.}(2003)\citenamefont {Krauss},
  \citenamefont {Nasri},\ and\ \citenamefont {Trodden}}]{Krauss:2002px}%
  \BibitemOpen
  \bibfield  {author} {\bibinfo {author} {\bibfnamefont {L.~M.}\ \bibnamefont
  {Krauss}}, \bibinfo {author} {\bibfnamefont {S.}~\bibnamefont {Nasri}}, \
  and\ \bibinfo {author} {\bibfnamefont {M.}~\bibnamefont {Trodden}},\ }\href
  {\doibase 10.1103/PhysRevD.67.085002} {\bibfield  {journal} {\bibinfo
  {journal} {Phys. Rev.}\ }\textbf {\bibinfo {volume} {D67}},\ \bibinfo {pages}
  {085002} (\bibinfo {year} {2003})},\ \Eprint
  {http://arxiv.org/abs/hep-ph/0210389} {arXiv:hep-ph/0210389 [hep-ph]}
  \BibitemShut {NoStop}%
%%CITATION = HEP-PH/0210389;%%
\bibitem [{\citenamefont {Ma}(2006{\natexlab{a}})}]{Ma:2006km}%
  \BibitemOpen
  \bibfield  {author} {\bibinfo {author} {\bibfnamefont {E.}~\bibnamefont
  {Ma}},\ }\href {\doibase 10.1103/PhysRevD.73.077301} {\bibfield  {journal}
  {\bibinfo  {journal} {Phys. Rev.}\ }\textbf {\bibinfo {volume} {D73}},\
  \bibinfo {pages} {077301} (\bibinfo {year} {2006}{\natexlab{a}})},\ \Eprint
  {http://arxiv.org/abs/hep-ph/0601225} {arXiv:hep-ph/0601225 [hep-ph]}
  \BibitemShut {NoStop}%
%%CITATION = HEP-PH/0601225;%%
\bibitem [{\citenamefont {Asaka}\ \emph {et~al.}(2005)\citenamefont {Asaka},
  \citenamefont {Blanchet},\ and\ \citenamefont {Shaposhnikov}}]{Asaka:2005an}%
  \BibitemOpen
  \bibfield  {author} {\bibinfo {author} {\bibfnamefont {T.}~\bibnamefont
  {Asaka}}, \bibinfo {author} {\bibfnamefont {S.}~\bibnamefont {Blanchet}}, \
  and\ \bibinfo {author} {\bibfnamefont {M.}~\bibnamefont {Shaposhnikov}},\
  }\href {\doibase 10.1016/j.physletb.2005.09.070} {\bibfield  {journal}
  {\bibinfo  {journal} {Phys. Lett.}\ }\textbf {\bibinfo {volume} {B631}},\
  \bibinfo {pages} {151} (\bibinfo {year} {2005})},\ \Eprint
  {http://arxiv.org/abs/hep-ph/0503065} {arXiv:hep-ph/0503065 [hep-ph]}
  \BibitemShut {NoStop}%
%%CITATION = HEP-PH/0503065;%%
\bibitem [{\citenamefont {Boehm}\ \emph {et~al.}(2008)\citenamefont {Boehm},
  \citenamefont {Farzan}, \citenamefont {Hambye}, \citenamefont
  {Palomares-Ruiz},\ and\ \citenamefont {Pascoli}}]{Boehm:2006mi}%
  \BibitemOpen
  \bibfield  {author} {\bibinfo {author} {\bibfnamefont {C.}~\bibnamefont
  {Boehm}}, \bibinfo {author} {\bibfnamefont {Y.}~\bibnamefont {Farzan}},
  \bibinfo {author} {\bibfnamefont {T.}~\bibnamefont {Hambye}}, \bibinfo
  {author} {\bibfnamefont {S.}~\bibnamefont {Palomares-Ruiz}}, \ and\ \bibinfo
  {author} {\bibfnamefont {S.}~\bibnamefont {Pascoli}},\ }\href {\doibase
  10.1103/PhysRevD.77.043516} {\bibfield  {journal} {\bibinfo  {journal} {Phys.
  Rev.}\ }\textbf {\bibinfo {volume} {D77}},\ \bibinfo {pages} {043516}
  (\bibinfo {year} {2008})},\ \Eprint {http://arxiv.org/abs/hep-ph/0612228}
  {arXiv:hep-ph/0612228 [hep-ph]} \BibitemShut {NoStop}%
%%CITATION = HEP-PH/0612228;%%
\bibitem [{\citenamefont {Kubo}\ \emph {et~al.}(2006)\citenamefont {Kubo},
  \citenamefont {Ma},\ and\ \citenamefont {Suematsu}}]{Kubo:2006yx}%
  \BibitemOpen
  \bibfield  {author} {\bibinfo {author} {\bibfnamefont {J.}~\bibnamefont
  {Kubo}}, \bibinfo {author} {\bibfnamefont {E.}~\bibnamefont {Ma}}, \ and\
  \bibinfo {author} {\bibfnamefont {D.}~\bibnamefont {Suematsu}},\ }\href
  {\doibase 10.1016/j.physletb.2006.08.085} {\bibfield  {journal} {\bibinfo
  {journal} {Phys. Lett.}\ }\textbf {\bibinfo {volume} {B642}},\ \bibinfo
  {pages} {18} (\bibinfo {year} {2006})},\ \Eprint
  {http://arxiv.org/abs/hep-ph/0604114} {arXiv:hep-ph/0604114 [hep-ph]}
  \BibitemShut {NoStop}%
%%CITATION = HEP-PH/0604114;%%
\bibitem [{\citenamefont {Ma}(2006{\natexlab{b}})}]{Ma:2006fn}%
  \BibitemOpen
  \bibfield  {author} {\bibinfo {author} {\bibfnamefont {E.}~\bibnamefont
  {Ma}},\ }\href {\doibase 10.1142/S0217732306021141} {\bibfield  {journal}
  {\bibinfo  {journal} {Mod. Phys. Lett.}\ }\textbf {\bibinfo {volume} {A21}},\
  \bibinfo {pages} {1777} (\bibinfo {year} {2006}{\natexlab{b}})},\ \Eprint
  {http://arxiv.org/abs/hep-ph/0605180} {arXiv:hep-ph/0605180 [hep-ph]}
  \BibitemShut {NoStop}%
%%CITATION = HEP-PH/0605180;%%
\bibitem [{\citenamefont {Hambye}\ \emph {et~al.}(2007)\citenamefont {Hambye},
  \citenamefont {Kannike}, \citenamefont {Ma},\ and\ \citenamefont
  {Raidal}}]{Hambye:2006zn}%
  \BibitemOpen
  \bibfield  {author} {\bibinfo {author} {\bibfnamefont {T.}~\bibnamefont
  {Hambye}}, \bibinfo {author} {\bibfnamefont {K.}~\bibnamefont {Kannike}},
  \bibinfo {author} {\bibfnamefont {E.}~\bibnamefont {Ma}}, \ and\ \bibinfo
  {author} {\bibfnamefont {M.}~\bibnamefont {Raidal}},\ }\href {\doibase
  10.1103/PhysRevD.75.095003} {\bibfield  {journal} {\bibinfo  {journal} {Phys.
  Rev.}\ }\textbf {\bibinfo {volume} {D75}},\ \bibinfo {pages} {095003}
  (\bibinfo {year} {2007})},\ \Eprint {http://arxiv.org/abs/hep-ph/0609228}
  {arXiv:hep-ph/0609228 [hep-ph]} \BibitemShut {NoStop}%
%%CITATION = HEP-PH/0609228;%%
\bibitem [{\citenamefont {Lattanzi}\ and\ \citenamefont
  {Valle}(2007)}]{Lattanzi:2007ux}%
  \BibitemOpen
  \bibfield  {author} {\bibinfo {author} {\bibfnamefont {M.}~\bibnamefont
  {Lattanzi}}\ and\ \bibinfo {author} {\bibfnamefont {J.~W.~F.}\ \bibnamefont
  {Valle}},\ }\href {\doibase 10.1103/PhysRevLett.99.121301} {\bibfield
  {journal} {\bibinfo  {journal} {Phys. Rev. Lett.}\ }\textbf {\bibinfo
  {volume} {99}},\ \bibinfo {pages} {121301} (\bibinfo {year} {2007})},\
  \Eprint {http://arxiv.org/abs/0705.2406} {arXiv:0705.2406 [astro-ph]}
  \BibitemShut {NoStop}%
%%CITATION = ARXIV:0705.2406;%%
\bibitem [{\citenamefont {Ma}(2008)}]{Ma:2007gq}%
  \BibitemOpen
  \bibfield  {author} {\bibinfo {author} {\bibfnamefont {E.}~\bibnamefont
  {Ma}},\ }\href {\doibase 10.1016/j.physletb.2008.02.053} {\bibfield
  {journal} {\bibinfo  {journal} {Phys. Lett.}\ }\textbf {\bibinfo {volume}
  {B662}},\ \bibinfo {pages} {49} (\bibinfo {year} {2008})},\ \Eprint
  {http://arxiv.org/abs/0708.3371} {arXiv:0708.3371 [hep-ph]} \BibitemShut
  {NoStop}%
%%CITATION = ARXIV:0708.3371;%%
\bibitem [{\citenamefont {Allahverdi}\ \emph {et~al.}(2007)\citenamefont
  {Allahverdi}, \citenamefont {Dutta},\ and\ \citenamefont
  {Mazumdar}}]{Allahverdi:2007wt}%
  \BibitemOpen
  \bibfield  {author} {\bibinfo {author} {\bibfnamefont {R.}~\bibnamefont
  {Allahverdi}}, \bibinfo {author} {\bibfnamefont {B.}~\bibnamefont {Dutta}}, \
  and\ \bibinfo {author} {\bibfnamefont {A.}~\bibnamefont {Mazumdar}},\ }\href
  {\doibase 10.1103/PhysRevLett.99.261301} {\bibfield  {journal} {\bibinfo
  {journal} {Phys. Rev. Lett.}\ }\textbf {\bibinfo {volume} {99}},\ \bibinfo
  {pages} {261301} (\bibinfo {year} {2007})},\ \Eprint
  {http://arxiv.org/abs/0708.3983} {arXiv:0708.3983 [hep-ph]} \BibitemShut
  {NoStop}%
%%CITATION = ARXIV:0708.3983;%%
\bibitem [{\citenamefont {Gu}\ and\ \citenamefont {Sarkar}(2008)}]{Gu:2007ug}%
  \BibitemOpen
  \bibfield  {author} {\bibinfo {author} {\bibfnamefont {P.-H.}\ \bibnamefont
  {Gu}}\ and\ \bibinfo {author} {\bibfnamefont {U.}~\bibnamefont {Sarkar}},\
  }\href {\doibase 10.1103/PhysRevD.77.105031} {\bibfield  {journal} {\bibinfo
  {journal} {Phys. Rev.}\ }\textbf {\bibinfo {volume} {D77}},\ \bibinfo {pages}
  {105031} (\bibinfo {year} {2008})},\ \Eprint {http://arxiv.org/abs/0712.2933}
  {arXiv:0712.2933 [hep-ph]} \BibitemShut {NoStop}%
%%CITATION = ARXIV:0712.2933;%%
\bibitem [{\citenamefont {Sahu}\ and\ \citenamefont
  {Sarkar}(2008)}]{Sahu:2008aw}%
  \BibitemOpen
  \bibfield  {author} {\bibinfo {author} {\bibfnamefont {N.}~\bibnamefont
  {Sahu}}\ and\ \bibinfo {author} {\bibfnamefont {U.}~\bibnamefont {Sarkar}},\
  }\href {\doibase 10.1103/PhysRevD.78.115013} {\bibfield  {journal} {\bibinfo
  {journal} {Phys. Rev.}\ }\textbf {\bibinfo {volume} {D78}},\ \bibinfo {pages}
  {115013} (\bibinfo {year} {2008})},\ \Eprint {http://arxiv.org/abs/0804.2072}
  {arXiv:0804.2072 [hep-ph]} \BibitemShut {NoStop}%
%%CITATION = ARXIV:0804.2072;%%
\bibitem [{\citenamefont {Arina}\ \emph {et~al.}(2008)\citenamefont {Arina},
  \citenamefont {Bazzocchi}, \citenamefont {Fornengo}, \citenamefont {Romao},\
  and\ \citenamefont {Valle}}]{Arina:2008bb}%
  \BibitemOpen
  \bibfield  {author} {\bibinfo {author} {\bibfnamefont {C.}~\bibnamefont
  {Arina}}, \bibinfo {author} {\bibfnamefont {F.}~\bibnamefont {Bazzocchi}},
  \bibinfo {author} {\bibfnamefont {N.}~\bibnamefont {Fornengo}}, \bibinfo
  {author} {\bibfnamefont {J.~C.}\ \bibnamefont {Romao}}, \ and\ \bibinfo
  {author} {\bibfnamefont {J.~W.~F.}\ \bibnamefont {Valle}},\ }\href {\doibase
  10.1103/PhysRevLett.101.161802} {\bibfield  {journal} {\bibinfo  {journal}
  {Phys. Rev. Lett.}\ }\textbf {\bibinfo {volume} {101}},\ \bibinfo {pages}
  {161802} (\bibinfo {year} {2008})},\ \Eprint {http://arxiv.org/abs/0806.3225}
  {arXiv:0806.3225 [hep-ph]} \BibitemShut {NoStop}%
%%CITATION = ARXIV:0806.3225;%%
\bibitem [{\citenamefont {Aoki}\ \emph
  {et~al.}(2009{\natexlab{a}})\citenamefont {Aoki}, \citenamefont {Kanemura},\
  and\ \citenamefont {Seto}}]{Aoki:2008av}%
  \BibitemOpen
  \bibfield  {author} {\bibinfo {author} {\bibfnamefont {M.}~\bibnamefont
  {Aoki}}, \bibinfo {author} {\bibfnamefont {S.}~\bibnamefont {Kanemura}}, \
  and\ \bibinfo {author} {\bibfnamefont {O.}~\bibnamefont {Seto}},\ }\href
  {\doibase 10.1103/PhysRevLett.102.051805} {\bibfield  {journal} {\bibinfo
  {journal} {Phys. Rev. Lett.}\ }\textbf {\bibinfo {volume} {102}},\ \bibinfo
  {pages} {051805} (\bibinfo {year} {2009}{\natexlab{a}})},\ \Eprint
  {http://arxiv.org/abs/0807.0361} {arXiv:0807.0361 [hep-ph]} \BibitemShut
  {NoStop}%
%%CITATION = ARXIV:0807.0361;%%
\bibitem [{\citenamefont {Ma}\ and\ \citenamefont
  {Suematsu}(2009)}]{Ma:2008cu}%
  \BibitemOpen
  \bibfield  {author} {\bibinfo {author} {\bibfnamefont {E.}~\bibnamefont
  {Ma}}\ and\ \bibinfo {author} {\bibfnamefont {D.}~\bibnamefont {Suematsu}},\
  }\href {\doibase 10.1142/S021773230903059X} {\bibfield  {journal} {\bibinfo
  {journal} {Mod. Phys. Lett.}\ }\textbf {\bibinfo {volume} {A24}},\ \bibinfo
  {pages} {583} (\bibinfo {year} {2009})},\ \Eprint
  {http://arxiv.org/abs/0809.0942} {arXiv:0809.0942 [hep-ph]} \BibitemShut
  {NoStop}%
%%CITATION = ARXIV:0809.0942;%%
\bibitem [{\citenamefont {Gu}\ \emph {et~al.}(2009)\citenamefont {Gu},
  \citenamefont {Hirsch}, \citenamefont {Sarkar},\ and\ \citenamefont
  {Valle}}]{Gu:2008yj}%
  \BibitemOpen
  \bibfield  {author} {\bibinfo {author} {\bibfnamefont {P.-H.}\ \bibnamefont
  {Gu}}, \bibinfo {author} {\bibfnamefont {M.}~\bibnamefont {Hirsch}}, \bibinfo
  {author} {\bibfnamefont {U.}~\bibnamefont {Sarkar}}, \ and\ \bibinfo {author}
  {\bibfnamefont {J.~W.~F.}\ \bibnamefont {Valle}},\ }\href {\doibase
  10.1103/PhysRevD.79.033010} {\bibfield  {journal} {\bibinfo  {journal} {Phys.
  Rev.}\ }\textbf {\bibinfo {volume} {D79}},\ \bibinfo {pages} {033010}
  (\bibinfo {year} {2009})},\ \Eprint {http://arxiv.org/abs/0811.0953}
  {arXiv:0811.0953 [hep-ph]} \BibitemShut {NoStop}%
%%CITATION = ARXIV:0811.0953;%%
\bibitem [{\citenamefont {Aoki}\ \emph
  {et~al.}(2009{\natexlab{b}})\citenamefont {Aoki}, \citenamefont {Kanemura},\
  and\ \citenamefont {Seto}}]{Aoki:2009vf}%
  \BibitemOpen
  \bibfield  {author} {\bibinfo {author} {\bibfnamefont {M.}~\bibnamefont
  {Aoki}}, \bibinfo {author} {\bibfnamefont {S.}~\bibnamefont {Kanemura}}, \
  and\ \bibinfo {author} {\bibfnamefont {O.}~\bibnamefont {Seto}},\ }\href
  {\doibase 10.1103/PhysRevD.80.033007} {\bibfield  {journal} {\bibinfo
  {journal} {Phys. Rev.}\ }\textbf {\bibinfo {volume} {D80}},\ \bibinfo {pages}
  {033007} (\bibinfo {year} {2009}{\natexlab{b}})},\ \Eprint
  {http://arxiv.org/abs/0904.3829} {arXiv:0904.3829 [hep-ph]} \BibitemShut
  {NoStop}%
%%CITATION = ARXIV:0904.3829;%%
\bibitem [{\citenamefont {Gu}(2010)}]{Gu:2010yf}%
  \BibitemOpen
  \bibfield  {author} {\bibinfo {author} {\bibfnamefont {P.-H.}\ \bibnamefont
  {Gu}},\ }\href {\doibase 10.1103/PhysRevD.81.095002} {\bibfield  {journal}
  {\bibinfo  {journal} {Phys. Rev.}\ }\textbf {\bibinfo {volume} {D81}},\
  \bibinfo {pages} {095002} (\bibinfo {year} {2010})},\ \Eprint
  {http://arxiv.org/abs/1001.1341} {arXiv:1001.1341 [hep-ph]} \BibitemShut
  {NoStop}%
%%CITATION = ARXIV:1001.1341;%%
\bibitem [{\citenamefont {Hirsch}\ \emph {et~al.}(2010)\citenamefont {Hirsch},
  \citenamefont {Morisi}, \citenamefont {Peinado},\ and\ \citenamefont
  {Valle}}]{Hirsch:2010ru}%
  \BibitemOpen
  \bibfield  {author} {\bibinfo {author} {\bibfnamefont {M.}~\bibnamefont
  {Hirsch}}, \bibinfo {author} {\bibfnamefont {S.}~\bibnamefont {Morisi}},
  \bibinfo {author} {\bibfnamefont {E.}~\bibnamefont {Peinado}}, \ and\
  \bibinfo {author} {\bibfnamefont {J.~W.~F.}\ \bibnamefont {Valle}},\ }\href
  {\doibase 10.1103/PhysRevD.82.116003} {\bibfield  {journal} {\bibinfo
  {journal} {Phys. Rev.}\ }\textbf {\bibinfo {volume} {D82}},\ \bibinfo {pages}
  {116003} (\bibinfo {year} {2010})},\ \Eprint {http://arxiv.org/abs/1007.0871}
  {arXiv:1007.0871 [hep-ph]} \BibitemShut {NoStop}%
%%CITATION = ARXIV:1007.0871;%%
\bibitem [{\citenamefont {Esteves}\ \emph {et~al.}(2010)\citenamefont
  {Esteves}, \citenamefont {Joaquim}, \citenamefont {Joshipura}, \citenamefont
  {Romao}, \citenamefont {Tortola},\ and\ \citenamefont
  {Valle}}]{Esteves:2010sh}%
  \BibitemOpen
  \bibfield  {author} {\bibinfo {author} {\bibfnamefont {J.~N.}\ \bibnamefont
  {Esteves}}, \bibinfo {author} {\bibfnamefont {F.~R.}\ \bibnamefont
  {Joaquim}}, \bibinfo {author} {\bibfnamefont {A.~S.}\ \bibnamefont
  {Joshipura}}, \bibinfo {author} {\bibfnamefont {J.~C.}\ \bibnamefont
  {Romao}}, \bibinfo {author} {\bibfnamefont {M.~A.}\ \bibnamefont {Tortola}},
  \ and\ \bibinfo {author} {\bibfnamefont {J.~W.~F.}\ \bibnamefont {Valle}},\
  }\href {\doibase 10.1103/PhysRevD.82.073008} {\bibfield  {journal} {\bibinfo
  {journal} {Phys. Rev.}\ }\textbf {\bibinfo {volume} {D82}},\ \bibinfo {pages}
  {073008} (\bibinfo {year} {2010})},\ \Eprint {http://arxiv.org/abs/1007.0898}
  {arXiv:1007.0898 [hep-ph]} \BibitemShut {NoStop}%
%%CITATION = ARXIV:1007.0898;%%
\bibitem [{\citenamefont {Kanemura}\ \emph {et~al.}(2011)\citenamefont
  {Kanemura}, \citenamefont {Seto},\ and\ \citenamefont
  {Shimomura}}]{Kanemura:2011vm}%
  \BibitemOpen
  \bibfield  {author} {\bibinfo {author} {\bibfnamefont {S.}~\bibnamefont
  {Kanemura}}, \bibinfo {author} {\bibfnamefont {O.}~\bibnamefont {Seto}}, \
  and\ \bibinfo {author} {\bibfnamefont {T.}~\bibnamefont {Shimomura}},\ }\href
  {\doibase 10.1103/PhysRevD.84.016004} {\bibfield  {journal} {\bibinfo
  {journal} {Phys. Rev.}\ }\textbf {\bibinfo {volume} {D84}},\ \bibinfo {pages}
  {016004} (\bibinfo {year} {2011})},\ \Eprint {http://arxiv.org/abs/1101.5713}
  {arXiv:1101.5713 [hep-ph]} \BibitemShut {NoStop}%
%%CITATION = ARXIV:1101.5713;%%
\bibitem [{\citenamefont {Lindner}\ \emph {et~al.}(2011)\citenamefont
  {Lindner}, \citenamefont {Schmidt},\ and\ \citenamefont
  {Schwetz}}]{Lindner:2011it}%
  \BibitemOpen
  \bibfield  {author} {\bibinfo {author} {\bibfnamefont {M.}~\bibnamefont
  {Lindner}}, \bibinfo {author} {\bibfnamefont {D.}~\bibnamefont {Schmidt}}, \
  and\ \bibinfo {author} {\bibfnamefont {T.}~\bibnamefont {Schwetz}},\ }\href
  {\doibase 10.1016/j.physletb.2011.10.022} {\bibfield  {journal} {\bibinfo
  {journal} {Phys. Lett.}\ }\textbf {\bibinfo {volume} {B705}},\ \bibinfo
  {pages} {324} (\bibinfo {year} {2011})},\ \Eprint
  {http://arxiv.org/abs/1105.4626} {arXiv:1105.4626 [hep-ph]} \BibitemShut
  {NoStop}%
%%CITATION = ARXIV:1105.4626;%%
\bibitem [{\citenamefont {Josse-Michaux}\ and\ \citenamefont
  {Molinaro}(2011)}]{JosseMichaux:2011ba}%
  \BibitemOpen
  \bibfield  {author} {\bibinfo {author} {\bibfnamefont {F.-X.}\ \bibnamefont
  {Josse-Michaux}}\ and\ \bibinfo {author} {\bibfnamefont {E.}~\bibnamefont
  {Molinaro}},\ }\href {\doibase 10.1103/PhysRevD.84.125021} {\bibfield
  {journal} {\bibinfo  {journal} {Phys. Rev.}\ }\textbf {\bibinfo {volume}
  {D84}},\ \bibinfo {pages} {125021} (\bibinfo {year} {2011})},\ \Eprint
  {http://arxiv.org/abs/1108.0482} {arXiv:1108.0482 [hep-ph]} \BibitemShut
  {NoStop}%
%%CITATION = ARXIV:1108.0482;%%
\bibitem [{\citenamefont {Schmidt}\ \emph {et~al.}(2012)\citenamefont
  {Schmidt}, \citenamefont {Schwetz},\ and\ \citenamefont
  {Toma}}]{Schmidt:2012yg}%
  \BibitemOpen
  \bibfield  {author} {\bibinfo {author} {\bibfnamefont {D.}~\bibnamefont
  {Schmidt}}, \bibinfo {author} {\bibfnamefont {T.}~\bibnamefont {Schwetz}}, \
  and\ \bibinfo {author} {\bibfnamefont {T.}~\bibnamefont {Toma}},\ }\href
  {\doibase 10.1103/PhysRevD.85.073009} {\bibfield  {journal} {\bibinfo
  {journal} {Phys. Rev.}\ }\textbf {\bibinfo {volume} {D85}},\ \bibinfo {pages}
  {073009} (\bibinfo {year} {2012})},\ \Eprint {http://arxiv.org/abs/1201.0906}
  {arXiv:1201.0906 [hep-ph]} \BibitemShut {NoStop}%
%%CITATION = ARXIV:1201.0906;%%
\bibitem [{\citenamefont {Borah}\ and\ \citenamefont
  {Adhikari}(2012)}]{Borah:2012qr}%
  \BibitemOpen
  \bibfield  {author} {\bibinfo {author} {\bibfnamefont {D.}~\bibnamefont
  {Borah}}\ and\ \bibinfo {author} {\bibfnamefont {R.}~\bibnamefont
  {Adhikari}},\ }\href {\doibase 10.1103/PhysRevD.85.095002} {\bibfield
  {journal} {\bibinfo  {journal} {Phys. Rev.}\ }\textbf {\bibinfo {volume}
  {D85}},\ \bibinfo {pages} {095002} (\bibinfo {year} {2012})},\ \Eprint
  {http://arxiv.org/abs/1202.2718} {arXiv:1202.2718 [hep-ph]} \BibitemShut
  {NoStop}%
%%CITATION = ARXIV:1202.2718;%%
\bibitem [{\citenamefont {Farzan}\ and\ \citenamefont
  {Ma}(2012)}]{Farzan:2012sa}%
  \BibitemOpen
  \bibfield  {author} {\bibinfo {author} {\bibfnamefont {Y.}~\bibnamefont
  {Farzan}}\ and\ \bibinfo {author} {\bibfnamefont {E.}~\bibnamefont {Ma}},\
  }\href {\doibase 10.1103/PhysRevD.86.033007} {\bibfield  {journal} {\bibinfo
  {journal} {Phys. Rev.}\ }\textbf {\bibinfo {volume} {D86}},\ \bibinfo {pages}
  {033007} (\bibinfo {year} {2012})},\ \Eprint {http://arxiv.org/abs/1204.4890}
  {arXiv:1204.4890 [hep-ph]} \BibitemShut {NoStop}%
%%CITATION = ARXIV:1204.4890;%%
\bibitem [{\citenamefont {Chao}\ \emph {et~al.}(2012)\citenamefont {Chao},
  \citenamefont {Gonderinger},\ and\ \citenamefont
  {Ramsey-Musolf}}]{Chao:2012mx}%
  \BibitemOpen
  \bibfield  {author} {\bibinfo {author} {\bibfnamefont {W.}~\bibnamefont
  {Chao}}, \bibinfo {author} {\bibfnamefont {M.}~\bibnamefont {Gonderinger}}, \
  and\ \bibinfo {author} {\bibfnamefont {M.~J.}\ \bibnamefont
  {Ramsey-Musolf}},\ }\href {\doibase 10.1103/PhysRevD.86.113017} {\bibfield
  {journal} {\bibinfo  {journal} {Phys. Rev.}\ }\textbf {\bibinfo {volume}
  {D86}},\ \bibinfo {pages} {113017} (\bibinfo {year} {2012})},\ \Eprint
  {http://arxiv.org/abs/1210.0491} {arXiv:1210.0491 [hep-ph]} \BibitemShut
  {NoStop}%
%%CITATION = ARXIV:1210.0491;%%
\bibitem [{\citenamefont {Gustafsson}\ \emph {et~al.}(2013)\citenamefont
  {Gustafsson}, \citenamefont {No},\ and\ \citenamefont
  {Rivera}}]{Gustafsson:2012vj}%
  \BibitemOpen
  \bibfield  {author} {\bibinfo {author} {\bibfnamefont {M.}~\bibnamefont
  {Gustafsson}}, \bibinfo {author} {\bibfnamefont {J.~M.}\ \bibnamefont {No}},
  \ and\ \bibinfo {author} {\bibfnamefont {M.~A.}\ \bibnamefont {Rivera}},\
  }\href {\doibase 10.1103/PhysRevLett.110.211802,
  10.1103/PhysRevLett.112.259902} {\bibfield  {journal} {\bibinfo  {journal}
  {Phys. Rev. Lett.}\ }\textbf {\bibinfo {volume} {110}},\ \bibinfo {pages}
  {211802} (\bibinfo {year} {2013})},\ \bibinfo {note} {[Erratum: Phys. Rev.
  Lett.112,no.25,259902(2014)]},\ \Eprint {http://arxiv.org/abs/1212.4806}
  {arXiv:1212.4806 [hep-ph]} \BibitemShut {NoStop}%
%%CITATION = ARXIV:1212.4806;%%
\bibitem [{\citenamefont {Blennow}\ \emph {et~al.}(2013)\citenamefont
  {Blennow}, \citenamefont {Carrigan},\ and\ \citenamefont
  {Fernandez~Martinez}}]{Blennow:2013pya}%
  \BibitemOpen
  \bibfield  {author} {\bibinfo {author} {\bibfnamefont {M.}~\bibnamefont
  {Blennow}}, \bibinfo {author} {\bibfnamefont {M.}~\bibnamefont {Carrigan}}, \
  and\ \bibinfo {author} {\bibfnamefont {E.}~\bibnamefont
  {Fernandez~Martinez}},\ }\href {\doibase 10.1088/1475-7516/2013/06/038}
  {\bibfield  {journal} {\bibinfo  {journal} {JCAP}\ }\textbf {\bibinfo
  {volume} {1306}},\ \bibinfo {pages} {038} (\bibinfo {year} {2013})},\ \Eprint
  {http://arxiv.org/abs/1303.4530} {arXiv:1303.4530 [hep-ph]} \BibitemShut
  {NoStop}%
%%CITATION = ARXIV:1303.4530;%%
\bibitem [{\citenamefont {Law}\ and\ \citenamefont
  {McDonald}(2013)}]{Law:2013saa}%
  \BibitemOpen
  \bibfield  {author} {\bibinfo {author} {\bibfnamefont {S.~S.~C.}\
  \bibnamefont {Law}}\ and\ \bibinfo {author} {\bibfnamefont {K.~L.}\
  \bibnamefont {McDonald}},\ }\href {\doibase 10.1007/JHEP09(2013)092}
  {\bibfield  {journal} {\bibinfo  {journal} {JHEP}\ }\textbf {\bibinfo
  {volume} {09}},\ \bibinfo {pages} {092} (\bibinfo {year} {2013})},\ \Eprint
  {http://arxiv.org/abs/1305.6467} {arXiv:1305.6467 [hep-ph]} \BibitemShut
  {NoStop}%
%%CITATION = ARXIV:1305.6467;%%
\bibitem [{\citenamefont {Carcamo~Hernandez}\ \emph {et~al.}(2013)\citenamefont
  {Carcamo~Hernandez}, \citenamefont {de~Medeiros~Varzielas}, \citenamefont
  {Kovalenko}, \citenamefont {Päs},\ and\ \citenamefont
  {Schmidt}}]{Hernandez:2013dta}%
  \BibitemOpen
  \bibfield  {author} {\bibinfo {author} {\bibfnamefont {A.~E.}\ \bibnamefont
  {Carcamo~Hernandez}}, \bibinfo {author} {\bibfnamefont {I.}~\bibnamefont
  {de~Medeiros~Varzielas}}, \bibinfo {author} {\bibfnamefont {S.~G.}\
  \bibnamefont {Kovalenko}}, \bibinfo {author} {\bibfnamefont {H.}~\bibnamefont
  {Päs}}, \ and\ \bibinfo {author} {\bibfnamefont {I.}~\bibnamefont
  {Schmidt}},\ }\href {\doibase 10.1103/PhysRevD.88.076014} {\bibfield
  {journal} {\bibinfo  {journal} {Phys. Rev.}\ }\textbf {\bibinfo {volume}
  {D88}},\ \bibinfo {pages} {076014} (\bibinfo {year} {2013})},\ \Eprint
  {http://arxiv.org/abs/1307.6499} {arXiv:1307.6499 [hep-ph]} \BibitemShut
  {NoStop}%
%%CITATION = ARXIV:1307.6499;%%
\bibitem [{\citenamefont {Restrepo}\ \emph {et~al.}(2013)\citenamefont
  {Restrepo}, \citenamefont {Zapata},\ and\ \citenamefont
  {Yaguna}}]{Restrepo:2013aga}%
  \BibitemOpen
  \bibfield  {author} {\bibinfo {author} {\bibfnamefont {D.}~\bibnamefont
  {Restrepo}}, \bibinfo {author} {\bibfnamefont {O.}~\bibnamefont {Zapata}}, \
  and\ \bibinfo {author} {\bibfnamefont {C.~E.}\ \bibnamefont {Yaguna}},\
  }\href {\doibase 10.1007/JHEP11(2013)011} {\bibfield  {journal} {\bibinfo
  {journal} {JHEP}\ }\textbf {\bibinfo {volume} {11}},\ \bibinfo {pages} {011}
  (\bibinfo {year} {2013})},\ \Eprint {http://arxiv.org/abs/1308.3655}
  {arXiv:1308.3655 [hep-ph]} \BibitemShut {NoStop}%
%%CITATION = ARXIV:1308.3655;%%
\bibitem [{\citenamefont {Chakraborty}\ and\ \citenamefont
  {Roy}(2014)}]{Chakraborty:2013gea}%
  \BibitemOpen
  \bibfield  {author} {\bibinfo {author} {\bibfnamefont {S.}~\bibnamefont
  {Chakraborty}}\ and\ \bibinfo {author} {\bibfnamefont {S.}~\bibnamefont
  {Roy}},\ }\href {\doibase 10.1007/JHEP01(2014)101} {\bibfield  {journal}
  {\bibinfo  {journal} {JHEP}\ }\textbf {\bibinfo {volume} {01}},\ \bibinfo
  {pages} {101} (\bibinfo {year} {2014})},\ \Eprint
  {http://arxiv.org/abs/1309.6538} {arXiv:1309.6538 [hep-ph]} \BibitemShut
  {NoStop}%
%%CITATION = ARXIV:1309.6538;%%
\bibitem [{\citenamefont {Ahriche}\ \emph {et~al.}(2014)\citenamefont
  {Ahriche}, \citenamefont {Chen}, \citenamefont {McDonald},\ and\
  \citenamefont {Nasri}}]{Ahriche:2014cda}%
  \BibitemOpen
  \bibfield  {author} {\bibinfo {author} {\bibfnamefont {A.}~\bibnamefont
  {Ahriche}}, \bibinfo {author} {\bibfnamefont {C.-S.}\ \bibnamefont {Chen}},
  \bibinfo {author} {\bibfnamefont {K.~L.}\ \bibnamefont {McDonald}}, \ and\
  \bibinfo {author} {\bibfnamefont {S.}~\bibnamefont {Nasri}},\ }\href
  {\doibase 10.1103/PhysRevD.90.015024} {\bibfield  {journal} {\bibinfo
  {journal} {Phys. Rev.}\ }\textbf {\bibinfo {volume} {D90}},\ \bibinfo {pages}
  {015024} (\bibinfo {year} {2014})},\ \Eprint {http://arxiv.org/abs/1404.2696}
  {arXiv:1404.2696 [hep-ph]} \BibitemShut {NoStop}%
%%CITATION = ARXIV:1404.2696;%%
\bibitem [{\citenamefont {Kanemura}\ \emph {et~al.}(2014)\citenamefont
  {Kanemura}, \citenamefont {Matsui},\ and\ \citenamefont
  {Sugiyama}}]{Kanemura:2014rpa}%
  \BibitemOpen
  \bibfield  {author} {\bibinfo {author} {\bibfnamefont {S.}~\bibnamefont
  {Kanemura}}, \bibinfo {author} {\bibfnamefont {T.}~\bibnamefont {Matsui}}, \
  and\ \bibinfo {author} {\bibfnamefont {H.}~\bibnamefont {Sugiyama}},\ }\href
  {\doibase 10.1103/PhysRevD.90.013001} {\bibfield  {journal} {\bibinfo
  {journal} {Phys. Rev.}\ }\textbf {\bibinfo {volume} {D90}},\ \bibinfo {pages}
  {013001} (\bibinfo {year} {2014})},\ \Eprint {http://arxiv.org/abs/1405.1935}
  {arXiv:1405.1935 [hep-ph]} \BibitemShut {NoStop}%
%%CITATION = ARXIV:1405.1935;%%
\bibitem [{\citenamefont {Huang}\ and\ \citenamefont
  {Deppisch}(2015)}]{Huang:2014bva}%
  \BibitemOpen
  \bibfield  {author} {\bibinfo {author} {\bibfnamefont {W.-C.}\ \bibnamefont
  {Huang}}\ and\ \bibinfo {author} {\bibfnamefont {F.~F.}\ \bibnamefont
  {Deppisch}},\ }\href {\doibase 10.1103/PhysRevD.91.093011} {\bibfield
  {journal} {\bibinfo  {journal} {Phys. Rev.}\ }\textbf {\bibinfo {volume}
  {D91}},\ \bibinfo {pages} {093011} (\bibinfo {year} {2015})},\ \Eprint
  {http://arxiv.org/abs/1412.2027} {arXiv:1412.2027 [hep-ph]} \BibitemShut
  {NoStop}%
%%CITATION = ARXIV:1412.2027;%%
\bibitem [{\citenamefont {de~Medeiros~Varzielas}\ \emph
  {et~al.}(2015)\citenamefont {de~Medeiros~Varzielas}, \citenamefont
  {Fischer},\ and\ \citenamefont {Maurer}}]{Varzielas:2015joa}%
  \BibitemOpen
  \bibfield  {author} {\bibinfo {author} {\bibfnamefont {I.}~\bibnamefont
  {de~Medeiros~Varzielas}}, \bibinfo {author} {\bibfnamefont {O.}~\bibnamefont
  {Fischer}}, \ and\ \bibinfo {author} {\bibfnamefont {V.}~\bibnamefont
  {Maurer}},\ }\href {\doibase 10.1007/JHEP08(2015)080} {\bibfield  {journal}
  {\bibinfo  {journal} {JHEP}\ }\textbf {\bibinfo {volume} {08}},\ \bibinfo
  {pages} {080} (\bibinfo {year} {2015})},\ \Eprint
  {http://arxiv.org/abs/1504.03955} {arXiv:1504.03955 [hep-ph]} \BibitemShut
  {NoStop}%
%%CITATION = ARXIV:1504.03955;%%
\bibitem [{\citenamefont {Sánchez-Vega}\ and\ \citenamefont
  {Schmitz}(2015)}]{Sanchez-Vega:2015qva}%
  \BibitemOpen
  \bibfield  {author} {\bibinfo {author} {\bibfnamefont {B.~L.}\ \bibnamefont
  {Sánchez-Vega}}\ and\ \bibinfo {author} {\bibfnamefont {E.~R.}\ \bibnamefont
  {Schmitz}},\ }\href {\doibase 10.1103/PhysRevD.92.053007} {\bibfield
  {journal} {\bibinfo  {journal} {Phys. Rev.}\ }\textbf {\bibinfo {volume}
  {D92}},\ \bibinfo {pages} {053007} (\bibinfo {year} {2015})},\ \Eprint
  {http://arxiv.org/abs/1505.03595} {arXiv:1505.03595 [hep-ph]} \BibitemShut
  {NoStop}%
%%CITATION = ARXIV:1505.03595;%%
\bibitem [{\citenamefont {Fraser}\ \emph {et~al.}(2016)\citenamefont {Fraser},
  \citenamefont {Kownacki}, \citenamefont {Ma},\ and\ \citenamefont
  {Popov}}]{Fraser:2015mhb}%
  \BibitemOpen
  \bibfield  {author} {\bibinfo {author} {\bibfnamefont {S.}~\bibnamefont
  {Fraser}}, \bibinfo {author} {\bibfnamefont {C.}~\bibnamefont {Kownacki}},
  \bibinfo {author} {\bibfnamefont {E.}~\bibnamefont {Ma}}, \ and\ \bibinfo
  {author} {\bibfnamefont {O.}~\bibnamefont {Popov}},\ }\href {\doibase
  10.1103/PhysRevD.93.013021} {\bibfield  {journal} {\bibinfo  {journal} {Phys.
  Rev.}\ }\textbf {\bibinfo {volume} {D93}},\ \bibinfo {pages} {013021}
  (\bibinfo {year} {2016})},\ \Eprint {http://arxiv.org/abs/1511.06375}
  {arXiv:1511.06375 [hep-ph]} \BibitemShut {NoStop}%
%%CITATION = ARXIV:1511.06375;%%
\bibitem [{\citenamefont {Adhikari}\ \emph {et~al.}(2016)\citenamefont
  {Adhikari}, \citenamefont {Borah},\ and\ \citenamefont
  {Ma}}]{Adhikari:2015woo}%
  \BibitemOpen
  \bibfield  {author} {\bibinfo {author} {\bibfnamefont {R.}~\bibnamefont
  {Adhikari}}, \bibinfo {author} {\bibfnamefont {D.}~\bibnamefont {Borah}}, \
  and\ \bibinfo {author} {\bibfnamefont {E.}~\bibnamefont {Ma}},\ }\href
  {\doibase 10.1016/j.physletb.2016.02.039} {\bibfield  {journal} {\bibinfo
  {journal} {Phys. Lett.}\ }\textbf {\bibinfo {volume} {B755}},\ \bibinfo
  {pages} {414} (\bibinfo {year} {2016})},\ \Eprint
  {http://arxiv.org/abs/1512.05491} {arXiv:1512.05491 [hep-ph]} \BibitemShut
  {NoStop}%
%%CITATION = ARXIV:1512.05491;%%
\bibitem [{\citenamefont {Ahriche}\ \emph {et~al.}(2016)\citenamefont
  {Ahriche}, \citenamefont {McDonald}, \citenamefont {Nasri},\ and\
  \citenamefont {Picek}}]{Ahriche:2016rgf}%
  \BibitemOpen
  \bibfield  {author} {\bibinfo {author} {\bibfnamefont {A.}~\bibnamefont
  {Ahriche}}, \bibinfo {author} {\bibfnamefont {K.~L.}\ \bibnamefont
  {McDonald}}, \bibinfo {author} {\bibfnamefont {S.}~\bibnamefont {Nasri}}, \
  and\ \bibinfo {author} {\bibfnamefont {I.}~\bibnamefont {Picek}},\ }\href
  {\doibase 10.1016/j.physletb.2016.04.022} {\bibfield  {journal} {\bibinfo
  {journal} {Phys. Lett.}\ }\textbf {\bibinfo {volume} {B757}},\ \bibinfo
  {pages} {399} (\bibinfo {year} {2016})},\ \Eprint
  {http://arxiv.org/abs/1603.01247} {arXiv:1603.01247 [hep-ph]} \BibitemShut
  {NoStop}%
%%CITATION = ARXIV:1603.01247;%%
\bibitem [{\citenamefont {Aristizabal~Sierra}\ \emph
  {et~al.}(2016)\citenamefont {Aristizabal~Sierra}, \citenamefont {Simoes},\
  and\ \citenamefont {Wegman}}]{Sierra:2016qfa}%
  \BibitemOpen
  \bibfield  {author} {\bibinfo {author} {\bibfnamefont {D.}~\bibnamefont
  {Aristizabal~Sierra}}, \bibinfo {author} {\bibfnamefont {C.}~\bibnamefont
  {Simoes}}, \ and\ \bibinfo {author} {\bibfnamefont {D.}~\bibnamefont
  {Wegman}},\ }\href {\doibase 10.1007/JHEP06(2016)108} {\bibfield  {journal}
  {\bibinfo  {journal} {JHEP}\ }\textbf {\bibinfo {volume} {06}},\ \bibinfo
  {pages} {108} (\bibinfo {year} {2016})},\ \Eprint
  {http://arxiv.org/abs/1603.04723} {arXiv:1603.04723 [hep-ph]} \BibitemShut
  {NoStop}%
%%CITATION = ARXIV:1603.04723;%%
\bibitem [{\citenamefont {Lu}\ and\ \citenamefont {Gu}(2016)}]{Lu:2016ucn}%
  \BibitemOpen
  \bibfield  {author} {\bibinfo {author} {\bibfnamefont {W.-B.}\ \bibnamefont
  {Lu}}\ and\ \bibinfo {author} {\bibfnamefont {P.-H.}\ \bibnamefont {Gu}},\
  }\href {\doibase 10.1088/1475-7516/2016/05/040} {\bibfield  {journal}
  {\bibinfo  {journal} {JCAP}\ }\textbf {\bibinfo {volume} {1605}},\ \bibinfo
  {pages} {040} (\bibinfo {year} {2016})},\ \Eprint
  {http://arxiv.org/abs/1603.05074} {arXiv:1603.05074 [hep-ph]} \BibitemShut
  {NoStop}%
%%CITATION = ARXIV:1603.05074;%%
\bibitem [{\citenamefont {Batell}\ \emph {et~al.}(2016)\citenamefont {Batell},
  \citenamefont {Pospelov},\ and\ \citenamefont {Shuve}}]{Batell:2016zod}%
  \BibitemOpen
  \bibfield  {author} {\bibinfo {author} {\bibfnamefont {B.}~\bibnamefont
  {Batell}}, \bibinfo {author} {\bibfnamefont {M.}~\bibnamefont {Pospelov}}, \
  and\ \bibinfo {author} {\bibfnamefont {B.}~\bibnamefont {Shuve}},\ }\href
  {\doibase 10.1007/JHEP08(2016)052} {\bibfield  {journal} {\bibinfo  {journal}
  {JHEP}\ }\textbf {\bibinfo {volume} {08}},\ \bibinfo {pages} {052} (\bibinfo
  {year} {2016})},\ \Eprint {http://arxiv.org/abs/1604.06099} {arXiv:1604.06099
  [hep-ph]} \BibitemShut {NoStop}%
%%CITATION = ARXIV:1604.06099;%%
\bibitem [{\citenamefont {Ho}\ \emph {et~al.}(2016)\citenamefont {Ho},
  \citenamefont {Toma},\ and\ \citenamefont {Tsumura}}]{Ho:2016aye}%
  \BibitemOpen
  \bibfield  {author} {\bibinfo {author} {\bibfnamefont {S.-Y.}\ \bibnamefont
  {Ho}}, \bibinfo {author} {\bibfnamefont {T.}~\bibnamefont {Toma}}, \ and\
  \bibinfo {author} {\bibfnamefont {K.}~\bibnamefont {Tsumura}},\ }\href
  {\doibase 10.1103/PhysRevD.94.033007} {\bibfield  {journal} {\bibinfo
  {journal} {Phys. Rev.}\ }\textbf {\bibinfo {volume} {D94}},\ \bibinfo {pages}
  {033007} (\bibinfo {year} {2016})},\ \Eprint
  {http://arxiv.org/abs/1604.07894} {arXiv:1604.07894 [hep-ph]} \BibitemShut
  {NoStop}%
%%CITATION = ARXIV:1604.07894;%%
\bibitem [{\citenamefont {Escudero}\ \emph {et~al.}(2017)\citenamefont
  {Escudero}, \citenamefont {Rius},\ and\ \citenamefont
  {Sanz}}]{Escudero:2016ksa}%
  \BibitemOpen
  \bibfield  {author} {\bibinfo {author} {\bibfnamefont {M.}~\bibnamefont
  {Escudero}}, \bibinfo {author} {\bibfnamefont {N.}~\bibnamefont {Rius}}, \
  and\ \bibinfo {author} {\bibfnamefont {V.}~\bibnamefont {Sanz}},\ }\href
  {\doibase 10.1140/epjc/s10052-017-4963-x} {\bibfield  {journal} {\bibinfo
  {journal} {Eur. Phys. J.}\ }\textbf {\bibinfo {volume} {C77}},\ \bibinfo
  {pages} {397} (\bibinfo {year} {2017})},\ \Eprint
  {http://arxiv.org/abs/1607.02373} {arXiv:1607.02373 [hep-ph]} \BibitemShut
  {NoStop}%
%%CITATION = ARXIV:1607.02373;%%
\bibitem [{\citenamefont {Bonilla}\ \emph {et~al.}(2016)\citenamefont
  {Bonilla}, \citenamefont {Ma}, \citenamefont {Peinado},\ and\ \citenamefont
  {Valle}}]{Bonilla:2016diq}%
  \BibitemOpen
  \bibfield  {author} {\bibinfo {author} {\bibfnamefont {C.}~\bibnamefont
  {Bonilla}}, \bibinfo {author} {\bibfnamefont {E.}~\bibnamefont {Ma}},
  \bibinfo {author} {\bibfnamefont {E.}~\bibnamefont {Peinado}}, \ and\
  \bibinfo {author} {\bibfnamefont {J.~W.~F.}\ \bibnamefont {Valle}},\ }\href
  {\doibase 10.1016/j.physletb.2016.09.027} {\bibfield  {journal} {\bibinfo
  {journal} {Phys. Lett.}\ }\textbf {\bibinfo {volume} {B762}},\ \bibinfo
  {pages} {214} (\bibinfo {year} {2016})},\ \Eprint
  {http://arxiv.org/abs/1607.03931} {arXiv:1607.03931 [hep-ph]} \BibitemShut
  {NoStop}%
%%CITATION = ARXIV:1607.03931;%%
\bibitem [{\citenamefont {Borah}\ and\ \citenamefont
  {Dasgupta}(2016)}]{Borah:2016zbd}%
  \BibitemOpen
  \bibfield  {author} {\bibinfo {author} {\bibfnamefont {D.}~\bibnamefont
  {Borah}}\ and\ \bibinfo {author} {\bibfnamefont {A.}~\bibnamefont
  {Dasgupta}},\ }\href {\doibase 10.1088/1475-7516/2016/12/034} {\bibfield
  {journal} {\bibinfo  {journal} {JCAP}\ }\textbf {\bibinfo {volume} {1612}},\
  \bibinfo {pages} {034} (\bibinfo {year} {2016})},\ \Eprint
  {http://arxiv.org/abs/1608.03872} {arXiv:1608.03872 [hep-ph]} \BibitemShut
  {NoStop}%
%%CITATION = ARXIV:1608.03872;%%
\bibitem [{\citenamefont {Biswas}\ \emph {et~al.}(2016)\citenamefont {Biswas},
  \citenamefont {Choubey},\ and\ \citenamefont {Khan}}]{Biswas:2016yan}%
  \BibitemOpen
  \bibfield  {author} {\bibinfo {author} {\bibfnamefont {A.}~\bibnamefont
  {Biswas}}, \bibinfo {author} {\bibfnamefont {S.}~\bibnamefont {Choubey}}, \
  and\ \bibinfo {author} {\bibfnamefont {S.}~\bibnamefont {Khan}},\ }\href
  {\doibase 10.1007/JHEP09(2016)147} {\bibfield  {journal} {\bibinfo  {journal}
  {JHEP}\ }\textbf {\bibinfo {volume} {09}},\ \bibinfo {pages} {147} (\bibinfo
  {year} {2016})},\ \Eprint {http://arxiv.org/abs/1608.04194} {arXiv:1608.04194
  [hep-ph]} \BibitemShut {NoStop}%
%%CITATION = ARXIV:1608.04194;%%
\bibitem [{\citenamefont {Hierro}\ \emph {et~al.}(2017)\citenamefont {Hierro},
  \citenamefont {King},\ and\ \citenamefont {Rigolin}}]{Hierro:2016nwm}%
  \BibitemOpen
  \bibfield  {author} {\bibinfo {author} {\bibfnamefont {I.~M.}\ \bibnamefont
  {Hierro}}, \bibinfo {author} {\bibfnamefont {S.~F.}\ \bibnamefont {King}}, \
  and\ \bibinfo {author} {\bibfnamefont {S.}~\bibnamefont {Rigolin}},\ }\href
  {\doibase 10.1016/j.physletb.2017.03.037} {\bibfield  {journal} {\bibinfo
  {journal} {Phys. Lett.}\ }\textbf {\bibinfo {volume} {B769}},\ \bibinfo
  {pages} {121} (\bibinfo {year} {2017})},\ \Eprint
  {http://arxiv.org/abs/1609.02872} {arXiv:1609.02872 [hep-ph]} \BibitemShut
  {NoStop}%
%%CITATION = ARXIV:1609.02872;%%
\bibitem [{\citenamefont {Bhattacharya}\ \emph
  {et~al.}(2017{\natexlab{a}})\citenamefont {Bhattacharya}, \citenamefont
  {Jana},\ and\ \citenamefont {Nandi}}]{Bhattacharya:2016qsg}%
  \BibitemOpen
  \bibfield  {author} {\bibinfo {author} {\bibfnamefont {S.}~\bibnamefont
  {Bhattacharya}}, \bibinfo {author} {\bibfnamefont {S.}~\bibnamefont {Jana}},
  \ and\ \bibinfo {author} {\bibfnamefont {S.}~\bibnamefont {Nandi}},\ }\href
  {\doibase 10.1103/PhysRevD.95.055003} {\bibfield  {journal} {\bibinfo
  {journal} {Phys. Rev.}\ }\textbf {\bibinfo {volume} {D95}},\ \bibinfo {pages}
  {055003} (\bibinfo {year} {2017}{\natexlab{a}})},\ \Eprint
  {http://arxiv.org/abs/1609.03274} {arXiv:1609.03274 [hep-ph]} \BibitemShut
  {NoStop}%
%%CITATION = ARXIV:1609.03274;%%
\bibitem [{\citenamefont {Chakraborty}\ and\ \citenamefont
  {Chakrabortty}(2017)}]{Chakraborty:2017dfg}%
  \BibitemOpen
  \bibfield  {author} {\bibinfo {author} {\bibfnamefont {S.}~\bibnamefont
  {Chakraborty}}\ and\ \bibinfo {author} {\bibfnamefont {J.}~\bibnamefont
  {Chakrabortty}},\ }\href {\doibase 10.1007/JHEP10(2017)012} {\bibfield
  {journal} {\bibinfo  {journal} {JHEP}\ }\textbf {\bibinfo {volume} {10}},\
  \bibinfo {pages} {012} (\bibinfo {year} {2017})},\ \Eprint
  {http://arxiv.org/abs/1701.04566} {arXiv:1701.04566 [hep-ph]} \BibitemShut
  {NoStop}%
%%CITATION = ARXIV:1701.04566;%%
\bibitem [{\citenamefont {Bhattacharya}\ \emph
  {et~al.}(2017{\natexlab{b}})\citenamefont {Bhattacharya}, \citenamefont
  {Sahoo},\ and\ \citenamefont {Sahu}}]{Bhattacharya:2017sml}%
  \BibitemOpen
  \bibfield  {author} {\bibinfo {author} {\bibfnamefont {S.}~\bibnamefont
  {Bhattacharya}}, \bibinfo {author} {\bibfnamefont {N.}~\bibnamefont {Sahoo}},
  \ and\ \bibinfo {author} {\bibfnamefont {N.}~\bibnamefont {Sahu}},\ }\href
  {\doibase 10.1103/PhysRevD.96.035010} {\bibfield  {journal} {\bibinfo
  {journal} {Phys. Rev.}\ }\textbf {\bibinfo {volume} {D96}},\ \bibinfo {pages}
  {035010} (\bibinfo {year} {2017}{\natexlab{b}})},\ \Eprint
  {http://arxiv.org/abs/1704.03417} {arXiv:1704.03417 [hep-ph]} \BibitemShut
  {NoStop}%
%%CITATION = ARXIV:1704.03417;%%
\bibitem [{\citenamefont {Ho}\ \emph {et~al.}(2017)\citenamefont {Ho},
  \citenamefont {Toma},\ and\ \citenamefont {Tsumura}}]{Ho:2017fte}%
  \BibitemOpen
  \bibfield  {author} {\bibinfo {author} {\bibfnamefont {S.-Y.}\ \bibnamefont
  {Ho}}, \bibinfo {author} {\bibfnamefont {T.}~\bibnamefont {Toma}}, \ and\
  \bibinfo {author} {\bibfnamefont {K.}~\bibnamefont {Tsumura}},\ }\href
  {\doibase 10.1007/JHEP07(2017)101} {\bibfield  {journal} {\bibinfo  {journal}
  {JHEP}\ }\textbf {\bibinfo {volume} {07}},\ \bibinfo {pages} {101} (\bibinfo
  {year} {2017})},\ \Eprint {http://arxiv.org/abs/1705.00592} {arXiv:1705.00592
  [hep-ph]} \BibitemShut {NoStop}%
%%CITATION = ARXIV:1705.00592;%%
\bibitem [{\citenamefont {Ghosh}\ \emph {et~al.}(2018)\citenamefont {Ghosh},
  \citenamefont {Saha},\ and\ \citenamefont {Sil}}]{Ghosh:2017fmr}%
  \BibitemOpen
  \bibfield  {author} {\bibinfo {author} {\bibfnamefont {P.}~\bibnamefont
  {Ghosh}}, \bibinfo {author} {\bibfnamefont {A.~K.}\ \bibnamefont {Saha}}, \
  and\ \bibinfo {author} {\bibfnamefont {A.}~\bibnamefont {Sil}},\ }\href
  {\doibase 10.1103/PhysRevD.97.075034} {\bibfield  {journal} {\bibinfo
  {journal} {Phys. Rev.}\ }\textbf {\bibinfo {volume} {D97}},\ \bibinfo {pages}
  {075034} (\bibinfo {year} {2018})},\ \Eprint
  {http://arxiv.org/abs/1706.04931} {arXiv:1706.04931 [hep-ph]} \BibitemShut
  {NoStop}%
%%CITATION = ARXIV:1706.04931;%%
\bibitem [{\citenamefont {Nanda}\ and\ \citenamefont
  {Borah}(2017)}]{Nanda:2017bmi}%
  \BibitemOpen
  \bibfield  {author} {\bibinfo {author} {\bibfnamefont {D.}~\bibnamefont
  {Nanda}}\ and\ \bibinfo {author} {\bibfnamefont {D.}~\bibnamefont {Borah}},\
  }\href {\doibase 10.1103/PhysRevD.96.115014} {\bibfield  {journal} {\bibinfo
  {journal} {Phys. Rev.}\ }\textbf {\bibinfo {volume} {D96}},\ \bibinfo {pages}
  {115014} (\bibinfo {year} {2017})},\ \Eprint
  {http://arxiv.org/abs/1709.08417} {arXiv:1709.08417 [hep-ph]} \BibitemShut
  {NoStop}%
%%CITATION = ARXIV:1709.08417;%%
\bibitem [{\citenamefont {Narendra}\ \emph {et~al.}(2017)\citenamefont
  {Narendra}, \citenamefont {Sahoo},\ and\ \citenamefont
  {Sahu}}]{Narendra:2017uxl}%
  \BibitemOpen
  \bibfield  {author} {\bibinfo {author} {\bibfnamefont {N.}~\bibnamefont
  {Narendra}}, \bibinfo {author} {\bibfnamefont {N.}~\bibnamefont {Sahoo}}, \
  and\ \bibinfo {author} {\bibfnamefont {N.}~\bibnamefont {Sahu}},\ }\href@noop
  {} {\  (\bibinfo {year} {2017})},\ \Eprint {http://arxiv.org/abs/1712.02960}
  {arXiv:1712.02960 [hep-ph]} \BibitemShut {NoStop}%
%%CITATION = ARXIV:1712.02960;%%
\bibitem [{\citenamefont {Bernal}\ \emph
  {et~al.}(2018{\natexlab{a}})\citenamefont {Bernal}, \citenamefont
  {Cárcamo~Hernández}, \citenamefont {de~Medeiros~Varzielas},\ and\
  \citenamefont {Kovalenko}}]{Bernal:2017xat}%
  \BibitemOpen
  \bibfield  {author} {\bibinfo {author} {\bibfnamefont {N.}~\bibnamefont
  {Bernal}}, \bibinfo {author} {\bibfnamefont {A.~E.}\ \bibnamefont
  {Cárcamo~Hernández}}, \bibinfo {author} {\bibfnamefont {I.}~\bibnamefont
  {de~Medeiros~Varzielas}}, \ and\ \bibinfo {author} {\bibfnamefont
  {S.}~\bibnamefont {Kovalenko}},\ }\href {\doibase 10.1007/JHEP05(2018)053}
  {\bibfield  {journal} {\bibinfo  {journal} {JHEP}\ }\textbf {\bibinfo
  {volume} {05}},\ \bibinfo {pages} {053} (\bibinfo {year}
  {2018}{\natexlab{a}})},\ \Eprint {http://arxiv.org/abs/1712.02792}
  {arXiv:1712.02792 [hep-ph]} \BibitemShut {NoStop}%
%%CITATION = ARXIV:1712.02792;%%
\bibitem [{\citenamefont {Borah}\ \emph {et~al.}(2018)\citenamefont {Borah},
  \citenamefont {Karmakar},\ and\ \citenamefont {Nanda}}]{Borah:2018gjk}%
  \BibitemOpen
  \bibfield  {author} {\bibinfo {author} {\bibfnamefont {D.}~\bibnamefont
  {Borah}}, \bibinfo {author} {\bibfnamefont {B.}~\bibnamefont {Karmakar}}, \
  and\ \bibinfo {author} {\bibfnamefont {D.}~\bibnamefont {Nanda}},\
  }\href@noop {} {\  (\bibinfo {year} {2018})},\ \Eprint
  {http://arxiv.org/abs/1805.11115} {arXiv:1805.11115 [hep-ph]} \BibitemShut
  {NoStop}%
%%CITATION = ARXIV:1805.11115;%%
\bibitem [{\citenamefont {Batell}\ \emph {et~al.}(2018)\citenamefont {Batell},
  \citenamefont {Han}, \citenamefont {McKeen},\ and\ \citenamefont {Shams
  Es~Haghi}}]{Batell:2017cmf}%
  \BibitemOpen
  \bibfield  {author} {\bibinfo {author} {\bibfnamefont {B.}~\bibnamefont
  {Batell}}, \bibinfo {author} {\bibfnamefont {T.}~\bibnamefont {Han}},
  \bibinfo {author} {\bibfnamefont {D.}~\bibnamefont {McKeen}}, \ and\ \bibinfo
  {author} {\bibfnamefont {B.}~\bibnamefont {Shams Es~Haghi}},\ }\href
  {\doibase 10.1103/PhysRevD.97.075016} {\bibfield  {journal} {\bibinfo
  {journal} {Phys. Rev.}\ }\textbf {\bibinfo {volume} {D97}},\ \bibinfo {pages}
  {075016} (\bibinfo {year} {2018})},\ \Eprint
  {http://arxiv.org/abs/1709.07001} {arXiv:1709.07001 [hep-ph]} \BibitemShut
  {NoStop}%
%%CITATION = ARXIV:1709.07001;%%
\bibitem [{\citenamefont {Pospelov}\ \emph {et~al.}(2008)\citenamefont
  {Pospelov}, \citenamefont {Ritz},\ and\ \citenamefont
  {Voloshin}}]{Pospelov:2007mp}%
  \BibitemOpen
  \bibfield  {author} {\bibinfo {author} {\bibfnamefont {M.}~\bibnamefont
  {Pospelov}}, \bibinfo {author} {\bibfnamefont {A.}~\bibnamefont {Ritz}}, \
  and\ \bibinfo {author} {\bibfnamefont {M.~B.}\ \bibnamefont {Voloshin}},\
  }\href {\doibase 10.1016/j.physletb.2008.02.052} {\bibfield  {journal}
  {\bibinfo  {journal} {Phys. Lett.}\ }\textbf {\bibinfo {volume} {B662}},\
  \bibinfo {pages} {53} (\bibinfo {year} {2008})},\ \Eprint
  {http://arxiv.org/abs/0711.4866} {arXiv:0711.4866 [hep-ph]} \BibitemShut
  {NoStop}%
%%CITATION = ARXIV:0711.4866;%%
\bibitem [{\citenamefont {Falkowski}\ \emph {et~al.}(2009)\citenamefont
  {Falkowski}, \citenamefont {Juknevich},\ and\ \citenamefont
  {Shelton}}]{Falkowski:2009yz}%
  \BibitemOpen
  \bibfield  {author} {\bibinfo {author} {\bibfnamefont {A.}~\bibnamefont
  {Falkowski}}, \bibinfo {author} {\bibfnamefont {J.}~\bibnamefont
  {Juknevich}}, \ and\ \bibinfo {author} {\bibfnamefont {J.}~\bibnamefont
  {Shelton}},\ }\href@noop {} {\  (\bibinfo {year} {2009})},\ \Eprint
  {http://arxiv.org/abs/0908.1790} {arXiv:0908.1790 [hep-ph]} \BibitemShut
  {NoStop}%
%%CITATION = ARXIV:0908.1790;%%
\bibitem [{\citenamefont {Falkowski}\ \emph {et~al.}(2011)\citenamefont
  {Falkowski}, \citenamefont {Ruderman},\ and\ \citenamefont
  {Volansky}}]{Falkowski:2011xh}%
  \BibitemOpen
  \bibfield  {author} {\bibinfo {author} {\bibfnamefont {A.}~\bibnamefont
  {Falkowski}}, \bibinfo {author} {\bibfnamefont {J.~T.}\ \bibnamefont
  {Ruderman}}, \ and\ \bibinfo {author} {\bibfnamefont {T.}~\bibnamefont
  {Volansky}},\ }\href {\doibase 10.1007/JHEP05(2011)106} {\bibfield  {journal}
  {\bibinfo  {journal} {JHEP}\ }\textbf {\bibinfo {volume} {05}},\ \bibinfo
  {pages} {106} (\bibinfo {year} {2011})},\ \Eprint
  {http://arxiv.org/abs/1101.4936} {arXiv:1101.4936 [hep-ph]} \BibitemShut
  {NoStop}%
%%CITATION = ARXIV:1101.4936;%%
\bibitem [{\citenamefont {Cherry}\ \emph {et~al.}(2014)\citenamefont {Cherry},
  \citenamefont {Friedland},\ and\ \citenamefont {Shoemaker}}]{Cherry:2014xra}%
  \BibitemOpen
  \bibfield  {author} {\bibinfo {author} {\bibfnamefont {J.~F.}\ \bibnamefont
  {Cherry}}, \bibinfo {author} {\bibfnamefont {A.}~\bibnamefont {Friedland}}, \
  and\ \bibinfo {author} {\bibfnamefont {I.~M.}\ \bibnamefont {Shoemaker}},\
  }\href@noop {} {\  (\bibinfo {year} {2014})},\ \Eprint
  {http://arxiv.org/abs/1411.1071} {arXiv:1411.1071 [hep-ph]} \BibitemShut
  {NoStop}%
%%CITATION = ARXIV:1411.1071;%%
\bibitem [{\citenamefont {Bertoni}\ \emph {et~al.}(2015)\citenamefont
  {Bertoni}, \citenamefont {Ipek}, \citenamefont {McKeen},\ and\ \citenamefont
  {Nelson}}]{Bertoni:2014mva}%
  \BibitemOpen
  \bibfield  {author} {\bibinfo {author} {\bibfnamefont {B.}~\bibnamefont
  {Bertoni}}, \bibinfo {author} {\bibfnamefont {S.}~\bibnamefont {Ipek}},
  \bibinfo {author} {\bibfnamefont {D.}~\bibnamefont {McKeen}}, \ and\ \bibinfo
  {author} {\bibfnamefont {A.~E.}\ \bibnamefont {Nelson}},\ }\href {\doibase
  10.1007/JHEP04(2015)170} {\bibfield  {journal} {\bibinfo  {journal} {JHEP}\
  }\textbf {\bibinfo {volume} {04}},\ \bibinfo {pages} {170} (\bibinfo {year}
  {2015})},\ \Eprint {http://arxiv.org/abs/1412.3113} {arXiv:1412.3113
  [hep-ph]} \BibitemShut {NoStop}%
%%CITATION = ARXIV:1412.3113;%%
\bibitem [{\citenamefont {Allahverdi}\ \emph {et~al.}(2017)\citenamefont
  {Allahverdi}, \citenamefont {Gao}, \citenamefont {Knockel},\ and\
  \citenamefont {Shalgar}}]{Allahverdi:2016fvl}%
  \BibitemOpen
  \bibfield  {author} {\bibinfo {author} {\bibfnamefont {R.}~\bibnamefont
  {Allahverdi}}, \bibinfo {author} {\bibfnamefont {Y.}~\bibnamefont {Gao}},
  \bibinfo {author} {\bibfnamefont {B.}~\bibnamefont {Knockel}}, \ and\
  \bibinfo {author} {\bibfnamefont {S.}~\bibnamefont {Shalgar}},\ }\href
  {\doibase 10.1103/PhysRevD.95.075001} {\bibfield  {journal} {\bibinfo
  {journal} {Phys. Rev.}\ }\textbf {\bibinfo {volume} {D95}},\ \bibinfo {pages}
  {075001} (\bibinfo {year} {2017})},\ \Eprint
  {http://arxiv.org/abs/1612.03110} {arXiv:1612.03110 [hep-ph]} \BibitemShut
  {NoStop}%
%%CITATION = ARXIV:1612.03110;%%
\bibitem [{\citenamefont {Karam}\ and\ \citenamefont
  {Tamvakis}(2015)}]{Karam:2015jta}%
  \BibitemOpen
  \bibfield  {author} {\bibinfo {author} {\bibfnamefont {A.}~\bibnamefont
  {Karam}}\ and\ \bibinfo {author} {\bibfnamefont {K.}~\bibnamefont
  {Tamvakis}},\ }\href {\doibase 10.1103/PhysRevD.92.075010} {\bibfield
  {journal} {\bibinfo  {journal} {Phys. Rev.}\ }\textbf {\bibinfo {volume}
  {D92}},\ \bibinfo {pages} {075010} (\bibinfo {year} {2015})},\ \Eprint
  {http://arxiv.org/abs/1508.03031} {arXiv:1508.03031 [hep-ph]} \BibitemShut
  {NoStop}%
%%CITATION = ARXIV:1508.03031;%%
\bibitem [{\citenamefont {Bhattacharya}\ \emph {et~al.}(2018)\citenamefont
  {Bhattacharya}, \citenamefont {de~Medeiros~Varzielas}, \citenamefont
  {Karmakar}, \citenamefont {King},\ and\ \citenamefont
  {Sil}}]{Bhattacharya:2018ljs}%
  \BibitemOpen
  \bibfield  {author} {\bibinfo {author} {\bibfnamefont {S.}~\bibnamefont
  {Bhattacharya}}, \bibinfo {author} {\bibfnamefont {I.}~\bibnamefont
  {de~Medeiros~Varzielas}}, \bibinfo {author} {\bibfnamefont {B.}~\bibnamefont
  {Karmakar}}, \bibinfo {author} {\bibfnamefont {S.~F.}\ \bibnamefont {King}},
  \ and\ \bibinfo {author} {\bibfnamefont {A.}~\bibnamefont {Sil}},\
  }\href@noop {} {\  (\bibinfo {year} {2018})},\ \Eprint
  {http://arxiv.org/abs/1806.00490} {arXiv:1806.00490 [hep-ph]} \BibitemShut
  {NoStop}%
%%CITATION = ARXIV:1806.00490;%%
\bibitem [{\citenamefont {Biswas}\ \emph {et~al.}(2019)\citenamefont {Biswas},
  \citenamefont {Choubey}, \citenamefont {Covi},\ and\ \citenamefont
  {Khan}}]{Biswas:2018sib}%
  \BibitemOpen
  \bibfield  {author} {\bibinfo {author} {\bibfnamefont {A.}~\bibnamefont
  {Biswas}}, \bibinfo {author} {\bibfnamefont {S.}~\bibnamefont {Choubey}},
  \bibinfo {author} {\bibfnamefont {L.}~\bibnamefont {Covi}}, \ and\ \bibinfo
  {author} {\bibfnamefont {S.}~\bibnamefont {Khan}},\ }\href {\doibase
  10.1007/JHEP05(2019)193} {\bibfield  {journal} {\bibinfo  {journal} {JHEP}\
  }\textbf {\bibinfo {volume} {05}},\ \bibinfo {pages} {193} (\bibinfo {year}
  {2019})},\ \Eprint {http://arxiv.org/abs/1812.06122} {arXiv:1812.06122
  [hep-ph]} \BibitemShut {NoStop}%
\bibitem [{\citenamefont {Gehrlein}\ and\ \citenamefont
  {Pierre}(2020)}]{Gehrlein:2019iwl}%
  \BibitemOpen
  \bibfield  {author} {\bibinfo {author} {\bibfnamefont {J.}~\bibnamefont
  {Gehrlein}}\ and\ \bibinfo {author} {\bibfnamefont {M.}~\bibnamefont
  {Pierre}},\ }\href {\doibase 10.1007/JHEP02(2020)068} {\bibfield  {journal}
  {\bibinfo  {journal} {JHEP}\ }\textbf {\bibinfo {volume} {02}},\ \bibinfo
  {pages} {068} (\bibinfo {year} {2020})},\ \Eprint
  {http://arxiv.org/abs/1912.06661} {arXiv:1912.06661 [hep-ph]} \BibitemShut
  {NoStop}%
\bibitem [{\citenamefont {Hashiba}\ and\ \citenamefont
  {Yokoyama}(2019{\natexlab{a}})}]{Hashiba:2019mzm}%
  \BibitemOpen
  \bibfield  {author} {\bibinfo {author} {\bibfnamefont {S.}~\bibnamefont
  {Hashiba}}\ and\ \bibinfo {author} {\bibfnamefont {J.}~\bibnamefont
  {Yokoyama}},\ }\href {\doibase 10.1016/j.physletb.2019.135024} {\bibfield
  {journal} {\bibinfo  {journal} {Phys. Lett. B}\ }\textbf {\bibinfo {volume}
  {798}},\ \bibinfo {pages} {135024} (\bibinfo {year} {2019}{\natexlab{a}})},\
  \Eprint {http://arxiv.org/abs/1905.12423} {arXiv:1905.12423 [hep-ph]}
  \BibitemShut {NoStop}%
\bibitem [{\citenamefont {Dasgupta}\ \emph {et~al.}(2019)\citenamefont
  {Dasgupta}, \citenamefont {Bhupal~Dev}, \citenamefont {Kang},\ and\
  \citenamefont {Zhang}}]{Dasgupta:2019lha}%
  \BibitemOpen
  \bibfield  {author} {\bibinfo {author} {\bibfnamefont {A.}~\bibnamefont
  {Dasgupta}}, \bibinfo {author} {\bibfnamefont {P.}~\bibnamefont
  {Bhupal~Dev}}, \bibinfo {author} {\bibfnamefont {S.~K.}\ \bibnamefont
  {Kang}}, \ and\ \bibinfo {author} {\bibfnamefont {Y.}~\bibnamefont {Zhang}},\
  }\href@noop {} {\  (\bibinfo {year} {2019})},\ \Eprint
  {http://arxiv.org/abs/1911.03013} {arXiv:1911.03013 [hep-ph]} \BibitemShut
  {NoStop}%
\bibitem [{\citenamefont {Samanta}\ \emph {et~al.}(2020)\citenamefont
  {Samanta}, \citenamefont {Biswas},\ and\ \citenamefont
  {Bhattacharya}}]{Samanta:2020gdw}%
  \BibitemOpen
  \bibfield  {author} {\bibinfo {author} {\bibfnamefont {R.}~\bibnamefont
  {Samanta}}, \bibinfo {author} {\bibfnamefont {A.}~\bibnamefont {Biswas}}, \
  and\ \bibinfo {author} {\bibfnamefont {S.}~\bibnamefont {Bhattacharya}},\
  }\href@noop {} {\  (\bibinfo {year} {2020})},\ \Eprint
  {http://arxiv.org/abs/2006.02960} {arXiv:2006.02960 [hep-ph]} \BibitemShut
  {NoStop}%
\bibitem [{\citenamefont {Chianese}\ and\ \citenamefont
  {King}(2018)}]{Chianese:2018dsz}%
  \BibitemOpen
  \bibfield  {author} {\bibinfo {author} {\bibfnamefont {M.}~\bibnamefont
  {Chianese}}\ and\ \bibinfo {author} {\bibfnamefont {S.~F.}\ \bibnamefont
  {King}},\ }\href {\doibase 10.1088/1475-7516/2018/09/027} {\bibfield
  {journal} {\bibinfo  {journal} {JCAP}\ }\textbf {\bibinfo {volume} {1809}},\
  \bibinfo {pages} {027} (\bibinfo {year} {2018})},\ \Eprint
  {http://arxiv.org/abs/1806.10606} {arXiv:1806.10606 [hep-ph]} \BibitemShut
  {NoStop}%
%%CITATION = ARXIV:1806.10606;%%
\bibitem [{\citenamefont {Chianese}\ \emph
  {et~al.}(2020{\natexlab{a}})\citenamefont {Chianese}, \citenamefont {Fu},\
  and\ \citenamefont {King}}]{Chianese:2019epo}%
  \BibitemOpen
  \bibfield  {author} {\bibinfo {author} {\bibfnamefont {M.}~\bibnamefont
  {Chianese}}, \bibinfo {author} {\bibfnamefont {B.}~\bibnamefont {Fu}}, \ and\
  \bibinfo {author} {\bibfnamefont {S.~F.}\ \bibnamefont {King}},\ }\href
  {\doibase 10.1088/1475-7516/2020/03/030} {\bibfield  {journal} {\bibinfo
  {journal} {JCAP}\ }\textbf {\bibinfo {volume} {03}},\ \bibinfo {pages} {030}
  (\bibinfo {year} {2020}{\natexlab{a}})},\ \Eprint
  {http://arxiv.org/abs/1910.12916} {arXiv:1910.12916 [hep-ph]} \BibitemShut
  {NoStop}%
\bibitem [{\citenamefont {Becker}(2019)}]{Becker:2018rve}%
  \BibitemOpen
  \bibfield  {author} {\bibinfo {author} {\bibfnamefont {M.}~\bibnamefont
  {Becker}},\ }\href {\doibase 10.1140/epjc/s10052-019-7095-7} {\bibfield
  {journal} {\bibinfo  {journal} {Eur. Phys. J.}\ }\textbf {\bibinfo {volume}
  {C79}},\ \bibinfo {pages} {611} (\bibinfo {year} {2019})},\ \Eprint
  {http://arxiv.org/abs/1806.08579} {arXiv:1806.08579 [hep-ph]} \BibitemShut
  {NoStop}%
%%CITATION = ARXIV:1806.08579;%%
\bibitem [{\citenamefont {Bian}\ and\ \citenamefont
  {Tang}(2018)}]{Bian:2018mkl}%
  \BibitemOpen
  \bibfield  {author} {\bibinfo {author} {\bibfnamefont {L.}~\bibnamefont
  {Bian}}\ and\ \bibinfo {author} {\bibfnamefont {Y.-L.}\ \bibnamefont
  {Tang}},\ }\href {\doibase 10.1007/JHEP12(2018)006} {\bibfield  {journal}
  {\bibinfo  {journal} {JHEP}\ }\textbf {\bibinfo {volume} {12}},\ \bibinfo
  {pages} {006} (\bibinfo {year} {2018})},\ \Eprint
  {http://arxiv.org/abs/1810.03172} {arXiv:1810.03172 [hep-ph]} \BibitemShut
  {NoStop}%
\bibitem [{\citenamefont {Bandyopadhyay}\ \emph {et~al.}(2019)\citenamefont
  {Bandyopadhyay}, \citenamefont {Chun}, \citenamefont {Mandal},\ and\
  \citenamefont {Queiroz}}]{Bandyopadhyay:2018qcv}%
  \BibitemOpen
  \bibfield  {author} {\bibinfo {author} {\bibfnamefont {P.}~\bibnamefont
  {Bandyopadhyay}}, \bibinfo {author} {\bibfnamefont {E.~J.}\ \bibnamefont
  {Chun}}, \bibinfo {author} {\bibfnamefont {R.}~\bibnamefont {Mandal}}, \ and\
  \bibinfo {author} {\bibfnamefont {F.~S.}\ \bibnamefont {Queiroz}},\ }\href
  {\doibase 10.1016/j.physletb.2018.12.003} {\bibfield  {journal} {\bibinfo
  {journal} {Phys. Lett. B}\ }\textbf {\bibinfo {volume} {788}},\ \bibinfo
  {pages} {530} (\bibinfo {year} {2019})},\ \Eprint
  {http://arxiv.org/abs/1807.05122} {arXiv:1807.05122 [hep-ph]} \BibitemShut
  {NoStop}%
\bibitem [{\citenamefont {Liu}\ \emph {et~al.}(2020)\citenamefont {Liu},
  \citenamefont {Han}, \citenamefont {Jin},\ and\ \citenamefont
  {Yang}}]{Liu:2020mxj}%
  \BibitemOpen
  \bibfield  {author} {\bibinfo {author} {\bibfnamefont {A.}~\bibnamefont
  {Liu}}, \bibinfo {author} {\bibfnamefont {Z.-L.}\ \bibnamefont {Han}},
  \bibinfo {author} {\bibfnamefont {Y.}~\bibnamefont {Jin}}, \ and\ \bibinfo
  {author} {\bibfnamefont {F.-X.}\ \bibnamefont {Yang}},\ }\href {\doibase
  10.1103/PhysRevD.101.095005} {\bibfield  {journal} {\bibinfo  {journal}
  {Phys. Rev. D}\ }\textbf {\bibinfo {volume} {101}},\ \bibinfo {pages}
  {095005} (\bibinfo {year} {2020})},\ \Eprint
  {http://arxiv.org/abs/2001.04085} {arXiv:2001.04085 [hep-ph]} \BibitemShut
  {NoStop}%
\bibitem [{\citenamefont {Cosme}\ \emph {et~al.}(2020)\citenamefont {Cosme},
  \citenamefont {Dutra}, \citenamefont {Ma}, \citenamefont {Wu},\ and\
  \citenamefont {Yang}}]{Cosme:2020mck}%
  \BibitemOpen
  \bibfield  {author} {\bibinfo {author} {\bibfnamefont {C.}~\bibnamefont
  {Cosme}}, \bibinfo {author} {\bibfnamefont {M.}~\bibnamefont {Dutra}},
  \bibinfo {author} {\bibfnamefont {T.}~\bibnamefont {Ma}}, \bibinfo {author}
  {\bibfnamefont {Y.}~\bibnamefont {Wu}}, \ and\ \bibinfo {author}
  {\bibfnamefont {L.}~\bibnamefont {Yang}},\ }\href@noop {} {\  (\bibinfo
  {year} {2020})},\ \Eprint {http://arxiv.org/abs/2003.01723} {arXiv:2003.01723
  [hep-ph]} \BibitemShut {NoStop}%
\bibitem [{\citenamefont {Du}\ \emph {et~al.}(2020)\citenamefont {Du},
  \citenamefont {Huang}, \citenamefont {Li},\ and\ \citenamefont
  {Yu}}]{Du:2020avz}%
  \BibitemOpen
  \bibfield  {author} {\bibinfo {author} {\bibfnamefont {Y.}~\bibnamefont
  {Du}}, \bibinfo {author} {\bibfnamefont {F.}~\bibnamefont {Huang}}, \bibinfo
  {author} {\bibfnamefont {H.-L.}\ \bibnamefont {Li}}, \ and\ \bibinfo {author}
  {\bibfnamefont {J.-H.}\ \bibnamefont {Yu}},\ }\href@noop {} {\  (\bibinfo
  {year} {2020})},\ \Eprint {http://arxiv.org/abs/2005.01717} {arXiv:2005.01717
  [hep-ph]} \BibitemShut {NoStop}%
\bibitem [{\citenamefont {Bandyopadhyay}\ \emph {et~al.}(2020)\citenamefont
  {Bandyopadhyay}, \citenamefont {Chun},\ and\ \citenamefont
  {Mandal}}]{Bandyopadhyay:2020qpn}%
  \BibitemOpen
  \bibfield  {author} {\bibinfo {author} {\bibfnamefont {P.}~\bibnamefont
  {Bandyopadhyay}}, \bibinfo {author} {\bibfnamefont {E.~J.}\ \bibnamefont
  {Chun}}, \ and\ \bibinfo {author} {\bibfnamefont {R.}~\bibnamefont
  {Mandal}},\ }\href {\doibase 10.1088/1475-7516/2020/08/019} {\bibfield
  {journal} {\bibinfo  {journal} {JCAP}\ }\textbf {\bibinfo {volume} {08}},\
  \bibinfo {pages} {019} (\bibinfo {year} {2020})},\ \Eprint
  {http://arxiv.org/abs/2005.13933} {arXiv:2005.13933 [hep-ph]} \BibitemShut
  {NoStop}%
\bibitem [{\citenamefont {McDonald}(2002)}]{McDonald:2001vt}%
  \BibitemOpen
  \bibfield  {author} {\bibinfo {author} {\bibfnamefont {J.}~\bibnamefont
  {McDonald}},\ }\href {\doibase 10.1103/PhysRevLett.88.091304} {\bibfield
  {journal} {\bibinfo  {journal} {Phys. Rev. Lett.}\ }\textbf {\bibinfo
  {volume} {88}},\ \bibinfo {pages} {091304} (\bibinfo {year} {2002})},\
  \Eprint {http://arxiv.org/abs/hep-ph/0106249} {arXiv:hep-ph/0106249}
  \BibitemShut {NoStop}%
\bibitem [{\citenamefont {Hall}\ \emph {et~al.}(2010)\citenamefont {Hall},
  \citenamefont {Jedamzik}, \citenamefont {March-Russell},\ and\ \citenamefont
  {West}}]{Hall:2009bx}%
  \BibitemOpen
  \bibfield  {author} {\bibinfo {author} {\bibfnamefont {L.~J.}\ \bibnamefont
  {Hall}}, \bibinfo {author} {\bibfnamefont {K.}~\bibnamefont {Jedamzik}},
  \bibinfo {author} {\bibfnamefont {J.}~\bibnamefont {March-Russell}}, \ and\
  \bibinfo {author} {\bibfnamefont {S.~M.}\ \bibnamefont {West}},\ }\href
  {\doibase 10.1007/JHEP03(2010)080} {\bibfield  {journal} {\bibinfo  {journal}
  {JHEP}\ }\textbf {\bibinfo {volume} {03}},\ \bibinfo {pages} {080} (\bibinfo
  {year} {2010})},\ \Eprint {http://arxiv.org/abs/0911.1120} {arXiv:0911.1120
  [hep-ph]} \BibitemShut {NoStop}%
%%CITATION = ARXIV:0911.1120;%%
\bibitem [{\citenamefont {Bernal}\ \emph {et~al.}(2017)\citenamefont {Bernal},
  \citenamefont {Heikinheimo}, \citenamefont {Tenkanen}, \citenamefont
  {Tuominen},\ and\ \citenamefont {Vaskonen}}]{Bernal:2017kxu}%
  \BibitemOpen
  \bibfield  {author} {\bibinfo {author} {\bibfnamefont {N.}~\bibnamefont
  {Bernal}}, \bibinfo {author} {\bibfnamefont {M.}~\bibnamefont {Heikinheimo}},
  \bibinfo {author} {\bibfnamefont {T.}~\bibnamefont {Tenkanen}}, \bibinfo
  {author} {\bibfnamefont {K.}~\bibnamefont {Tuominen}}, \ and\ \bibinfo
  {author} {\bibfnamefont {V.}~\bibnamefont {Vaskonen}},\ }\href {\doibase
  10.1142/S0217751X1730023X} {\bibfield  {journal} {\bibinfo  {journal} {Int.
  J. Mod. Phys.}\ }\textbf {\bibinfo {volume} {A32}},\ \bibinfo {pages}
  {1730023} (\bibinfo {year} {2017})},\ \Eprint
  {http://arxiv.org/abs/1706.07442} {arXiv:1706.07442 [hep-ph]} \BibitemShut
  {NoStop}%
%%CITATION = ARXIV:1706.07442;%%
\bibitem [{\citenamefont {Garny}\ \emph {et~al.}(2016)\citenamefont {Garny},
  \citenamefont {Sandora},\ and\ \citenamefont {Sloth}}]{Garny:2015sjg}%
  \BibitemOpen
  \bibfield  {author} {\bibinfo {author} {\bibfnamefont {M.}~\bibnamefont
  {Garny}}, \bibinfo {author} {\bibfnamefont {M.}~\bibnamefont {Sandora}}, \
  and\ \bibinfo {author} {\bibfnamefont {M.~S.}\ \bibnamefont {Sloth}},\ }\href
  {\doibase 10.1103/PhysRevLett.116.101302} {\bibfield  {journal} {\bibinfo
  {journal} {Phys. Rev. Lett.}\ }\textbf {\bibinfo {volume} {116}},\ \bibinfo
  {pages} {101302} (\bibinfo {year} {2016})},\ \Eprint
  {http://arxiv.org/abs/1511.03278} {arXiv:1511.03278 [hep-ph]} \BibitemShut
  {NoStop}%
%%CITATION = ARXIV:1511.03278;%%
\bibitem [{\citenamefont {Tang}\ and\ \citenamefont {Wu}(2016)}]{Tang:2016vch}%
  \BibitemOpen
  \bibfield  {author} {\bibinfo {author} {\bibfnamefont {Y.}~\bibnamefont
  {Tang}}\ and\ \bibinfo {author} {\bibfnamefont {Y.-L.}\ \bibnamefont {Wu}},\
  }\href {\doibase 10.1016/j.physletb.2016.05.045} {\bibfield  {journal}
  {\bibinfo  {journal} {Phys. Lett.}\ }\textbf {\bibinfo {volume} {B758}},\
  \bibinfo {pages} {402} (\bibinfo {year} {2016})},\ \Eprint
  {http://arxiv.org/abs/1604.04701} {arXiv:1604.04701 [hep-ph]} \BibitemShut
  {NoStop}%
%%CITATION = ARXIV:1604.04701;%%
\bibitem [{\citenamefont {Tang}\ and\ \citenamefont {Wu}(2017)}]{Tang:2017hvq}%
  \BibitemOpen
  \bibfield  {author} {\bibinfo {author} {\bibfnamefont {Y.}~\bibnamefont
  {Tang}}\ and\ \bibinfo {author} {\bibfnamefont {Y.-L.}\ \bibnamefont {Wu}},\
  }\href {\doibase 10.1016/j.physletb.2017.10.034} {\bibfield  {journal}
  {\bibinfo  {journal} {Phys. Lett.}\ }\textbf {\bibinfo {volume} {B774}},\
  \bibinfo {pages} {676} (\bibinfo {year} {2017})},\ \Eprint
  {http://arxiv.org/abs/1708.05138} {arXiv:1708.05138 [hep-ph]} \BibitemShut
  {NoStop}%
%%CITATION = ARXIV:1708.05138;%%
\bibitem [{\citenamefont {Garny}\ \emph {et~al.}(2018)\citenamefont {Garny},
  \citenamefont {Palessandro}, \citenamefont {Sandora},\ and\ \citenamefont
  {Sloth}}]{Garny:2017kha}%
  \BibitemOpen
  \bibfield  {author} {\bibinfo {author} {\bibfnamefont {M.}~\bibnamefont
  {Garny}}, \bibinfo {author} {\bibfnamefont {A.}~\bibnamefont {Palessandro}},
  \bibinfo {author} {\bibfnamefont {M.}~\bibnamefont {Sandora}}, \ and\
  \bibinfo {author} {\bibfnamefont {M.~S.}\ \bibnamefont {Sloth}},\ }\href
  {\doibase 10.1088/1475-7516/2018/02/027} {\bibfield  {journal} {\bibinfo
  {journal} {JCAP}\ }\textbf {\bibinfo {volume} {1802}},\ \bibinfo {pages}
  {027} (\bibinfo {year} {2018})},\ \Eprint {http://arxiv.org/abs/1709.09688}
  {arXiv:1709.09688 [hep-ph]} \BibitemShut {NoStop}%
%%CITATION = ARXIV:1709.09688;%%
\bibitem [{\citenamefont {Bernal}\ \emph
  {et~al.}(2018{\natexlab{b}})\citenamefont {Bernal}, \citenamefont {Dutra},
  \citenamefont {Mambrini}, \citenamefont {Olive}, \citenamefont {Peloso},\
  and\ \citenamefont {Pierre}}]{Bernal:2018qlk}%
  \BibitemOpen
  \bibfield  {author} {\bibinfo {author} {\bibfnamefont {N.}~\bibnamefont
  {Bernal}}, \bibinfo {author} {\bibfnamefont {M.}~\bibnamefont {Dutra}},
  \bibinfo {author} {\bibfnamefont {Y.}~\bibnamefont {Mambrini}}, \bibinfo
  {author} {\bibfnamefont {K.}~\bibnamefont {Olive}}, \bibinfo {author}
  {\bibfnamefont {M.}~\bibnamefont {Peloso}}, \ and\ \bibinfo {author}
  {\bibfnamefont {M.}~\bibnamefont {Pierre}},\ }\href {\doibase
  10.1103/PhysRevD.97.115020} {\bibfield  {journal} {\bibinfo  {journal} {Phys.
  Rev.}\ }\textbf {\bibinfo {volume} {D97}},\ \bibinfo {pages} {115020}
  (\bibinfo {year} {2018}{\natexlab{b}})},\ \Eprint
  {http://arxiv.org/abs/1803.01866} {arXiv:1803.01866 [hep-ph]} \BibitemShut
  {NoStop}%
%%CITATION = ARXIV:1803.01866;%%
\bibitem [{\citenamefont {Garny}\ \emph {et~al.}(2019)\citenamefont {Garny},
  \citenamefont {Palessandro}, \citenamefont {Sandora},\ and\ \citenamefont
  {Sloth}}]{Garny:2018grs}%
  \BibitemOpen
  \bibfield  {author} {\bibinfo {author} {\bibfnamefont {M.}~\bibnamefont
  {Garny}}, \bibinfo {author} {\bibfnamefont {A.}~\bibnamefont {Palessandro}},
  \bibinfo {author} {\bibfnamefont {M.}~\bibnamefont {Sandora}}, \ and\
  \bibinfo {author} {\bibfnamefont {M.~S.}\ \bibnamefont {Sloth}},\ }\href
  {\doibase 10.1088/1475-7516/2019/01/021} {\bibfield  {journal} {\bibinfo
  {journal} {JCAP}\ }\textbf {\bibinfo {volume} {1901}},\ \bibinfo {pages}
  {021} (\bibinfo {year} {2019})},\ \Eprint {http://arxiv.org/abs/1810.01428}
  {arXiv:1810.01428 [hep-ph]} \BibitemShut {NoStop}%
%%CITATION = ARXIV:1810.01428;%%
\bibitem [{\citenamefont {Hashiba}\ and\ \citenamefont
  {Yokoyama}(2019{\natexlab{b}})}]{Hashiba:2018tbu}%
  \BibitemOpen
  \bibfield  {author} {\bibinfo {author} {\bibfnamefont {S.}~\bibnamefont
  {Hashiba}}\ and\ \bibinfo {author} {\bibfnamefont {J.}~\bibnamefont
  {Yokoyama}},\ }\href {\doibase 10.1103/PhysRevD.99.043008} {\bibfield
  {journal} {\bibinfo  {journal} {Phys. Rev.}\ }\textbf {\bibinfo {volume}
  {D99}},\ \bibinfo {pages} {043008} (\bibinfo {year} {2019}{\natexlab{b}})},\
  \Eprint {http://arxiv.org/abs/1812.10032} {arXiv:1812.10032 [hep-ph]}
  \BibitemShut {NoStop}%
%%CITATION = ARXIV:1812.10032;%%
\bibitem [{\citenamefont {Lee}\ \emph {et~al.}(2014{\natexlab{a}})\citenamefont
  {Lee}, \citenamefont {Park},\ and\ \citenamefont {Sanz}}]{Lee:2013bua}%
  \BibitemOpen
  \bibfield  {author} {\bibinfo {author} {\bibfnamefont {H.~M.}\ \bibnamefont
  {Lee}}, \bibinfo {author} {\bibfnamefont {M.}~\bibnamefont {Park}}, \ and\
  \bibinfo {author} {\bibfnamefont {V.}~\bibnamefont {Sanz}},\ }\href {\doibase
  10.1140/epjc/s10052-014-2715-8} {\bibfield  {journal} {\bibinfo  {journal}
  {Eur. Phys. J. C}\ }\textbf {\bibinfo {volume} {74}},\ \bibinfo {pages}
  {2715} (\bibinfo {year} {2014}{\natexlab{a}})},\ \Eprint
  {http://arxiv.org/abs/1306.4107} {arXiv:1306.4107 [hep-ph]} \BibitemShut
  {NoStop}%
\bibitem [{\citenamefont {Lee}\ \emph {et~al.}(2014{\natexlab{b}})\citenamefont
  {Lee}, \citenamefont {Park},\ and\ \citenamefont {Sanz}}]{Lee:2014caa}%
  \BibitemOpen
  \bibfield  {author} {\bibinfo {author} {\bibfnamefont {H.~M.}\ \bibnamefont
  {Lee}}, \bibinfo {author} {\bibfnamefont {M.}~\bibnamefont {Park}}, \ and\
  \bibinfo {author} {\bibfnamefont {V.}~\bibnamefont {Sanz}},\ }\href {\doibase
  10.1007/JHEP05(2014)063} {\bibfield  {journal} {\bibinfo  {journal} {JHEP}\
  }\textbf {\bibinfo {volume} {05}},\ \bibinfo {pages} {063} (\bibinfo {year}
  {2014}{\natexlab{b}})},\ \Eprint {http://arxiv.org/abs/1401.5301}
  {arXiv:1401.5301 [hep-ph]} \BibitemShut {NoStop}%
\bibitem [{\citenamefont {Kraml}\ \emph {et~al.}(2017)\citenamefont {Kraml},
  \citenamefont {Laa}, \citenamefont {Mawatari},\ and\ \citenamefont
  {Yamashita}}]{Kraml:2017atm}%
  \BibitemOpen
  \bibfield  {author} {\bibinfo {author} {\bibfnamefont {S.}~\bibnamefont
  {Kraml}}, \bibinfo {author} {\bibfnamefont {U.}~\bibnamefont {Laa}}, \bibinfo
  {author} {\bibfnamefont {K.}~\bibnamefont {Mawatari}}, \ and\ \bibinfo
  {author} {\bibfnamefont {K.}~\bibnamefont {Yamashita}},\ }\href {\doibase
  10.1140/epjc/s10052-017-4871-0} {\bibfield  {journal} {\bibinfo  {journal}
  {Eur. Phys. J. C}\ }\textbf {\bibinfo {volume} {77}},\ \bibinfo {pages} {326}
  (\bibinfo {year} {2017})},\ \Eprint {http://arxiv.org/abs/1701.07008}
  {arXiv:1701.07008 [hep-ph]} \BibitemShut {NoStop}%
\bibitem [{\citenamefont {Carrillo-Monteverde}\ \emph
  {et~al.}(2018)\citenamefont {Carrillo-Monteverde}, \citenamefont {Kang},
  \citenamefont {Lee}, \citenamefont {Park},\ and\ \citenamefont
  {Sanz}}]{Carrillo-Monteverde:2018phy}%
  \BibitemOpen
  \bibfield  {author} {\bibinfo {author} {\bibfnamefont {A.}~\bibnamefont
  {Carrillo-Monteverde}}, \bibinfo {author} {\bibfnamefont {Y.-J.}\
  \bibnamefont {Kang}}, \bibinfo {author} {\bibfnamefont {H.~M.}\ \bibnamefont
  {Lee}}, \bibinfo {author} {\bibfnamefont {M.}~\bibnamefont {Park}}, \ and\
  \bibinfo {author} {\bibfnamefont {V.}~\bibnamefont {Sanz}},\ }\href {\doibase
  10.1007/JHEP06(2018)037} {\bibfield  {journal} {\bibinfo  {journal} {JHEP}\
  }\textbf {\bibinfo {volume} {06}},\ \bibinfo {pages} {037} (\bibinfo {year}
  {2018})},\ \Eprint {http://arxiv.org/abs/1803.02144} {arXiv:1803.02144
  [hep-ph]} \BibitemShut {NoStop}%
\bibitem [{\citenamefont {Kang}\ and\ \citenamefont
  {Lee}(2020{\natexlab{a}})}]{Kang:2020huh}%
  \BibitemOpen
  \bibfield  {author} {\bibinfo {author} {\bibfnamefont {Y.-J.}\ \bibnamefont
  {Kang}}\ and\ \bibinfo {author} {\bibfnamefont {H.~M.}\ \bibnamefont {Lee}},\
  }\href {\doibase 10.1140/epjc/s10052-020-8153-x} {\bibfield  {journal}
  {\bibinfo  {journal} {Eur. Phys. J. C}\ }\textbf {\bibinfo {volume} {80}},\
  \bibinfo {pages} {602} (\bibinfo {year} {2020}{\natexlab{a}})},\ \Eprint
  {http://arxiv.org/abs/2001.04868} {arXiv:2001.04868 [hep-ph]} \BibitemShut
  {NoStop}%
\bibitem [{\citenamefont {Kang}\ and\ \citenamefont
  {Lee}(2020{\natexlab{b}})}]{Kang:2020yul}%
  \BibitemOpen
  \bibfield  {author} {\bibinfo {author} {\bibfnamefont {Y.-J.}\ \bibnamefont
  {Kang}}\ and\ \bibinfo {author} {\bibfnamefont {H.~M.}\ \bibnamefont {Lee}},\
  }\href@noop {} {\  (\bibinfo {year} {2020}{\natexlab{b}})},\ \Eprint
  {http://arxiv.org/abs/2002.12779} {arXiv:2002.12779 [hep-ph]} \BibitemShut
  {NoStop}%
\bibitem [{\citenamefont {Kang}\ and\ \citenamefont
  {Lee}(2020{\natexlab{c}})}]{Kang:2020afi}%
  \BibitemOpen
  \bibfield  {author} {\bibinfo {author} {\bibfnamefont {Y.-J.}\ \bibnamefont
  {Kang}}\ and\ \bibinfo {author} {\bibfnamefont {H.~M.}\ \bibnamefont {Lee}},\
  }\href@noop {} {\  (\bibinfo {year} {2020}{\natexlab{c}})},\ \Eprint
  {http://arxiv.org/abs/2003.09290} {arXiv:2003.09290 [hep-ph]} \BibitemShut
  {NoStop}%
\bibitem [{\citenamefont {Chianese}\ \emph
  {et~al.}(2020{\natexlab{b}})\citenamefont {Chianese}, \citenamefont {Fu},\
  and\ \citenamefont {King}}]{Chianese:2020yjo}%
  \BibitemOpen
  \bibfield  {author} {\bibinfo {author} {\bibfnamefont {M.}~\bibnamefont
  {Chianese}}, \bibinfo {author} {\bibfnamefont {B.}~\bibnamefont {Fu}}, \ and\
  \bibinfo {author} {\bibfnamefont {S.~F.}\ \bibnamefont {King}},\ }\href
  {\doibase 10.1088/1475-7516/2020/06/019} {\bibfield  {journal} {\bibinfo
  {journal} {JCAP}\ }\textbf {\bibinfo {volume} {06}},\ \bibinfo {pages} {019}
  (\bibinfo {year} {2020}{\natexlab{b}})},\ \Eprint
  {http://arxiv.org/abs/2003.07366} {arXiv:2003.07366 [hep-ph]} \BibitemShut
  {NoStop}%
\bibitem [{\citenamefont {Capozzi}\ \emph {et~al.}(2018)\citenamefont
  {Capozzi}, \citenamefont {Lisi}, \citenamefont {Marrone},\ and\ \citenamefont
  {Palazzo}}]{Capozzi:2018ubv}%
  \BibitemOpen
  \bibfield  {author} {\bibinfo {author} {\bibfnamefont {F.}~\bibnamefont
  {Capozzi}}, \bibinfo {author} {\bibfnamefont {E.}~\bibnamefont {Lisi}},
  \bibinfo {author} {\bibfnamefont {A.}~\bibnamefont {Marrone}}, \ and\
  \bibinfo {author} {\bibfnamefont {A.}~\bibnamefont {Palazzo}},\ }\href@noop
  {} {\  (\bibinfo {year} {2018})},\ \Eprint {http://arxiv.org/abs/1804.09678}
  {arXiv:1804.09678 [hep-ph]} \BibitemShut {NoStop}%
%%CITATION = ARXIV:1804.09678;%%
\bibitem [{\citenamefont {Capozzi}\ \emph {et~al.}(2020)\citenamefont
  {Capozzi}, \citenamefont {Di~Valentino}, \citenamefont {Lisi}, \citenamefont
  {Marrone}, \citenamefont {Melchiorri},\ and\ \citenamefont
  {Palazzo}}]{Capozzi:2020qhw}%
  \BibitemOpen
  \bibfield  {author} {\bibinfo {author} {\bibfnamefont {F.}~\bibnamefont
  {Capozzi}}, \bibinfo {author} {\bibfnamefont {E.}~\bibnamefont
  {Di~Valentino}}, \bibinfo {author} {\bibfnamefont {E.}~\bibnamefont {Lisi}},
  \bibinfo {author} {\bibfnamefont {A.}~\bibnamefont {Marrone}}, \bibinfo
  {author} {\bibfnamefont {A.}~\bibnamefont {Melchiorri}}, \ and\ \bibinfo
  {author} {\bibfnamefont {A.}~\bibnamefont {Palazzo}},\ }\href {\doibase
  10.1103/PhysRevD.101.116013} {\  (\bibinfo {year} {2020}),\
  10.1103/PhysRevD.101.116013},\ \bibinfo {note} {[Addendum: Phys.Rev.D 101,
  116013 (2020)]},\ \Eprint {http://arxiv.org/abs/2003.08511} {arXiv:2003.08511
  [hep-ph]} \BibitemShut {NoStop}%
\bibitem [{\citenamefont {Gariazzo}\ \emph {et~al.}(2018)\citenamefont
  {Gariazzo}, \citenamefont {Archidiacono}, \citenamefont {de~Salas},
  \citenamefont {Mena}, \citenamefont {Ternes},\ and\ \citenamefont
  {Tórtola}}]{Gariazzo:2018pei}%
  \BibitemOpen
  \bibfield  {author} {\bibinfo {author} {\bibfnamefont {S.}~\bibnamefont
  {Gariazzo}}, \bibinfo {author} {\bibfnamefont {M.}~\bibnamefont
  {Archidiacono}}, \bibinfo {author} {\bibfnamefont {P.}~\bibnamefont
  {de~Salas}}, \bibinfo {author} {\bibfnamefont {O.}~\bibnamefont {Mena}},
  \bibinfo {author} {\bibfnamefont {C.}~\bibnamefont {Ternes}}, \ and\ \bibinfo
  {author} {\bibfnamefont {M.}~\bibnamefont {Tórtola}},\ }\href {\doibase
  10.1088/1475-7516/2018/03/011} {\bibfield  {journal} {\bibinfo  {journal}
  {JCAP}\ }\textbf {\bibinfo {volume} {03}},\ \bibinfo {pages} {011} (\bibinfo
  {year} {2018})},\ \Eprint {http://arxiv.org/abs/1801.04946} {arXiv:1801.04946
  [hep-ph]} \BibitemShut {NoStop}%
\bibitem [{\citenamefont {de~Salas}\ \emph {et~al.}(2020)\citenamefont
  {de~Salas}, \citenamefont {Forero}, \citenamefont {Gariazzo}, \citenamefont
  {Martínez-Miravé}, \citenamefont {Mena}, \citenamefont {Ternes},
  \citenamefont {Tórtola},\ and\ \citenamefont {Valle}}]{deSalas:2020pgw}%
  \BibitemOpen
  \bibfield  {author} {\bibinfo {author} {\bibfnamefont {P.}~\bibnamefont
  {de~Salas}}, \bibinfo {author} {\bibfnamefont {D.}~\bibnamefont {Forero}},
  \bibinfo {author} {\bibfnamefont {S.}~\bibnamefont {Gariazzo}}, \bibinfo
  {author} {\bibfnamefont {P.}~\bibnamefont {Martínez-Miravé}}, \bibinfo
  {author} {\bibfnamefont {O.}~\bibnamefont {Mena}}, \bibinfo {author}
  {\bibfnamefont {C.}~\bibnamefont {Ternes}}, \bibinfo {author} {\bibfnamefont
  {M.}~\bibnamefont {Tórtola}}, \ and\ \bibinfo {author} {\bibfnamefont
  {J.}~\bibnamefont {Valle}},\ }\href@noop {} {\  (\bibinfo {year} {2020})},\
  \Eprint {http://arxiv.org/abs/2006.11237} {arXiv:2006.11237 [hep-ph]}
  \BibitemShut {NoStop}%
\bibitem [{\citenamefont {Esteban}\ \emph {et~al.}(2019)\citenamefont
  {Esteban}, \citenamefont {Gonzalez-Garcia}, \citenamefont
  {Hernandez-Cabezudo}, \citenamefont {Maltoni},\ and\ \citenamefont
  {Schwetz}}]{Esteban:2018azc}%
  \BibitemOpen
  \bibfield  {author} {\bibinfo {author} {\bibfnamefont {I.}~\bibnamefont
  {Esteban}}, \bibinfo {author} {\bibfnamefont {M.}~\bibnamefont
  {Gonzalez-Garcia}}, \bibinfo {author} {\bibfnamefont {A.}~\bibnamefont
  {Hernandez-Cabezudo}}, \bibinfo {author} {\bibfnamefont {M.}~\bibnamefont
  {Maltoni}}, \ and\ \bibinfo {author} {\bibfnamefont {T.}~\bibnamefont
  {Schwetz}},\ }\href {\doibase 10.1007/JHEP01(2019)106} {\bibfield  {journal}
  {\bibinfo  {journal} {JHEP}\ }\textbf {\bibinfo {volume} {01}},\ \bibinfo
  {pages} {106} (\bibinfo {year} {2019})},\ \Eprint
  {http://arxiv.org/abs/1811.05487} {arXiv:1811.05487 [hep-ph]} \BibitemShut
  {NoStop}%
\bibitem [{\citenamefont {Davidson}\ \emph {et~al.}(2008)\citenamefont
  {Davidson}, \citenamefont {Nardi},\ and\ \citenamefont
  {Nir}}]{Davidson:2008bu}%
  \BibitemOpen
  \bibfield  {author} {\bibinfo {author} {\bibfnamefont {S.}~\bibnamefont
  {Davidson}}, \bibinfo {author} {\bibfnamefont {E.}~\bibnamefont {Nardi}}, \
  and\ \bibinfo {author} {\bibfnamefont {Y.}~\bibnamefont {Nir}},\ }\href
  {\doibase 10.1016/j.physrep.2008.06.002} {\bibfield  {journal} {\bibinfo
  {journal} {Phys. Rept.}\ }\textbf {\bibinfo {volume} {466}},\ \bibinfo
  {pages} {105} (\bibinfo {year} {2008})},\ \Eprint
  {http://arxiv.org/abs/0802.2962} {arXiv:0802.2962 [hep-ph]} \BibitemShut
  {NoStop}%
\bibitem [{\citenamefont {Silveira}\ and\ \citenamefont
  {Zee}(1985)}]{Silveira:1985rk}%
  \BibitemOpen
  \bibfield  {author} {\bibinfo {author} {\bibfnamefont {V.}~\bibnamefont
  {Silveira}}\ and\ \bibinfo {author} {\bibfnamefont {A.}~\bibnamefont {Zee}},\
  }\href {\doibase 10.1016/0370-2693(85)90624-0} {\bibfield  {journal}
  {\bibinfo  {journal} {Phys. Lett.}\ }\textbf {\bibinfo {volume} {161B}},\
  \bibinfo {pages} {136} (\bibinfo {year} {1985})}\BibitemShut {NoStop}%
%%CITATION = PHLTA,161B,136;%%
\bibitem [{\citenamefont {McDonald}(1994)}]{McDonald:1993ex}%
  \BibitemOpen
  \bibfield  {author} {\bibinfo {author} {\bibfnamefont {J.}~\bibnamefont
  {McDonald}},\ }\href {\doibase 10.1103/PhysRevD.50.3637} {\bibfield
  {journal} {\bibinfo  {journal} {Phys. Rev.}\ }\textbf {\bibinfo {volume}
  {D50}},\ \bibinfo {pages} {3637} (\bibinfo {year} {1994})},\ \Eprint
  {http://arxiv.org/abs/hep-ph/0702143} {arXiv:hep-ph/0702143 [HEP-PH]}
  \BibitemShut {NoStop}%
%%CITATION = HEP-PH/0702143;%%
\bibitem [{\citenamefont {Burgess}\ \emph {et~al.}(2001)\citenamefont
  {Burgess}, \citenamefont {Pospelov},\ and\ \citenamefont {ter
  Veldhuis}}]{Burgess:2000yq}%
  \BibitemOpen
  \bibfield  {author} {\bibinfo {author} {\bibfnamefont {C.~P.}\ \bibnamefont
  {Burgess}}, \bibinfo {author} {\bibfnamefont {M.}~\bibnamefont {Pospelov}}, \
  and\ \bibinfo {author} {\bibfnamefont {T.}~\bibnamefont {ter Veldhuis}},\
  }\href {\doibase 10.1016/S0550-3213(01)00513-2} {\bibfield  {journal}
  {\bibinfo  {journal} {Nucl. Phys.}\ }\textbf {\bibinfo {volume} {B619}},\
  \bibinfo {pages} {709} (\bibinfo {year} {2001})},\ \Eprint
  {http://arxiv.org/abs/hep-ph/0011335} {arXiv:hep-ph/0011335 [hep-ph]}
  \BibitemShut {NoStop}%
%%CITATION = HEP-PH/0011335;%%
\bibitem [{\citenamefont {Patt}\ and\ \citenamefont
  {Wilczek}(2006)}]{Patt:2006fw}%
  \BibitemOpen
  \bibfield  {author} {\bibinfo {author} {\bibfnamefont {B.}~\bibnamefont
  {Patt}}\ and\ \bibinfo {author} {\bibfnamefont {F.}~\bibnamefont {Wilczek}},\
  }\href@noop {} {\  (\bibinfo {year} {2006})},\ \Eprint
  {http://arxiv.org/abs/hep-ph/0605188} {arXiv:hep-ph/0605188} \BibitemShut
  {NoStop}%
\bibitem [{\citenamefont {Cline}\ and\ \citenamefont
  {Kainulainen}(2013)}]{Cline:2012hg}%
  \BibitemOpen
  \bibfield  {author} {\bibinfo {author} {\bibfnamefont {J.~M.}\ \bibnamefont
  {Cline}}\ and\ \bibinfo {author} {\bibfnamefont {K.}~\bibnamefont
  {Kainulainen}},\ }\href {\doibase 10.1088/1475-7516/2013/01/012} {\bibfield
  {journal} {\bibinfo  {journal} {JCAP}\ }\textbf {\bibinfo {volume} {1301}},\
  \bibinfo {pages} {012} (\bibinfo {year} {2013})},\ \Eprint
  {http://arxiv.org/abs/1210.4196} {arXiv:1210.4196 [hep-ph]} \BibitemShut
  {NoStop}%
%%CITATION = ARXIV:1210.4196;%%
\bibitem [{\citenamefont {Cline}\ \emph {et~al.}(2013)\citenamefont {Cline},
  \citenamefont {Kainulainen}, \citenamefont {Scott},\ and\ \citenamefont
  {Weniger}}]{Cline:2013gha}%
  \BibitemOpen
  \bibfield  {author} {\bibinfo {author} {\bibfnamefont {J.~M.}\ \bibnamefont
  {Cline}}, \bibinfo {author} {\bibfnamefont {K.}~\bibnamefont {Kainulainen}},
  \bibinfo {author} {\bibfnamefont {P.}~\bibnamefont {Scott}}, \ and\ \bibinfo
  {author} {\bibfnamefont {C.}~\bibnamefont {Weniger}},\ }\href {\doibase
  10.1103/PhysRevD.88.055025} {\bibfield  {journal} {\bibinfo  {journal} {Phys.
  Rev. D}\ }\textbf {\bibinfo {volume} {88}},\ \bibinfo {pages} {055025}
  (\bibinfo {year} {2013})},\ \bibinfo {note} {[Erratum: Phys.Rev.D 92, 039906
  (2015)]},\ \Eprint {http://arxiv.org/abs/1306.4710} {arXiv:1306.4710
  [hep-ph]} \BibitemShut {NoStop}%
\bibitem [{\citenamefont {Athron}\ \emph {et~al.}(2017)\citenamefont {Athron}
  \emph {et~al.}}]{Athron:2017kgt}%
  \BibitemOpen
  \bibfield  {author} {\bibinfo {author} {\bibfnamefont {P.}~\bibnamefont
  {Athron}} \emph {et~al.} (\bibinfo {collaboration} {GAMBIT}),\ }\href
  {\doibase 10.1140/epjc/s10052-017-5113-1} {\bibfield  {journal} {\bibinfo
  {journal} {Eur. Phys. J. C}\ }\textbf {\bibinfo {volume} {77}},\ \bibinfo
  {pages} {568} (\bibinfo {year} {2017})},\ \Eprint
  {http://arxiv.org/abs/1705.07931} {arXiv:1705.07931 [hep-ph]} \BibitemShut
  {NoStop}%
\bibitem [{\citenamefont {Chanda}\ \emph {et~al.}(2020)\citenamefont {Chanda},
  \citenamefont {Hamdan},\ and\ \citenamefont {Unwin}}]{Chanda:2019xyl}%
  \BibitemOpen
  \bibfield  {author} {\bibinfo {author} {\bibfnamefont {P.}~\bibnamefont
  {Chanda}}, \bibinfo {author} {\bibfnamefont {S.}~\bibnamefont {Hamdan}}, \
  and\ \bibinfo {author} {\bibfnamefont {J.}~\bibnamefont {Unwin}},\ }\href
  {\doibase 10.1088/1475-7516/2020/01/034} {\bibfield  {journal} {\bibinfo
  {journal} {JCAP}\ }\textbf {\bibinfo {volume} {01}},\ \bibinfo {pages} {034}
  (\bibinfo {year} {2020})},\ \Eprint {http://arxiv.org/abs/1911.02616}
  {arXiv:1911.02616 [hep-ph]} \BibitemShut {NoStop}%
\bibitem [{\citenamefont {Lebedev}\ and\ \citenamefont
  {Toma}(2019)}]{Lebedev:2019ton}%
  \BibitemOpen
  \bibfield  {author} {\bibinfo {author} {\bibfnamefont {O.}~\bibnamefont
  {Lebedev}}\ and\ \bibinfo {author} {\bibfnamefont {T.}~\bibnamefont {Toma}},\
  }\href {\doibase 10.1016/j.physletb.2019.134961} {\bibfield  {journal}
  {\bibinfo  {journal} {Phys. Lett. B}\ }\textbf {\bibinfo {volume} {798}},\
  \bibinfo {pages} {134961} (\bibinfo {year} {2019})},\ \Eprint
  {http://arxiv.org/abs/1908.05491} {arXiv:1908.05491 [hep-ph]} \BibitemShut
  {NoStop}%
\bibitem [{\citenamefont {Tanabashi}\ \emph {et~al.}(2018)\citenamefont
  {Tanabashi} \emph {et~al.}}]{Tanabashi:2018oca}%
  \BibitemOpen
  \bibfield  {author} {\bibinfo {author} {\bibfnamefont {M.}~\bibnamefont
  {Tanabashi}} \emph {et~al.} (\bibinfo {collaboration} {Particle Data
  Group}),\ }\href {\doibase 10.1103/PhysRevD.98.030001} {\bibfield  {journal}
  {\bibinfo  {journal} {Phys. Rev. D}\ }\textbf {\bibinfo {volume} {98}},\
  \bibinfo {pages} {030001} (\bibinfo {year} {2018})}\BibitemShut {NoStop}%
\bibitem [{\citenamefont {Gondolo}\ and\ \citenamefont
  {Gelmini}(1991)}]{Gondolo:1990dk}%
  \BibitemOpen
  \bibfield  {author} {\bibinfo {author} {\bibfnamefont {P.}~\bibnamefont
  {Gondolo}}\ and\ \bibinfo {author} {\bibfnamefont {G.}~\bibnamefont
  {Gelmini}},\ }\href {\doibase 10.1016/0550-3213(91)90438-4} {\bibfield
  {journal} {\bibinfo  {journal} {Nucl. Phys.}\ }\textbf {\bibinfo {volume}
  {B360}},\ \bibinfo {pages} {145} (\bibinfo {year} {1991})}\BibitemShut
  {NoStop}%
%%CITATION = NUPHA,B360,145;%%
\bibitem [{\citenamefont {Akrami}\ \emph {et~al.}(2018)\citenamefont {Akrami}
  \emph {et~al.}}]{Akrami:2018odb}%
  \BibitemOpen
  \bibfield  {author} {\bibinfo {author} {\bibfnamefont {Y.}~\bibnamefont
  {Akrami}} \emph {et~al.} (\bibinfo {collaboration} {Planck}),\ }\href@noop {}
  {\  (\bibinfo {year} {2018})},\ \Eprint {http://arxiv.org/abs/1807.06211}
  {arXiv:1807.06211 [astro-ph.CO]} \BibitemShut {NoStop}%
%%CITATION = ARXIV:1807.06211;%%
\bibitem [{\citenamefont {Ade}\ \emph {et~al.}(2016)\citenamefont {Ade} \emph
  {et~al.}}]{Ade:2015lrj}%
  \BibitemOpen
  \bibfield  {author} {\bibinfo {author} {\bibfnamefont {P.}~\bibnamefont
  {Ade}} \emph {et~al.} (\bibinfo {collaboration} {Planck}),\ }\href {\doibase
  10.1051/0004-6361/201525898} {\bibfield  {journal} {\bibinfo  {journal}
  {Astron. Astrophys.}\ }\textbf {\bibinfo {volume} {594}},\ \bibinfo {pages}
  {A20} (\bibinfo {year} {2016})},\ \Eprint {http://arxiv.org/abs/1502.02114}
  {arXiv:1502.02114 [astro-ph.CO]} \BibitemShut {NoStop}%
\bibitem [{\citenamefont {Ade}\ \emph {et~al.}(2018)\citenamefont {Ade} \emph
  {et~al.}}]{Ade:2018gkx}%
  \BibitemOpen
  \bibfield  {author} {\bibinfo {author} {\bibfnamefont {P.~A.~R.}\
  \bibnamefont {Ade}} \emph {et~al.} (\bibinfo {collaboration} {BICEP2, Keck
  Array}),\ }\href {\doibase 10.1103/PhysRevLett.121.221301} {\bibfield
  {journal} {\bibinfo  {journal} {Phys. Rev. Lett.}\ }\textbf {\bibinfo
  {volume} {121}},\ \bibinfo {pages} {221301} (\bibinfo {year} {2018})},\
  \Eprint {http://arxiv.org/abs/1810.05216} {arXiv:1810.05216 [astro-ph.CO]}
  \BibitemShut {NoStop}%
%%CITATION = ARXIV:1810.05216;%%
\bibitem [{\citenamefont {Elahi}\ \emph {et~al.}(2015)\citenamefont {Elahi},
  \citenamefont {Kolda},\ and\ \citenamefont {Unwin}}]{Elahi:2014fsa}%
  \BibitemOpen
  \bibfield  {author} {\bibinfo {author} {\bibfnamefont {F.}~\bibnamefont
  {Elahi}}, \bibinfo {author} {\bibfnamefont {C.}~\bibnamefont {Kolda}}, \ and\
  \bibinfo {author} {\bibfnamefont {J.}~\bibnamefont {Unwin}},\ }\href
  {\doibase 10.1007/JHEP03(2015)048} {\bibfield  {journal} {\bibinfo  {journal}
  {JHEP}\ }\textbf {\bibinfo {volume} {03}},\ \bibinfo {pages} {048} (\bibinfo
  {year} {2015})},\ \Eprint {http://arxiv.org/abs/1410.6157} {arXiv:1410.6157
  [hep-ph]} \BibitemShut {NoStop}%
\bibitem [{\citenamefont {Chung}\ \emph {et~al.}(2005)\citenamefont {Chung},
  \citenamefont {Kolb}, \citenamefont {Riotto},\ and\ \citenamefont
  {Senatore}}]{Chung:2004nh}%
  \BibitemOpen
  \bibfield  {author} {\bibinfo {author} {\bibfnamefont {D.~J.~H.}\
  \bibnamefont {Chung}}, \bibinfo {author} {\bibfnamefont {E.~W.}\ \bibnamefont
  {Kolb}}, \bibinfo {author} {\bibfnamefont {A.}~\bibnamefont {Riotto}}, \ and\
  \bibinfo {author} {\bibfnamefont {L.}~\bibnamefont {Senatore}},\ }\href
  {\doibase 10.1103/PhysRevD.72.023511} {\bibfield  {journal} {\bibinfo
  {journal} {Phys. Rev.}\ }\textbf {\bibinfo {volume} {D72}},\ \bibinfo {pages}
  {023511} (\bibinfo {year} {2005})},\ \Eprint
  {http://arxiv.org/abs/astro-ph/0411468} {arXiv:astro-ph/0411468 [astro-ph]}
  \BibitemShut {NoStop}%
%%CITATION = ASTRO-PH/0411468;%%
\bibitem [{\citenamefont {Nurmi}\ \emph {et~al.}(2015)\citenamefont {Nurmi},
  \citenamefont {Tenkanen},\ and\ \citenamefont {Tuominen}}]{Nurmi:2015ema}%
  \BibitemOpen
  \bibfield  {author} {\bibinfo {author} {\bibfnamefont {S.}~\bibnamefont
  {Nurmi}}, \bibinfo {author} {\bibfnamefont {T.}~\bibnamefont {Tenkanen}}, \
  and\ \bibinfo {author} {\bibfnamefont {K.}~\bibnamefont {Tuominen}},\ }\href
  {\doibase 10.1088/1475-7516/2015/11/001} {\bibfield  {journal} {\bibinfo
  {journal} {JCAP}\ }\textbf {\bibinfo {volume} {11}},\ \bibinfo {pages} {001}
  (\bibinfo {year} {2015})},\ \Eprint {http://arxiv.org/abs/1506.04048}
  {arXiv:1506.04048 [astro-ph.CO]} \BibitemShut {NoStop}%
\bibitem [{\citenamefont {Markkanen}\ \emph {et~al.}(2018)\citenamefont
  {Markkanen}, \citenamefont {Rajantie},\ and\ \citenamefont
  {Tenkanen}}]{Markkanen:2018gcw}%
  \BibitemOpen
  \bibfield  {author} {\bibinfo {author} {\bibfnamefont {T.}~\bibnamefont
  {Markkanen}}, \bibinfo {author} {\bibfnamefont {A.}~\bibnamefont {Rajantie}},
  \ and\ \bibinfo {author} {\bibfnamefont {T.}~\bibnamefont {Tenkanen}},\
  }\href {\doibase 10.1103/PhysRevD.98.123532} {\bibfield  {journal} {\bibinfo
  {journal} {Phys. Rev. D}\ }\textbf {\bibinfo {volume} {98}},\ \bibinfo
  {pages} {123532} (\bibinfo {year} {2018})},\ \Eprint
  {http://arxiv.org/abs/1811.02586} {arXiv:1811.02586 [astro-ph.CO]}
  \BibitemShut {NoStop}%
\bibitem [{\citenamefont {Tenkanen}(2019)}]{Tenkanen:2019aij}%
  \BibitemOpen
  \bibfield  {author} {\bibinfo {author} {\bibfnamefont {T.}~\bibnamefont
  {Tenkanen}},\ }\href {\doibase 10.1103/PhysRevLett.123.061302} {\bibfield
  {journal} {\bibinfo  {journal} {Phys. Rev. Lett.}\ }\textbf {\bibinfo
  {volume} {123}},\ \bibinfo {pages} {061302} (\bibinfo {year} {2019})},\
  \Eprint {http://arxiv.org/abs/1905.01214} {arXiv:1905.01214 [astro-ph.CO]}
  \BibitemShut {NoStop}%
\end{thebibliography}%

\end{document}